\documentclass[iop,numberedappendix,appendixfloats]{emulateapj}

\usepackage{xspace,graphics,graphicx,longtable,amsmath}
\usepackage{subfigure}
\usepackage{appendix}
\usepackage{natbib,amsbsy}
\def \lya {Ly$\alpha$}
\def \mkms {{\rm \; km\;s^{-1}}}
\def \cmsq {{~\rm cm^{-2}}}
\def \ldla {$\lambda_\mathrm{obs}^\mathrm{DLA}$}
\def \zdla {$z_\mathrm{DLA}$}
\def \wlya {$W_\mathrm{Ly\alpha}$}
\def \npairs {40 }
\def \mrperp {R_{\perp}}
\def \rperp {$\mrperp$}
\def \nhi {$N_\mathrm{HI}$}

\def \vesc {$v_\mathrm{esc}$}
\def \avgW  {$\langle W \rangle$}
\def \avgWCIV {$\langle W_{1548} \rangle$}
\def \avgWSiII {$\langle W_{1526} \rangle$}
\def \avgWSiIV {$\langle W_{1393} \rangle$}
\def \avgWCII {$\langle W_{1334} \rangle$}
\def \avgWLya {$\langle W_\mathrm{Ly\alpha} \rangle$}

\newcommand{\msunyr}{~M_{\odot}~\rm yr^{-1}}
\newcommand{\msun}{~M_{\odot}}

\slugcomment{Submitted to ApJ} 

\shorttitle{}
\shortauthors{Rubin et al.}

\begin{document}

\title{Dissecting the Gaseous Halos of \lowercase{$z\sim2$} Damped \lya\ Systems with Close Quasar Pairs}
\author{Kate H. R. Rubin \altaffilmark{1,2}, Joseph F. Hennawi\altaffilmark{2}, J. Xavier Prochaska\altaffilmark{2,3}, 
Robert A. Simcoe\altaffilmark{4}, Adam Myers\altaffilmark{5}, Marie Wingyee Lau\altaffilmark{3}}
\altaffiltext{1}{Harvard-Smithsonian Center for Astrophysics, 60 Garden St, Cambridge, MA 02138, USA; krubin@cfa.harvard.edu}
\altaffiltext{2}{Max-Planck-Institut f\"ur Astronomie, K\"onigstuhl 17, 69117 Heidelberg, Germany}
\altaffiltext{3}{Department of Astronomy \& Astrophysics, UCO/Lick Observatory, University of California, 1156 High St, Santa Cruz, CA 95064, USA}
\altaffiltext{4}{MIT-Kavli Institute for Astrophysics and Space
  Research, Massachusetts Institute of Technology, Cambridge, MA
  02139, USA}
\altaffiltext{5}{Department of Physics and Astronomy, University of
  Wyoming, Laramie, WY 82072, USA}

\begin{abstract}
We use spectroscopy of close pairs of quasars to study diffuse gas in the circumgalactic medium (CGM) surrounding a sample of 40 Damped Ly$\alpha$ systems (DLAs).
The primary sightline in each quasar pair probes an intervening DLA in the redshift range $1.6 < z_{\rm DLA} < 3.6$, 
such that the second quasar sightline then probes \lya, \ion{C}{2}, \ion{Si}{2}, and \ion{C}{4} absorption 
in the CGM transverse to the DLA  to projected distances $\mrperp <300$ kpc.
Analysis of the \lya\  profiles in these CGM sightlines constrains the covering fraction ($f_C$) of optically thick \ion{H}{1} (having column density \nhi\ $> 10^{17.2}\cmsq$) to be $\gtrsim30\%$ within $\mrperp < 200$ kpc of DLAs.  Strong \ion{Si}{2} $\lambda 1526$  absorption with equivalent width 
 $W_{1526} > 0.2$ \AA\ occurs with an incidence $f_C(W_{1526} > 0.2~\rm \AA) = 20_{-8}^{+12}\%$ 
in the closest CGM sightlines
 (at $\mrperp < 100$ kpc), indicating
that low-ionization metal absorption associated with DLAs probes material within a physical distance
$R_{\rm 3D} \lesssim 30$ kpc.  However, we find that strong \ion{C}{4} $\lambda 1548$ absorption is ubiquitous in these environments ($f_C(W_{1548} > 0.2~\rm \AA) = 57_{-13}^{+12}\%$ within $\mrperp < 100$ kpc), and in addition exhibits a high degree of kinematic coherence on scales up to $\sim175$ kpc.  We infer that this high-ionization material  arises
predominantly in large, quiescent structures extending beyond the scale of the DLA host dark matter halos
rather than in ongoing galactic winds.  
The  \lya\ equivalent width in the DLA-CGM is  anticorrelated with \rperp\ at $> 98\%$ confidence,
suggesting that DLAs arise close to the centers of their host halos rather than on their outskirts.
Finally, the average \lya, \ion{C}{2} and \ion{C}{4}  equivalent widths measured as a function of \rperp\
are consistent with those measured around $z\sim2$ Lyman Break Galaxies.
Assuming that DLAs trace a galaxy population at lower masses and luminosities, this finding implies
that the absorption strength of cool circumgalactic material has a weak dependence on dark matter halo mass 
at $M_h \lesssim10^{12}\msun$.
\end{abstract}

\section{Introduction}\label{sec.intro}

%{\bf Still need to update according to X's Dropbox comments.}
Damped \lya\ absorbers (DLAs) tracing \ion{H}{1} with column density \nhi $\ge 2\times 10^{20}\cmsq$ have contained 
most of the neutral gas since $z\sim5$ \citep{Wolfe1986,Storrie-LombardiWolfe2000}.  The significant decline in the neutral gas mass density between $z\sim3.5$ 
and today, concomitant with the buildup of over half the present-day mass in stars \citep{HopkinsBeacom2006}, suggests that DLAs dominate
the reservoir of fuel for star formation over cosmic time \citep{Wolfe2005}.

However, efforts to establish the direct link between DLAs
and the sites of active, ongoing star formation %, even during the peak in cosmic star formation activity at $z\sim1-3$ \citep{HopkinsBeacom2006}
have met with only partial success.  
The shape of the conjugate \lya\ emission and absorption profiles observed 
`down the barrel' toward luminous Lyman Break Galaxies (LBGs) at $z>2$
suggests that their galactic disks 
have \ion{H}{1} surface densities exceeding the DLA threshold
 \citep{Pettini2002,Shapley2003,Steidel2010}.  
%% DONE: JFH Cite Pettini et al. CB58 paper here, also maybe Shapley 2003. 
On the other hand, many of the observational programs targeting 
%stellar or ionized gas 
emission from counterpart galaxies
close to quasar sightlines probing DLAs have 
%% DONE: JFH stellar or ionized gas emission is confusing. I would just say galaixes or DLA-galaxies. 
yielded non-detections.  To date, these efforts have revealed only $\sim10$ associated galaxies at $z\sim2$ 
\citep[e.g.][]{Moller2004,Fynbo2010,Peroux2011,Bouche2012,Krogager2012,Fynbo2013,Jorgenson2014}, the majority of 
which were selected for study based on their relatively high metallicity \citep[e.g.,][]{Moller2004,Fynbo2013}.
These counterparts are typically within $\lesssim2\arcsec$ ($\lesssim20$ kpc) of the background QSO \citep{Krogager2012}, 
and in cases for which robust photometry is possible have magnitudes $R\sim24-25.5$, or $\sim 0.4-1.5 L^*$ \citep{Reddy2008}.
%% Here using ndens = reddy_lumfunc(2.3, mag=24.0, lum_ratio=lr, absM=absM), etc.

Systems with lower metallicities than those targeted in 
the aforementioned imaging studies, however, likely trace a much fainter, less massive galaxy population. 
The tight positive correlation between DLA metallicities and both the kinematic width of unsaturated low-ionization metal absorption 
and the equivalent width ($W$) of saturated transitions (i.e., \ion{Si}{2} $\lambda 1526$) is strongly evocative of
the mass-metallicity relation established for galaxies from the local universe out to $z\sim2$ \citep{Tremonti2004,Erb2006a,Moller2013}, inspiring the inference that 
$W_{1526}$ traces the kinematics (and hence the mass) of DLA host dark matter halos 
\citep{Prochaska2008,Neeleman2013}.  
Indeed, searches for individual DLA hosts selected without regard for metallicity
have resulted in very few detections, and a
recent statistical study of the rest-frame UV flux arising within $\lesssim10$ kpc of DLAs demands that the vast majority of 
these systems are forming stars at a rate $\leq2\msunyr$ \citep{Fumagalli2014DLAb}.
Such stringent limits strongly suggest that DLAs must be hosted by low-luminosity ($\sim0.1L^*$), low-mass galaxies.

At the same time, however, DLA velocity widths are too large to arise from the rotational motions of individual dwarf systems \citep[$> 60\mkms$;][]{Prochaska1997,Wolfe2005}.  Studies of DLAs in early cosmological simulations invoked multiple neutral gas `clumps'
virialized within a single dark matter halo to satisfy this latter constraint \citep[e.g.,][]{Haehnelt1998,Nagamine2004,Pontzen2008}.
More recent work has suggested that some fraction of DLAs arise in cold, dense inflowing streams extending over 
many tens of kpc and  feeding star formation in 
a massive central galaxy \citep{Razoumov2008,Fumagalli2011,Cen2012}, 
%% NOT DONE - JFH Relevant paper by Erkal as well. 
%% KHRR -- could not find this issue mentioned in either of his papers
and/or that they trace wind material lofted away from galactic disks by star formation driven outflows \citep{Pontzen2008,Razoumov2008}.
We note, however, that none of these cosmological simulations have been able to match the  full DLA velocity width distribution  
(although see \citealt{Bird2014b} for recent success along these lines).
Moreover, a constraint on the bias of DLAs from measurement of their 
cross-correlation with \lya\ forest absorption implies they must also arise in dark matter
halos with masses exceeding $\sim10^{12}\msun$ \citep{Font-Ribera2012}.  As clustering measurements 
suggest that such massive halos 
host luminous LBGs at $z\sim2$ \citep{Adelberger2005,Conroy2008,Rakic2013} 
with star formation rates (SFR) of $\sim20-50\msunyr$ \citep{Erb2006b}, this finding lies in apparent conflict with the stringent limits 
imposed on DLA-galaxy UV luminosities by direct imaging studies \citep[e.g.,][]{Fumagalli2014DLAb}.

%% DONE: JFH Elaborate this point to emphasize tension. 

%% DONE: JFH You have not establshed yet that there is tension. 
%The tension between 
%the kinematics and bias 
%% large bias
%of DLAs and the faintness of their emission 
This tension
leaves open a number of questions regarding the nature of DLAs.  To date, few experiments have assessed the incidence of 
DLAs as a function of halo mass (\citealt{QPQ6}; hereafter QPQ6),
%% DONE: JFH Cite QPQ6 here, we considered DLA abundance as a function of halo mass
the spatial extent of DLAs \citep[e.g.,][]{Cooke2010}, or the location of these systems within 
their host halos (e.g., in extended streams or in compact, central galaxy disks).
%remain sparse.  %and yet have the potential to 
%establish the processes by which DLAs fuel cosmic star formation.
%[{\bf Need another sentence here.}]
In principle, however, empirical constraints on these quantities can directly relate 
the cold gas content of DLAs with the star-forming regions they will feed.

One avenue toward meeting this goal is the measurement of the cool hydrogen and metal content in the environments surrounding DLAs; i.e., the study of their circumgalactic medium (CGM).
The more diffuse material in these regions must likewise compose the fuel for star formation at later epochs, and 
is likely enriched by the large-scale outflows driven by current and past star formation in nearby galaxies \citep{Heckman1990,Veilleux2003,Veilleux2005}.  %Observations of this phenomenon around local starbursts indicate that it may push cool, metal-enriched gas and dust to distances at least $\sim10$ kpc from their point of origin.  
%In previously-studied samples, there is a clear dependence of CGM absorption on galactic environment! (at both $z\sim2$ and locally!)
%Studies of the cool hydrogen and metal content of the galactic environment at $z\sim2-3$ 
Studies leveraging spectroscopy of lensed QSOs have recently begun to provide measurements of the CGM close to a small sample of DLAs ($\sim7$) over $\lesssim10$ kpc scales \citep{Smette1995,Lopez2005,Monier2009,Cooke2010}, 
%% JFH Isn't this scale a whole lot smaller for lenses, like 1 kpc? Not clear
%% I would call that CGM. 10 kpc seems extreme. 
with
the vast majority of these systems manifest as DLAs toward only one of the QSO images.  
Adopting a model assuming that the \ion{H}{1} column declines exponentially with projected distance, \citet{Cooke2010} found typical scalelengths for \nhi\ of $0.2-2.6$ kpc for this sample.  Their analysis suggests that such scale lengths imply total DLA sizes of $\sim10$ kpc,
%These authors also address the issue of the spatial coherence of metal-line absorption for one of the systems in their sample, noting that \ion{Si}{2} $\lambda 1260$ absorption is detected only toward the QSO image exhibiting the DLA, whereas highly-ionized transitions are strong and have strikingly similar velocity structure toward both QSO images (separated by 2.7 kpc).
supporting a picture in which the neutral material has a highly localized, compact structure.
In one of the only studies of the spatial distribution of cool gas absorption around DLAs on scales larger than $\sim 10$ kpc, \citet{Ellison2007} reported on spectroscopy of a $z\sim3$ binary QSO separated by $\sim100$ kpc (also included in the present analysis),  %; SDSSJ1116+4118),
identifying a $z=2.66$ absorption system having \nhi $> 10^{20.1}\cmsq$ in both QSO sightlines. %(see also Figure~\ref{fig.nhi}a).
From comparison with the \ion{H}{1} distribution in cosmological `zoom-in' simulations of two $10^{11.8}\msun$ halos \citep{Razoumov2006}, both of which have DLA-absorbing material distributed on scales $\ll 100$ kpc, they conclude that the coincidence is most likely due to a structure hosting more than one massive galaxy.

However, most studies of the $z\sim2$ CGM to date  have focused on the areas surrounding strongly star-forming or AGN-dominated systems which are identified with relative ease in deep imaging and spectroscopic surveys.
LBGs,
%\citep[LBGs;][]{Simcoe2006,Steidel2010,Rudie2012,Crighton2013}.  These systems,
photometrically selected 
from deep near-UV and optical imaging as described in \citet{Steidel2003,Steidel2004} and \citet{Adelberger2004}, %and having SFR $\sim20-50\msunyr$ %\citep{Erb2006b}, %, dynamical masses $\sim7\times10^{10}\msun$ \citep{Erb2006c}, and 
are now known to be surrounded by an envelop of \ion{H}{1} which is optically thick (with \nhi\ $>10^{17.2}\cmsq$)
in $\sim30$\% of sightlines to projected distances of $\mrperp <200$ kpc \citep{Rudie2012,Crighton2013,Crighton2014}.  
More recent work taking advantage of a large sample of close pairs of luminous quasars \citep{Hennawi2006,Hennawi2010}
%% DONE: JFH Shen 2010 is not the right reference here. It should be Hennawi 2010, which I changed. 
has explored the gaseous environments of $z\sim2$ quasar-host galaxies, revealing a
$\gtrsim60\%$ incidence of optically thick, metal-enriched material out to $\mrperp < 300$ kpc, with enhanced \ion{H}{1} absorption extending to even larger scales ($> 1$ Mpc; \citealt{QPQ1,QPQ2,QPQ4,QPQ5} or QPQ5 hereafter; QPQ6).  
%% DONE: JFH Please cite QPQ1, QPQ2 and QPQ4 here as well. These points were all made there. 
Taken together, these studies demonstrate clear, qualitative differences between the \ion{H}{1} and metal absorption 
properties of material tracing the high-mass dark matter halos hosting high-redshift quasars (with masses $> 10^{12.5}\msun$; 
\citealt{Wild2008,White2012,Font-Ribera2013}), and the gas in halos of more modest masses hosting LBGs 
($\sim10^{11.6-12}\msun$; \citealt{Adelberger2005,Conroy2008,Rakic2013}).
%% DONE: JFH Add a sentence here citing the status of the simulations, i.e. there are now papers 
%% by Fumagalli et al. 2013, and Faucher-Giguere et al. 2013 comparing these measurements to 
%% simulations. This makes the case for studying a different mass range (i.e. DLAS stronger). 
Detailed studies of the CGM around $z<1$ systems with a broad range of properties
similarly suggest a trend of increasing \ion{H}{1} and low-ionization metal absorption strength 
with halo mass (\citealt{Churchill2013,Werk2014}; R. Bordoloi et al.\ in prep; \citealt{QPQ7}  or QPQ7 hereafter).  
Characterization of these quantities in DLA environments thus offers a point of comparison with 
magnitude-selected samples, providing the opportunity to 
differentiate based on the properties of this CGM material.  
There has in addition been significant recent progress toward predicting the properties 
of the CGM using cosmological `zoom-in' simulations \citep[e.g.,][]{Fumagalli2014,CAFG2014} with a particular focus 
 on developing these predictions over a broad range in halo mass.
Study of the diffuse gas surrounding DLAs will directly address 
the degree of metal enrichment due to the effects of stellar feedback acting from nearby star-forming regions, 
potentially providing the only constraint on feedback physics 
in the lowest-mass halos studied in these simulations.

Using a subset of the large sample of close pairs of $z\gtrsim2$ quasars referenced
 above \citep{Hennawi2006,Hennawi2010}, 
%% DONE: JFH Again replace with Hennawi 2010, not Shen 2010
we have searched each quasar sightline for instances of damped \lya\ absorption in the foreground of both of the paired quasars.
%% DONE: JFH Add one ``foreground of one of the quasars in the pair''.   
Here we report our measurements of the \lya\ and metal-line absorption strength 
and kinematics in the CGM out to $\mrperp < 300$ kpc around $40$ of these systems, obtained from spectroscopy of the secondary quasar in each pair.  
Our sample selection and dataset are described in \S\ref{sec.data}, and our methods for assessing CGM absorption are described in \S\ref{sec.specanalysis}.  We present our results in \S\ref{sec.results}, and discuss their implications for the nature of DLAs and their relationship to luminous galaxies in \S\ref{sec.discussion}.
We adopt a $\Lambda$CDM cosmology with
$\Omega_M = 0.26, \Omega_\Lambda = 0.74$, and $H_0 = 70~{\rm km~
  s^{-1}~Mpc^{-1}}$ throughout.

\section{Data and Sample Selection}\label{sec.data}
Our DLA sample is drawn from an ongoing survey to obtain medium-resolution spectroscopy of close quasar pairs at $z\gtrsim2$ (QPQ6).  These pairs were initially identified via data mining techniques from SDSS photometry \citep{Bovy2011,Bovy2012}.
Pair candidates were then observed with low-resolution spectrographs on a suite of 3.5-6.5m telescopes 
at the APO, KPNO, MMTO and Calar Alto Observatory
as described in \citet{Hennawi2006,Hennawi2010} to develop a sample of 
confirmed QSO pairs having transverse separations $< 300$ kpc and minimum redshifts $> 1.6$. 
%% DONE: JFH Please cite the Bovy, Hennawi et al. 2010, and Bovy, Myers,
%% Hennawi 2011 XD-QSO papers here for the selection algorithms. There is a standard blurb that you 
%% can steal from QPQ4. 
We subsequently obtained deep, medium- and high-resolution spectroscopy of many of these quasars using a range of instruments, including LBT/MODS, Gemini/GMOS, Magellan/MagE, Magellan/MIKE, Keck/ESI and Keck/LRIS.  These observations and the data reduction procedures are described  in detail in \S2.2 of QPQ6.   

Following the publication of QPQ6 we added observations of three additional pairs to this survey.
Two of these pairs were observed with the Magellan Echellette Spectrograph (MagE; \citealt{Marshall2008})
on the 6.5m Magellan Clay telescope on the nights of UT 2014 February 1-4.  These data were collected with the $0.7\arcsec$-wide 
slit, and thus have a spectral resolution $R = 4000$ and a wavelength coverage $3050 - 10300$ \AA.
A single additional pair was observed with the Echellette Spectrometer and Imager (ESI; \citealt{Sheinis2002}) on the 10m Keck 2 telescope on the night of UT 2014 February 5 with the $0.75\arcsec$ slit.  These data provide a spectral resolution $R=5000$ and wavelength coverage $4000 - 10100$ \AA.  
We reduced these MagE and ESI data following the same procedures listed in QPQ6, making use of 
custom software available in the public XIDL software package\footnote{www.ucolick.org/${\sim}$xavier/IDL}.
%% DONE: JFH I think you need to briefly discuss/summarize the resolution you are working at etc, or perhaps
%% just reference a table. 

We further supplemented this spectroscopic sample with high-S/N SDSS and BOSS spectra where available \citep{Abazajian2009,Ahn2012}.  
%% JFH Maybe clarify which data relase you are using here. 
In the following analysis,  we use the highest spectral resolution data at hand for 
targets which have been observed with more than one instrument, preferring MIKE, ESI or MagE data, 
%% DONE: JFH I think you didn't necessarily prefer ESI over MagE right? I'd write, preferring ESI or MagE when
%% available but selecting....
but selecting LRIS, GMOS, or SDSS/BOSS spectra (in order of preference) when echelle or echellette coverage is not available. 
Quasar redshifts are calculated as described in QPQ6, and have uncertainties in the range 
$\delta z_{\rm QSO} \sim 270-770\mkms$.
%% DONE: JFH quote a typical error here. 

We  performed a by-eye search of each spectrum for the signature of a DLA with \lya\ absorption blueward of the \lya\ emission line in the foreground quasar in each pair and redward of the Lyman limit at the redshift of the corresponding background quasar.  
%% JFH complementary --> corresponding
The redshift of each DLA was 
set by an approximate centroid
of the associated low-ionization metal absorption.  We fitted a model Voigt profile to the \ion{H}{1} absorption in each DLA using custom routines (x$\_$fitdla, also available in the XIDL software package), obtaining $N_\mathrm{HI}$ constraints %to within $\pm XX$ dex.  
with typical uncertainties of $\lesssim 0.20$ dex dominated by continuum error
and line blending \citep{Prochaska2003}.
We then expunged all systems having $N_\mathrm{HI} < 10^{20.1}~\rm cm^{-2}$ and lying within $\delta v < 5000 \mkms$ of the foreground quasar redshift.  This 
liberal $N_\mathrm{HI}$ limit (slightly lower than the limit defining DLAs, $N_\mathrm{HI} \ge 10^{20.3}~\rm cm^{-2}$) increases our sample size while still selecting systems which are predominantly neutral (\ion{H}{1}/H $\gtrsim90\%$; \citealt{ProchaskaWolfe1996}).
Finally, both quasar spectra in every pair probing a DLA were continuum normalized using custom software as described in QPQ6.  All pairs with CGM sightlines having $\rm S/N < 4~\AA^{-1}$ at the wavelength corresponding to the \lya\ transition at the DLA redshift (henceforth \ldla) were then eliminated from the sample.  This leaves a total of \npairs pairs probing DLAs and with spectral S/N sufficient to constrain the \lya\ absorption $W$ in the CGM sightline.  
For three pairs exhibiting DLAs toward both QSOs, the sightlines were assigned to the DLA and CGM samples arbitrarily, and were treated as single systems.
%% DONE: JFH Maybe clarify that they were not counted twice anywhere. 
%% KHRR -- am ignoring this for now; hope that's ok
%% JFH Well this is a subtle point. For the covering factor I think you should be double counting these
%% objects, but then strictly speaking they are not independent. Perhaps this is the easiest to just
%% consider all once. Alteratnviely, rather than arbitrarily, you could call the DLA sightline the one
%% with the higher column density. 
The instrumentation, spectral resolution, and date of observation for each  of these \npairs pairs is listed in Table~\ref{tab.dlas}.
Representative spectroscopy of the \ion{H}{1} and metal-line absorption for 
three of our sample DLAs (red) and the corresponding CGM sightlines (black) is shown in Figure~\ref{fig.showspecs}.
We show spectroscopy of the full sample of \npairs pairs in the Appendix.

One of the QSO pairs in our sample, SDSSJ1029+2623, with an apparent angular separation $\theta_\mathrm{obs} =22.5\arcsec$, is not a physical pair but rather two images of a lensed QSO at $z=2.197$.  This system was first reported in \citet{Inada2006} and further analyzed by \citet{Oguri2013}, who obtained the redshift of the lensing cluster $z_\mathrm{lens} = 0.584$.  We use the relation derived by \citet{Cooke2010} and presented in their Eqn.\ 5 to calculate the transverse distance between the two light paths at \zdla\ $=1.97830$.  This distance, $\mrperp = 7.49$ kpc, is assumed throughout this work, and makes this system the closest QSO `pair' in our sample.

\begin{figure*}%[ht]
\begin{center}
\includegraphics[angle=90,width=\textwidth]{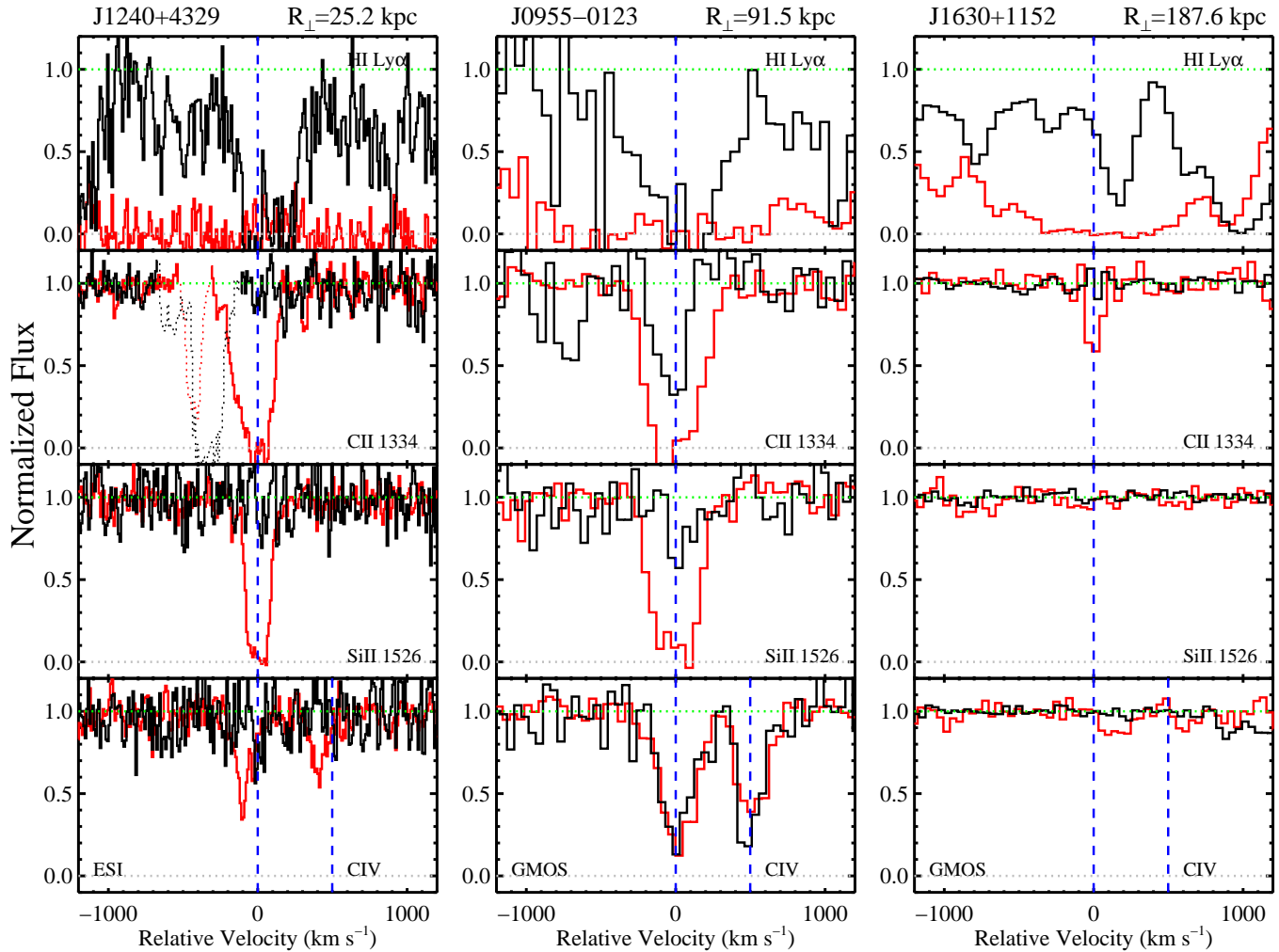}
\caption[]{QSO pair spectroscopy representative of our dataset.  
Each column shows the \lya, \ion{C}{2}, \ion{Si}{2}, and \ion{C}{4} absorption transitions due to a DLA (red histogram),
 with the QSO pair ID and its projected separation indicated above the topmost panel.  
The blue vertical dashed lines show the rest velocities of the corresponding transitions, with the velocities of both lines in the \ion{C}{4} doublet shown in the bottom panels.  
The black histogram shows the CGM absorption 
probed by the secondary QSO in each pair at the same redshift as the DLA.  The instrument used to obtain each spectrum is indicated in the \ion{C}{4} panels.  
Absorption due to material unrelated to the system at \zdla\ is shown with dotted histograms.
A subset of our 
sample was observed at high spectral resolution (FWHM$\lesssim50\mkms$) 
with, e.g., Magellan/MagE or Keck/ESI (left-hand column).
The majority of our pairs, however, were observed with medium-resolution setups (FWHM$\sim 125-180\mkms$; with Keck/LRIS, Gemini/GMOS, etc).  Similar figures showing each of the systems in our sample are provided in the Appendix.
\label{fig.showspecs}}
\end{center}
\end{figure*}
%% JFH I'm not sure what we gain by seeing the DLA damping trough in the top panels? 
%% KHRR -- ignoring this for now

\begin{figure*}%[ht]
%\begin{center}
\hskip 0.6in
\includegraphics[angle=0,width=4in]{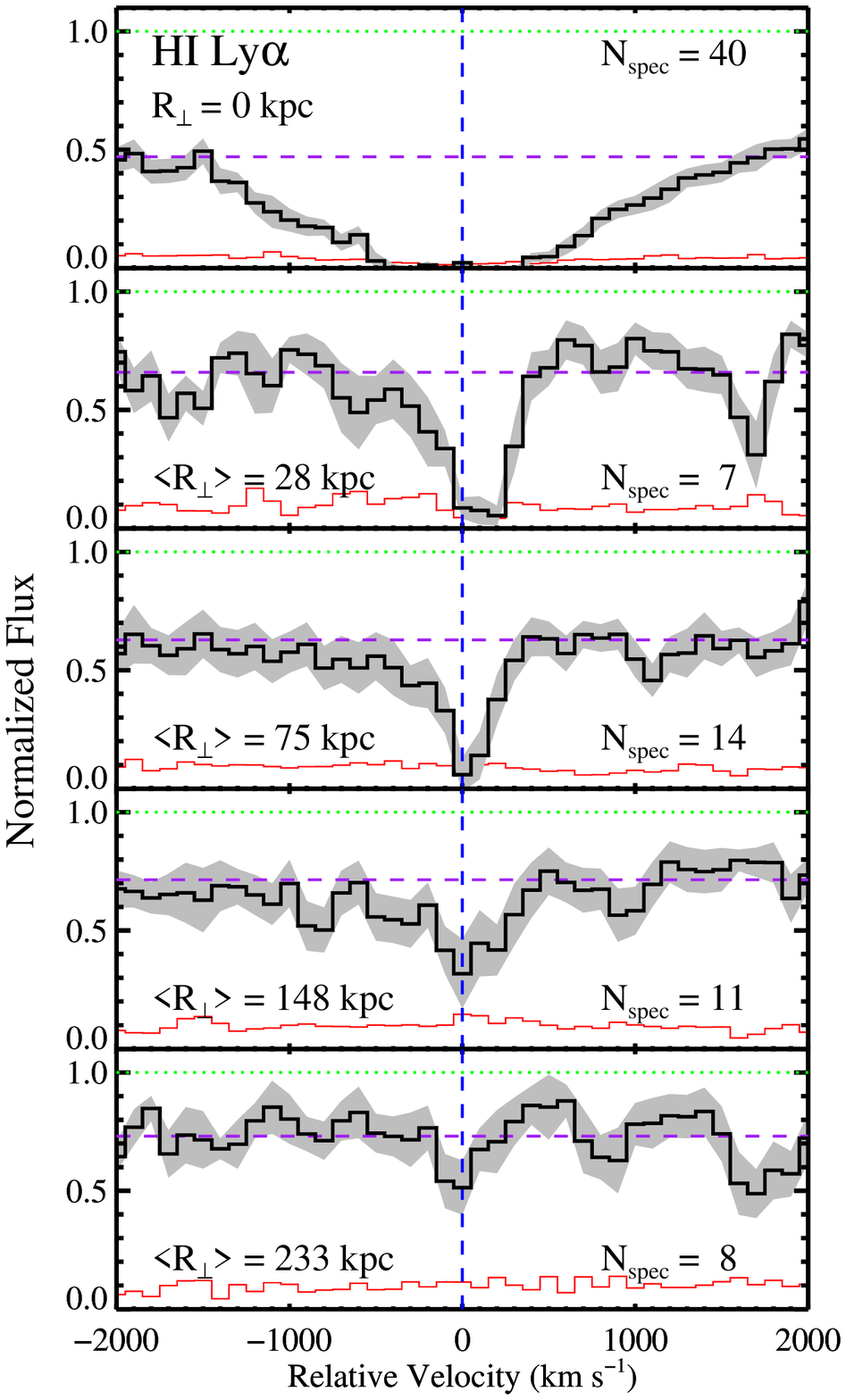}
\hskip -1in
\includegraphics[angle=0,width=4in]{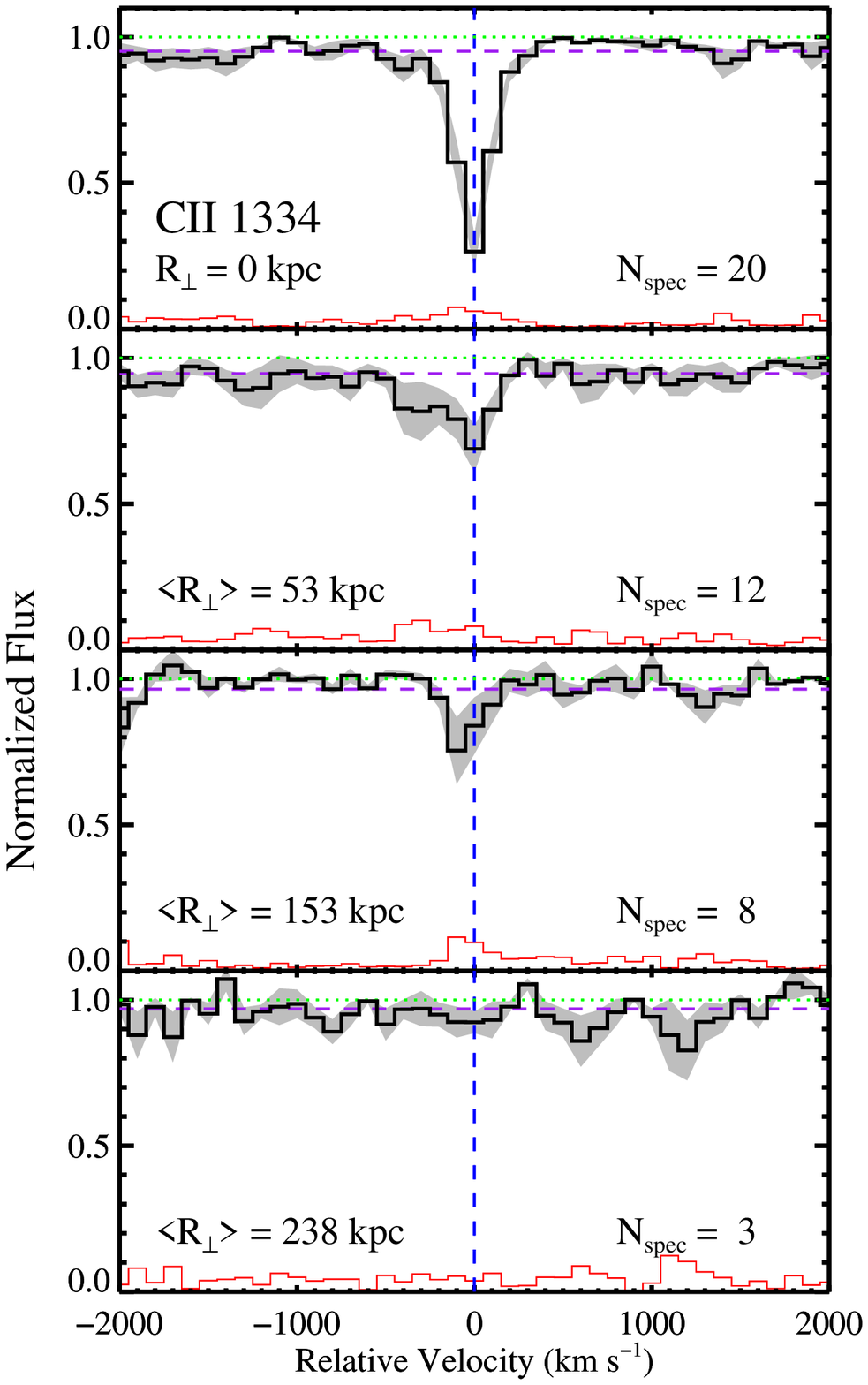}
\vskip -0.3in
\hskip 0.6in
\includegraphics[angle=0,width=4in]{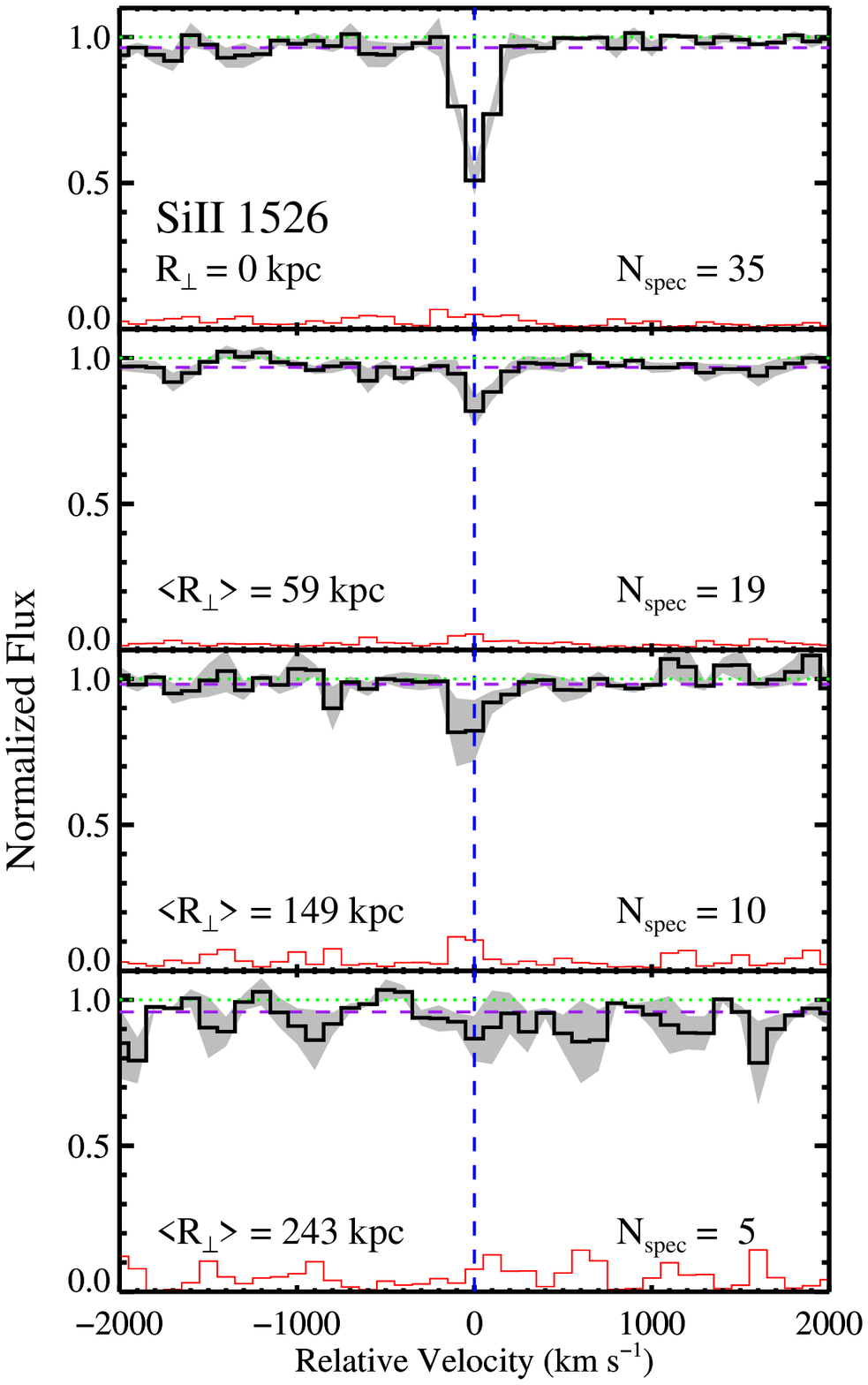}
\hskip -1in
\includegraphics[angle=0,width=4in]{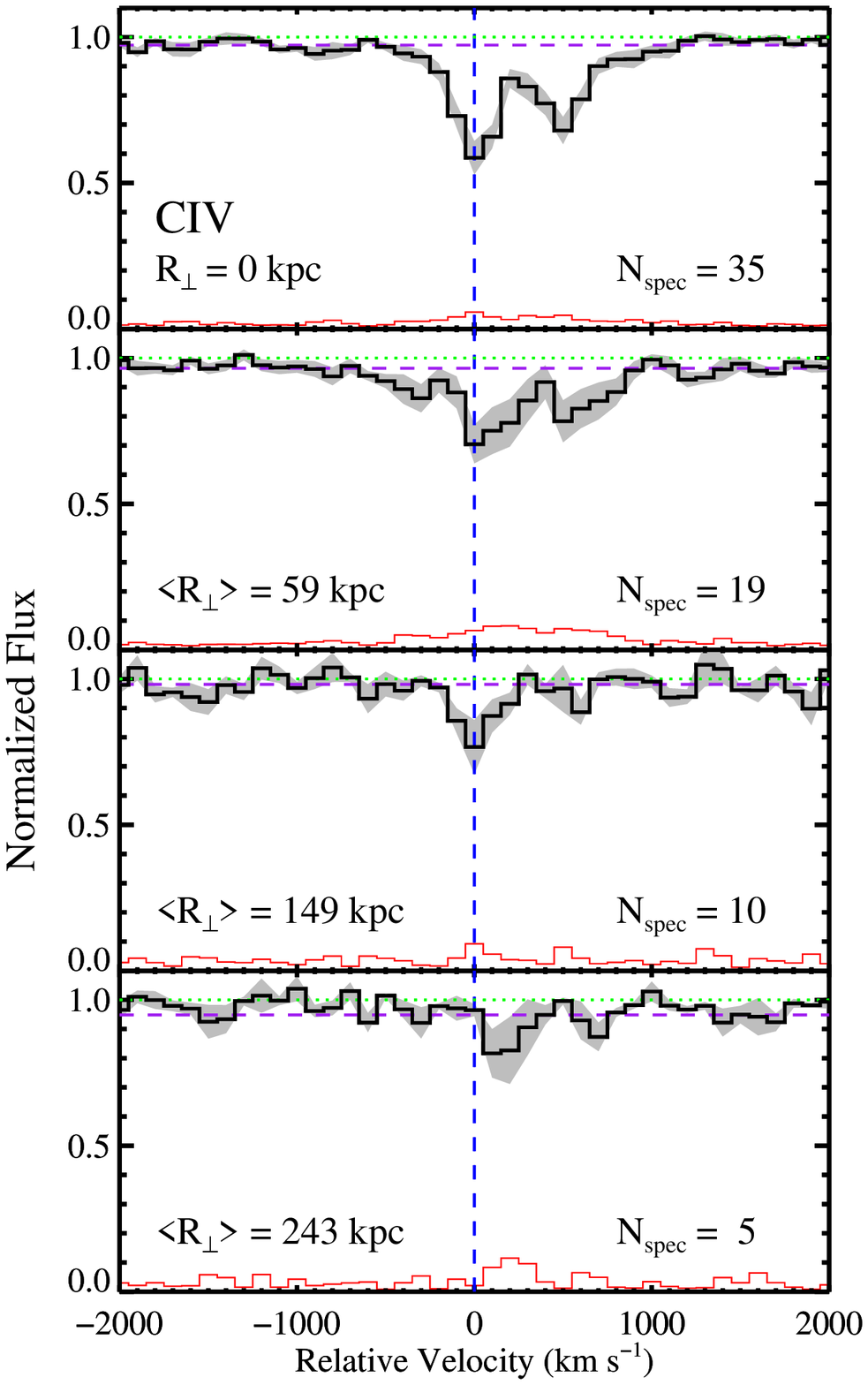}
\vskip -0.7in
\caption[]{Coadded spectra (black) of our DLA and CGM sightlines covering \lya\ (upper left), \ion{C}{2} 1334 (upper right), \ion{Si}{2} 1526 (lower left), and \ion{C}{4} 1548, 1550 (lower right).  The top panels for each transition show the coadds of our full sample of DLA spectra (at \rperp $= 0$ kpc), while the lower panels show coadds of CGM spectra divided into bins of increasing $R_{\perp}$ (indicated at the lower left).  The number of spectra included in each coadd is indicated at lower right.  The filled gray curves show the 
$\pm34$th-percentile interval for the flux values in our bootstrap sample in each pixel.
The red histogram shows this same $1\sigma$ error array.  The dashed purple curve shows a linear fit to the pseudo-continuum measured in the windows $-4000\mkms < \delta v < -3500\mkms$ and 
$3500\mkms < \delta v < 4000\mkms$.
\label{fig.stacks}}
%\end{center}
\end{figure*}
%% DONE: JFH Make this plot fill the page. It is too small. For the bootstrap errors, why don't we just
%% compute the 1-sigma region as the 16 and 84th precentils, and show that as light gray shaded. Then
%% we avoid all the craziness of these curves?  The point is when you plot many bootstraps, and don't
%% shade by density, one tends to overestimate the variance by eye, because all the lines overlap. 

\section{Line Profile Analysis}\label{sec.specanalysis}

%\subsection{Line Profile Analysis}

\subsection{Equivalent Widths}\label{sec.ews}

As a first step in our analysis, we measure boxcar $W$ of the 
\lya\ absorption at \zdla\ along the CGM sightline (\wlya).  
This measurement can be complicated by line-blending with intervening \lya\ forest absorbers, particularly for the subset of our sample observed at medium resolution and for systems at $z\gtrsim2.5$.  We search the spectral region within $\delta v \pm 600\mkms$ of \ldla\ by eye to find the ``single'' \ion{H}{1} absorption component closest to \ldla.   
%The identification of this component was also informed by the presence 
We choose the velocity range assigned to each absorber by hand, aiming to encompass the full velocity extent of this single component, and then perform a simple boxcar sum over this velocity range to obtain \wlya.  While this method is somewhat subjective, it at least provides a conservative lower bound on the \ion{H}{1} absorption strength along the sightline.

As we wish to characterize the possible enhancement of \lya\ absorption due to the presence of a nearby DLA, we 
must also assess the `background' strength of \lya\ absorption in randomly-selected regions of the intergalactic medium (IGM).  To do this, we draw from the 
much larger spectroscopic sample of QSO pairs described in QPQ6, which are similar in both S/N and spectral resolution 
to the present sample.  For each DLA, we select a QSO spectrum at random from all spectra 
for which \ldla\ is both redward of 
the Ly$\beta$ transition and blueward of the Ly$\alpha$ transition in the rest-frame of the QSO.
We also require that \ldla\ does not coincide with Ly$\alpha$ absorption from any close foreground QSO.
%$\lambda^{\rm QSO}_{\rm Ly\beta}$ and blueward of $\lambda^{\rm QSO}_{\rm Ly\alpha}$. 
We then search within a window $\delta v \pm 600\mkms$ around \ldla, again selecting the strongest \ion{H}{1} absorption 
component closest to \ldla.  This procedure results in a sampling of IGM \lya\ absorption with the same redshift distribution as our DLA sample, 
and which we verified to have a flat distribution of flux-weighted velocity centroids (measured relative to \ldla).  
%The distributions of \wlya\ in the CGM around DLAs and in this randomly-selected sample are shown in Figure~\ref{fig.wlya}a.

The strength of \ion{C}{2} 1334, \ion{Si}{2} 1526, and \ion{C}{4} 1548, 1550 absorption 
in the CGM sightlines was assessed in a similar manner, although the wavelength range chosen to span each  metal-line absorber was limited to within the velocity window previously determined for the corresponding \ion{H}{1} absorption.  We likewise measured the boxcar $W$ of each of these transitions in the DLAs themselves; here, because such metal-line absorption is nearly always strong and is used to determine \zdla, there is no ambiguity in line identification.  
These measurements, along with \wlya, are listed for each system in Table~\ref{tab.dlaews}.

%Finally, we calculate the velocity centroid of \ion{H}{1} and metal-line absorption components detected along the CGM sightlines.  We use a simple flux weighting (by $1 - f_i(\lambda)$, where $f_i (\lambda)$ is the continuum-normalized flux at each pixel) to determine the average observed-frame wavelength of each profile within the chosen velocity window.  
%EXPLAIN VELOCITY CENTROID MEASUREMENTS

\subsection{The Average CGM Absorption Strength}\label{sec.stacks}

We also wish to quantify the `average' absorption strength of the aforementioned transitions, both to track the change in the mean level of absorption with projected distance from DLAs and to facilitate comparisons with studies of the CGM around systems selected using complementary methods \citep[e.g., QPQ5;][]{Steidel2010,Crighton2013,Crighton2014,Turner2014}.  
%% DONE: JFH Cite Neil's papers here. 
To assess this average we coadd the continuum-normalized spectroscopy of our DLA and CGM sightlines covering 
\lya, \ion{C}{2}, \ion{Si}{4} $\lambda 1393$, \ion{Si}{2}, and \ion{C}{4} using the method described in \S3 of QPQ7.  
Briefly, we linearly interpolate each spectrum onto $100\mkms$-wide pixels, preserving the total normalized flux.  We then compute the average flux in each pixel, 
renormalize the resulting coadd via
a linear fit to the pseudo-continuum measured in the velocity windows $-4000\mkms < \delta v < -3500\mkms$ and 
$3500\mkms < \delta v < 4000\mkms$,
and measure the equivalent width (\avgW) of any resulting features.  
When generating coadded spectra for the metal-line transitions, we only include sightlines for which the transition of interest lies outside of the \lya\ forest (i.e., $\lambda > (1215.6701~\mathrm{\AA})(1 + z_{\rm QSO}) + 20~\rm \AA$).
For all of the transitions except for \ion{C}{4}, we use a relative velocity window $-500\mkms < \delta v < 500\mkms$ to measure \avgW.  
For the latter we choose a window $-500\mkms < \delta v < 249\mkms$, such that the red edge of the window falls at the midpoint between the two 
lines in the \ion{C}{4} $\lambda \lambda 1548.195, 1550.770$ doublet, and thus avoids absorption from the $\lambda = 1550.770$ \AA\ transition.
The uncertainty in this equivalent width is determined by generating 100 bootstrap samples of the spectra, coadding each sample in the same manner, and measuring the dispersion in the resultant mean absorption strength among these 100 samples.  The results of the coaddition of all DLA sightlines covering \lya, \ion{C}{2}, \ion{Si}{2}, and \ion{C}{4} are shown in Figure~\ref{fig.stacks}, along with coadds of the CGM sightlines sorted by \rperp.  \avgW\ measurements are listed in Table~\ref{tab.stacks} and discussed in \S\ref{sec.results}.

\subsection{$N_\mathrm{HI}$ along CGM Sightlines}

Finally, we make an effort to assess the column density of \ion{H}{1} detected at \ldla\ in each CGM sightline using 
detailed analysis of the line profile shapes and aided by our $W$ measurements of both \ion{H}{1} and metal absorption.  
 %Here we summarize our method, described in full in QPQ6.  
A significant fraction of the CGM \ion{H}{1} systems in our sample do not exhibit damping wings, and yet have \wlya\ values ($\sim 1-2$ \AA) placing them on the flat part of the curve of growth.  In such cases, 
the line profile shapes depend strongly on gas kinematics rather than gas column, and hence
%% DONE: JFH Remind people why, i.e. mention curve-of-growth. 
can only be used to 
place a lower limit on the amount of material along the sightline.  However, as we expect these limits to be constraining for galaxy formation models \citep[e.g.,][]{Shen2012,Fumagalli2014}, we move forward with the following approach (described in complete detail in QPQ6).  

For every spectrum with sufficient S/N ($> 9.5~\rm \AA^{-1}$ at \ldla), we first perform a by-eye Voigt profile fit to the \ion{H}{1} line profile using a custom interactive fitting code.  This code allows the user to adjust the model Doppler parameter and $N_\mathrm{HI}$ for an optimal fit.  In cases for which damping wings are clearly evident ($N_\mathrm{HI} \gtrsim 10^{19}\cmsq$), this method provides a relatively tight $N_\mathrm{HI}$ constraint with an uncertainty of $\approx 0.2$ dex.  
 For a handful of CGM sightline spectra obtained with MagE or ESI and which are sensitive to optically thin systems with \wlya\ $\lesssim 0.5$ \AA, we may likewise perform a direct comparison with model line profile shapes to obtain a tight column density constraint.
For the remaining systems, we use the absence of obvious damping wings to place an upper limit on the gas column by increasing the $N_\mathrm{HI}$ in the model profile until its shape is no longer consistent with the observed line. These measurements and limits are included in Table~\ref{tab.dlaews}.

% {\bf [Need to discuss uncertainties on $N_\mathrm{HI}$!]}

This latter category of absorbers makes up a substantial fraction of our sample, and we are therefore motivated to search for additional constraints on the gas column.  Systems having strong low-ionization metal absorption are very likely optically thick to ionizing radiation (with $N_\mathrm{HI} > 10^{17.2}\cmsq$; \citealt{Fumagalli2013}), and so we deem any system having low-ionization metal-line (\ion{C}{2} 1334 or \ion{Si}{2} 1526) $W> 0.3$ \AA\ `optically thick'.  
%% DONE: JFH Make it clear you also call it optically thick if you see evidence for damping wings. That is not
%% made explicit here. 
Systems with particularly high \wlya\ values ($> 1.8$ \AA, corresponding to a single absorber having a Doppler parameter of $40\mkms$ and $N_\mathrm{HI} > 10^{18.7}\cmsq$) or which exhibit damping wings are also assumed to be optically thick, even if the corresponding metal absorption is weak, blended with the \lya\ forest, or if we lack spectroscopic coverage of the  metal transitions of interest.  All other saturated systems which lack damping wings, however, are conservatively assumed to have `ambiguous' optical depths (below the previously-determined \nhi\ limit).

%{\bf [Need to show some HI profile fits.]}

\begin{figure}%[ht]
\begin{center}
\includegraphics[angle=0,width=\columnwidth]{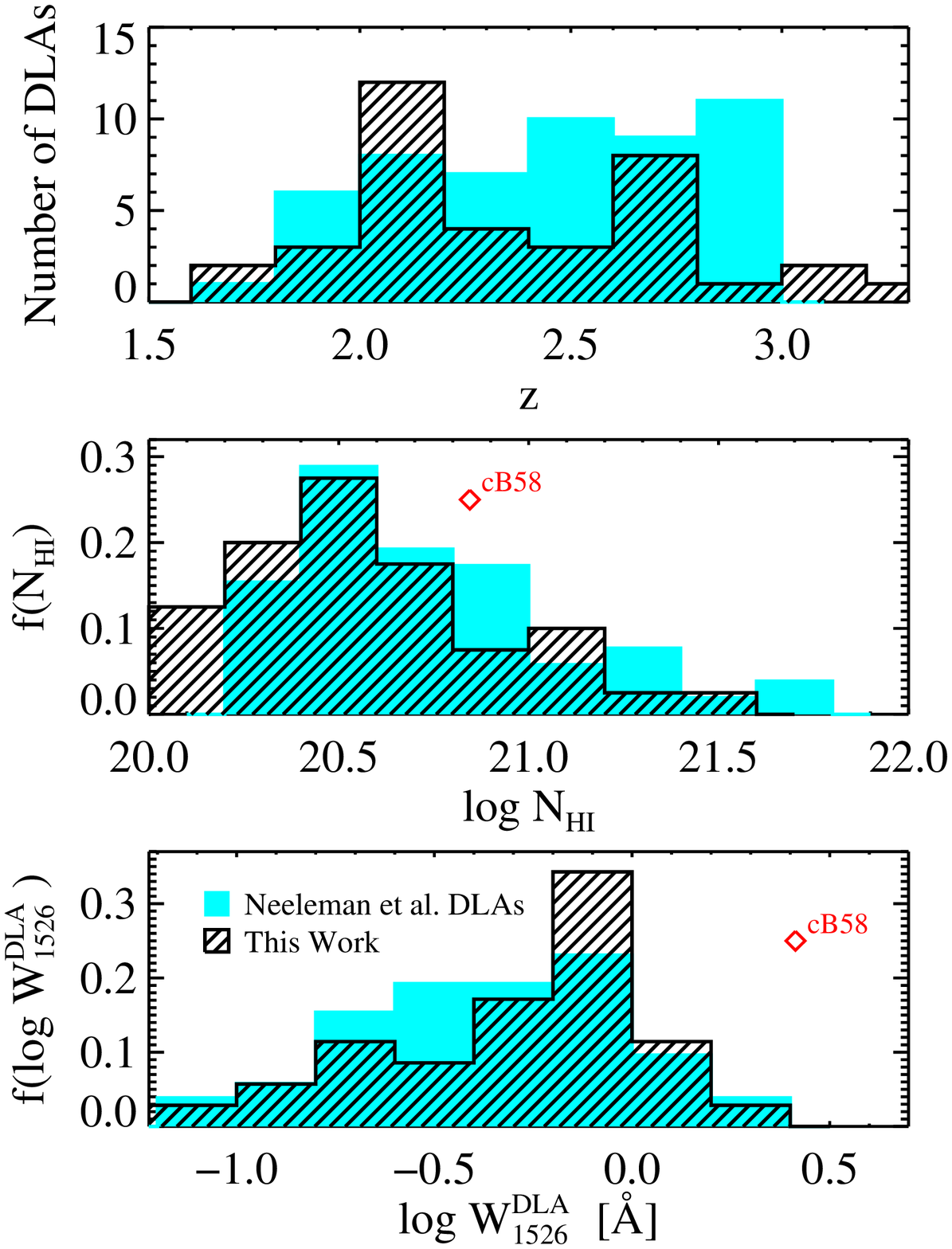}
\caption[]{\emph{Top:} Redshift distribution of our DLA sample (black).  The redshift distribution of a comparison sample of DLAs drawn from \citet{Neeleman2013} is shown in cyan.  The portion of the \citet{Neeleman2013} sample included here falls in the redshift range $1.5 < z < 3.0$ and has spectroscopic coverage of the \ion{Si}{2} $\lambda 1526$ transition.
\emph{Middle:}  The distribution of $\log N_\mathrm{HI}$ values for our DLA sample (black).
%DLAs for which we detect \ion{Si}{2} with a $>3\sigma$ significant EW are shown with filled black circles.  DLAs for which we place upper limits on $W_{1526}$ are shown with open circles and leftward arrows, and are located at their $3\sigma$ upper limit values along the x-axis.  
The cyan histogram shows the $N_\mathrm{HI}$ distribution for the \citet{Neeleman2013} subsample presented in the top panel.
The red diamond marks the $N_\mathrm{HI}$ measured toward the lensed LBG cB58 \citep{Pettini2002}.
\emph{Bottom:} The distribution of $\log W_{1526}^{\rm DLA}$ values 
among the 35 DLAs for which the \ion{Si}{2} transition does not 
fall in the Ly$\alpha$ forest of the corresponding QSO.  For one of these systems,
we do not detect significant \ion{Si}{2} absorption, but include the DLA in the bin containing the value of our $3\sigma$ upper limit on $\log W_{1526}^{\rm DLA}$.  The cyan histogram and red diamond show the $\log W_{1526}^{\rm DLA}$ distribution of the \citet{Neeleman2013} subsample and the $\log W_{1526}$ value measured toward cB58  as in the middle panel.
\label{fig.downthebarrel}}
\end{center}
\end{figure}

\begin{figure}%[ht]
\begin{center}
\vskip 0.2in
\includegraphics[angle=0,width=\columnwidth]{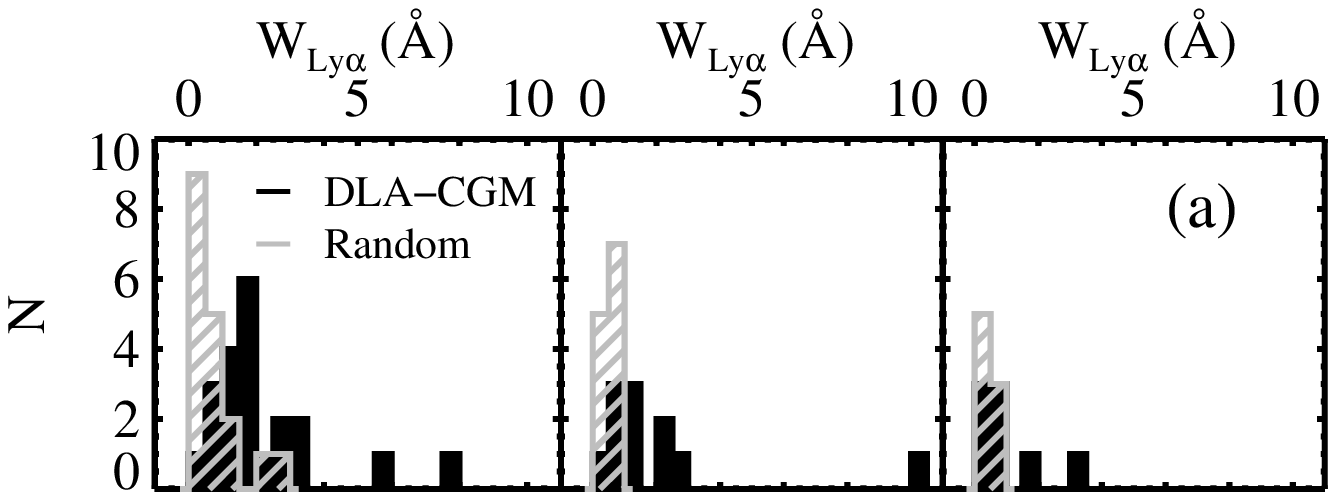}
\includegraphics[angle=0,width=\columnwidth]{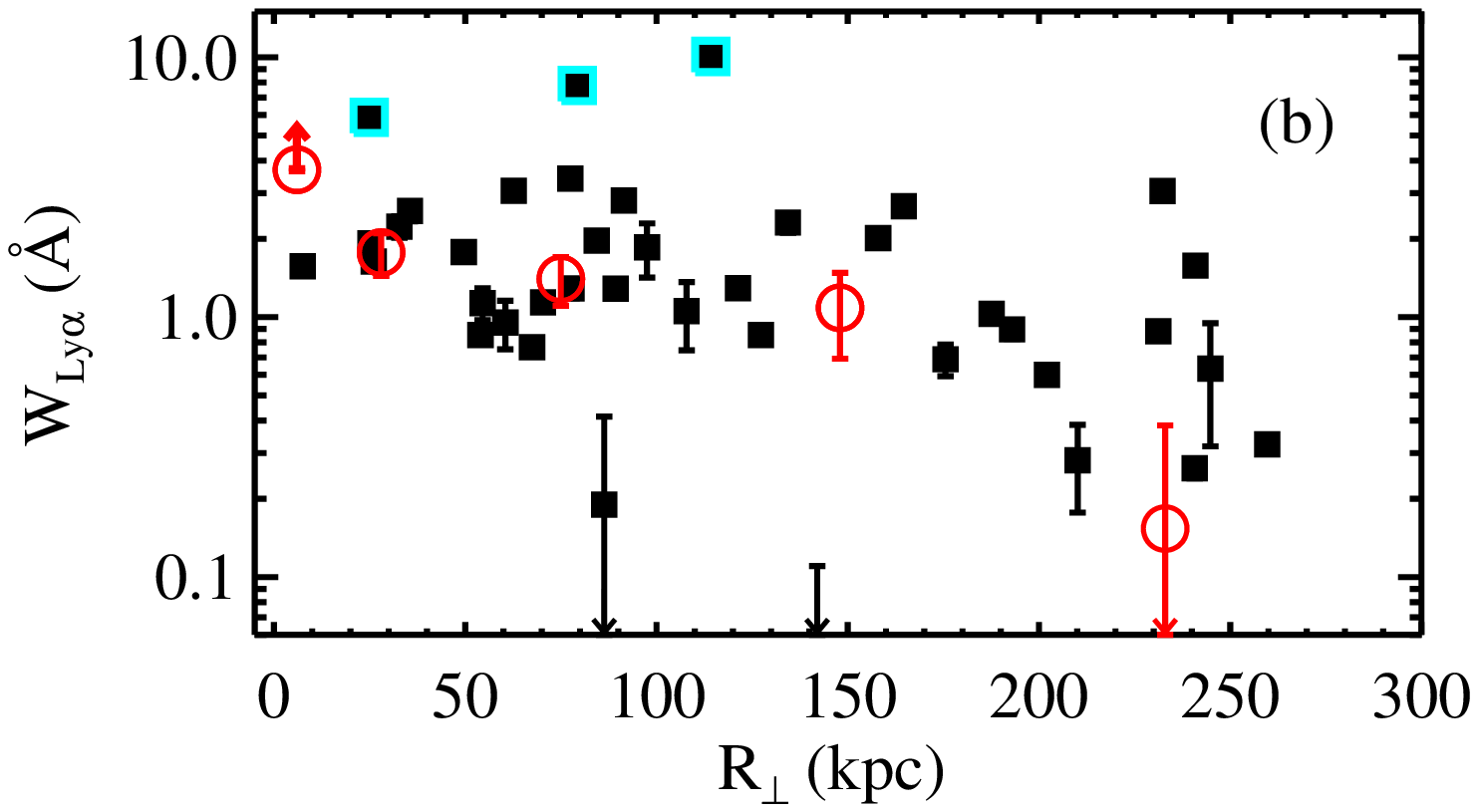}
\caption[]{\emph{(a)} Distribution of \wlya\ measured along CGM sightlines (filled black histograms) 
at $0~\mathrm{kpc} < \mrperp < 100$ kpc (left), 
$100~\mathrm{kpc} < \mrperp < 200$ kpc (middle), and $200~\mathrm{kpc} < \mrperp < 300$ kpc (right).  
The gray histograms show the distribution of \wlya\ measured in a randomly-selected sample of QSO spectra.  The distribution of 
redshifts at which these `control'
\wlya\ values are measured matches that of the DLA sample  in each panel.
\emph{(b)} \wlya\ measured along CGM sightlines as a function of projected separation (\rperp) from DLAs.
Downward arrows indicate \wlya\ values or $1\sigma$-uncertainty intervals which lie below the lower limit of the plot.  Points outlined in cyan indicate CGM sightlines with \nhi $\ge 10^{20.1}\cmsq$.  Red open circles show \avgWLya\ measured in coadded spectra of DLA sightlines (near $\mrperp = 0$ kpc) and of CGM sightlines divided into four bins in \rperp.
\label{fig.wlya}}
\end{center}
\end{figure}

\begin{figure}%[ht]
\begin{center}
\vskip 0.2in
\includegraphics[angle=0,width=\columnwidth]{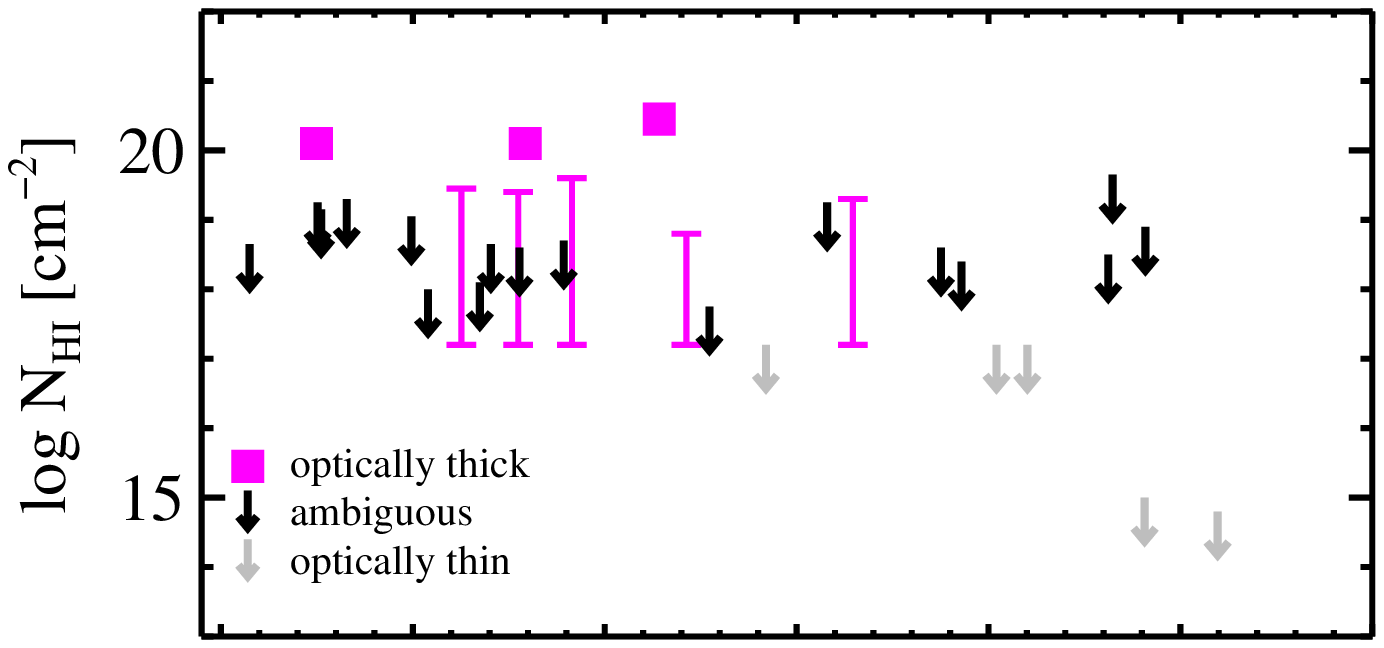}
\includegraphics[angle=0,width=\columnwidth]{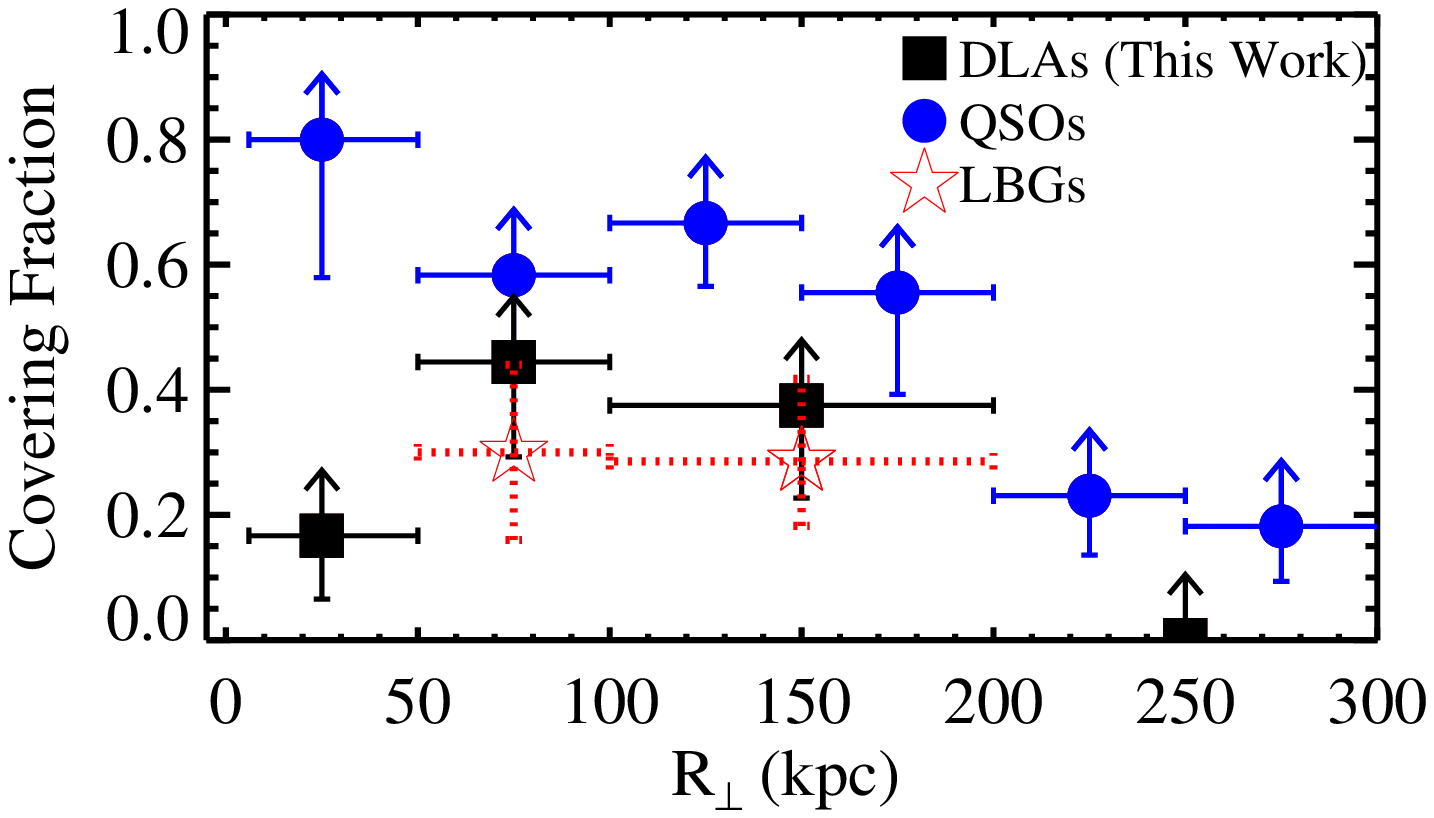}
\caption[]{\emph{Top:} Constraints on $\log N_\mathrm{HI}$ along CGM sightlines vs.\ \rperp.  Direct measurements of $N_\mathrm{HI}$ from Voigt-profile fitting 
to damping wings are shown with filled magenta squares. Systems deemed optically thick ($N_\mathrm{HI} > 10^{17.2}\cmsq$)
and for which our Voigt-profile fitting yields upper limits on \nhi\ are shown with magenta vertical bars.  
Black arrows show our upper limits
on $N_\mathrm{HI}$ for `ambiguous' systems, 
and gray arrows show upper limits for optically thin systems.  %, and optically thin systems are shown with open circles.  
\emph{Bottom:} Limits on the covering fraction of optically thick \ion{H}{1} (\nhi\ $>10^{17.2}\cmsq$) measured in several 
subsamples divided by projected separation from DLAs (\rperp; black).  The \rperp\ range for each subsample is 
 indicated with 
horizontal error bars.  Vertical error bars show the Wilson score 68\% confidence 
intervals.  The covering fraction of optically thick material measured around QSO host galaxies (QPQ5) and around LBGs \citep{Rudie2012} is shown in 
blue and red, respectively.  
\label{fig.nhi}}
\end{center}
\end{figure}

\section{Results}\label{sec.results}
\subsection{The DLA Sample in Context}

Figure~\ref{fig.downthebarrel} presents the redshift distribution of our DLA sample (top panel), along with 
the DLA \ion{H}{1} column densities (middle) and \ion{Si}{2} $\lambda 1526$ equivalent widths ($W_{1526}^{\rm DLA}$; bottom panel).  For comparison, we also show these properties for a random sample of DLAs drawn from the literature (\citealt{Neeleman2013}; cyan histograms) and selected solely on 
the basis of their \nhi.  %This comparison sample was selected in a heterogeneous manner (e.g., using a combination of criteria based on metal-line and \ion{H}{1} absorption strength), and thus is not a `random' sampling of damped \ion{H}{1} systems.  [{\bf Is this still true of the Neeleman sample?}]  
%% KHRR -- I think this is not a major issue
%% JFH I don't know why you are comparing to the NEeleman sample. We have an analytical formula for the
%% DLA column distribution from Xavier's work. The redshift distribution is also known, but integrating
%% this column density distribution over N_HI at varous redshifts. As for the W_1526, given that you are
%% using low-resolution data, I suppose that SDSS is just as valid a comparison. It is odd thta the 
%% Neeleman sample is the same size as yours. 
The kinematics and metal abundances of this comparison sample have been carefully analyzed in previous work, providing a rich set of ancillary measurements which will aid in later discussion.  

The median \zdla\ of our sample is 2.199,  similar to the mean redshift of the sample of LBGs discussed in \citet{Rudie2012} with $\langle z\rangle \sim2.3$.  Furthermore, both the $N_\mathrm{HI}$ and $W_{1526}^{\rm DLA}$ distributions of our DLA sample are similar to those in \citet{Neeleman2013}, although our $N_\mathrm{HI}$ distribution extends to slightly lower values due to our liberal DLA selection criterion.
%On the other hand, the $W_{1526}$ values measured for our sample are larger overall than those in the %`comparison' DLA sample.  
%% DONE: JFH This does not really look statistically significant? How different are the median values?
%This may be driven by the lower median \zdla\ of our DLAs: \citet{Neeleman2013} has demonstrated that DLA %$W_{1526}$ evolves to higher values at later times.   
The red diamond indicates $W_{1526}$ and $N_\mathrm{HI}$ measured `down the barrel' in high-resolution spectroscopy of the lensed LBG cB58 \citep{Pettini2002}.  The metal-line absorption observed toward the star-forming regions of LBGs have been shown to trace the kinematics of cool gas outflows \citep[e.g.,][]{Pettini2002,Shapley2003,Steidel2010}, and 
%% DONE: JFH Cite shapley et al. 2003 and Pettini paper here as well. 
these outflows may be driving the exceptionally large $W_{1526}$ observed along this unique sightline.  On the other hand, \citet{Prochaska2008} and \citet{Neeleman2013} have invoked the tight relationship between DLA metallicity and $W_{1526}$ to suggest that $W_{1526}$ traces the kinematics of a DLA's host dark matter halo, with larger $W_{1526}$ arising in more massive systems.  These issues will be discussed further in Section~\ref{sec.comparison}.

\subsection{\ion{H}{1} Absorption in DLA Environments}\label{sec.results_hi}

Here we present our measurements of the \ion{H}{1} absorption strength as a function of projected distance from DLA-host galaxies. 
The solid black histograms in Figure~\ref{fig.wlya}a show the distribution of \wlya\ in the DLA-CGM, divided into 3 bins according to \rperp\ ($\mrperp < 100$ kpc, $100~\mathrm{kpc} < \mrperp < 200$ kpc, etc.).  The distribution of \wlya\ in randomly-selected control sightlines, measured as described in \S\ref{sec.ews}, is shown in gray.  
The CGM \wlya\ distributions are skewed to higher equivalent widths relative to the control distributions in every \rperp\ bin.  
A Kolmogorov-Smirnov test indicates a very low probability that the control and CGM distributions are drawn from 
the same parent population in both the inner  ($\mrperp < 100$ kpc; $P=0.0007$) and middle
($100~\mathrm{kpc} < \mrperp < 200$ kpc; $P=0.0009$) bins.  The control 
and CGM \wlya\ distributions in the outer-most ($200~\mathrm{kpc} < \mrperp < 300$ kpc) bin, however, 
are relatively likely to have been drawn from the same parent population ($P=0.19$).  These probabilities 
point to a significant enhancement in \lya\ absorption within $\mrperp < 200$ kpc and $\delta v \pm 600\mkms$ of DLAs.

Figure~\ref{fig.wlya}b again shows our assessment of \wlya\ in each CGM sightline, 
now vs.\ \rperp.
%measured in the single strongest \ion{H}{1} component closest to %(and within $\pm 600\mkms$ of) \zdla.  
The three CGM systems exhibiting exceptionally strong \ion{H}{1} absorption (with \nhi\ $\ge 10^{20.1}\cmsq$) are 
%% DONE: JFH highlighted in cyan
highlighted in cyan, and are discussed in more detail below.  Almost every remaining sightline probes absorption stronger than \wlya\ $> 0.6$ \AA\ to nearly $\mrperp \sim 200$ kpc.  
The two-sided probability that \wlya\ is uncorrelated with \rperp\ indicated by Kendall's $\tau$ rank 
correlation test is $P=0.01$ with $\tau = -0.27$, bolstering our finding that \lya\ absorption is elevated 
significantly above the level in the ambient IGM close to DLAs.
The open red circles show our measurements of \avgWLya\ as described in \S\ref{sec.stacks}, and are similarly indicative of strong absorption extending to $\sim200$ kpc.  
%% DONE: JFH There should never be any upper limits in the upper panel of
%% Figure 4. The EW is always well defined in your stacks, it is just
%% an area under a curve. For individual measuerments, if the EW
%% widths are low such that the errors are large, then you could plot
%% these points with an error bar or just highlight that they thave larger errors somehow
%% with the symbol, but there is absolutely no process censoring these measurements. 
A Kendall's $\tau$ rank correlation test of the two-sided probability that \avgWLya\  is uncorrelated with \rperp\ yields $P=0.04$, demonstrating that 
the apparent decline in \avgWLya\ with \rperp\ is statistically significant (i.e., we reject a lack of correlation in favor of an anticorrelation with $98\%$ confidence).
We also include \avgWLya\ measured in the coadd of all DLA sightlines at $\mrperp = 6$ kpc in this panel.  The method we use to assess the continuum level in coadded spectra underestimates the true continuum in this case, as the \lya\ damping wings of DLAs extend well beyond $|\delta v| > 3500\mkms$.  Our value of \avgWLya\ therefore provides a lower bound on the average \lya\ absorption due to the DLAs themselves, and is shown here as a lower limit.  When this value is included in the Kendall's $\tau$ rank correlation test described above, the probability of no correlation decreases to $P=0.01$.
%{\bf Need some assessment of \wlya\ in random locations.}
%% NOT DONE: JFH I don't see why the inner bin is a lower limit here. I get that
%% you have systematics, but you have a column density fit to each
%% DLA. Why don't you just stack those, and then there are no
%% systematics. The EW of DLAs is not very informative anyway, but I guess it is good to have it here.
%% KHRR: I think this doesn't really matter in the end.

 Figure~\ref{fig.nhi} (top) shows our constraints on \nhi, with systems having 
\nhi $> 10^{17.2}\cmsq$ in magenta, with sightlines for which we place an `ambiguous' upper limit on \nhi\ in black, and with optically thin sightlines in gray. 
%For 3 of our CGM sightlines, the presence of damping wings allows us to measure \nhi\ directly (solid squares).  
Most of the CGM sightlines in our sample do not satisfy our DLA \nhi\ criterion, consistent with previous findings suggesting that DLAs have 
a covering fraction $f_C < 1$ for $\mrperp \gtrsim5$ kpc
%sizes $\lesssim 5$ kpc 
\citep{Cooke2010}.  However, three of these systems (J1026+0629 at \rperp\ $=79.3$ kpc,  J1116+4118 at \rperp\ $= 114$ kpc, and J1240+4329 at \rperp $=25$ kpc) have CGM \nhi\ $\ge 10^{20.1}\cmsq$ (solid magenta squares).  \citet{Ellison2007}, reporting on the J1116+4118 system, suggested that this QSO pair probes a relatively overdense environment, and the apparent paucity of such systems in our dataset lends further support to this interpretation.
Overall, our measurements and limits are indicative of a $\gtrsim30\%$  incidence of optically thick (\nhi $> 10^{17.2}~\rm cm^{-2}$) \ion{H}{1} 
%% DONE: JFH Reiterate optically thick means NHI > 17.2
out to \rperp\ $\sim200$ kpc.
%but we are not sensitive to changes in optical depth within this radius. 
%% DONE: JFH I don't know what you mean by ``we are not senstive to changes in optical depth within this
%% radius. That is obvious given that the statistic you consider is cumulative and rather instead
%% just causes confusion here. 
 It is only beyond \rperp\ $\gtrsim 200$ kpc that we may confidently rule out the presence of optically thick material in a handful of cases.  

We estimate a lower limit on the covering fraction of \nhi\ $> 10^{17.2}\cmsq$ material in several \rperp\ bins by dividing the number of \emph{bona fide} optically thick systems by the total number of sightlines in each bin.  These estimates are shown with black squares in Figure~\ref{fig.nhi} (bottom), with the horizontal error bars indicating the bin widths.  
%% DONE (RIGHT?): JFH These bins should be logarithmically spaced. 
The vertical error bars show the 68\% confidence Wilson score intervals.  We measure a covering fraction $f_C \sim 20-40\%$ extending to \rperp\ $\sim 200$ kpc, with our uncertainty intervals indicating $f_C$ is at least $\gtrsim30\%$ at 
$50~\mathrm{kpc} < \mrperp< 100$ kpc.  The true covering fraction may be significantly higher than these estimates due to the preponderance of sightlines with `ambiguous' 
\nhi\ absorption; however, 
the measured incidence is fully consistent with the estimate of the incidence of optically thick \ion{H}{1} in the CGM around LBGs from \citet{Rudie2012}, shown with red stars.  
 Measurements of \nhi\ in the host halos of massive QSOs, however, are suggestive of a higher $f_C$ 
in such environments (QPQ5; blue filled circles).  Although our limits on $f_C$ cannot formally rule out consistency with these constraints, the incidence of optically thick systems in the present study and in QPQ5 could be brought into agreement only if it was found that our dataset is significantly less complete for optically thick systems than that of QPQ5.  Because these datasets are of very similar quality and fidelity, we consider such a discrepancy unlikely.
%% DONE: JFH Clarify what you mean here, i.e. since you are comparing lower limits you cannot rule out 
%% consistency with QPQ, however this would require that you were missing a large number of optically
%% thick systems and QPQ is not, which seems unlikely. 

\begin{figure}%[ht]
\begin{center}
\vskip 0.2in
\includegraphics[angle=0,width=\columnwidth]{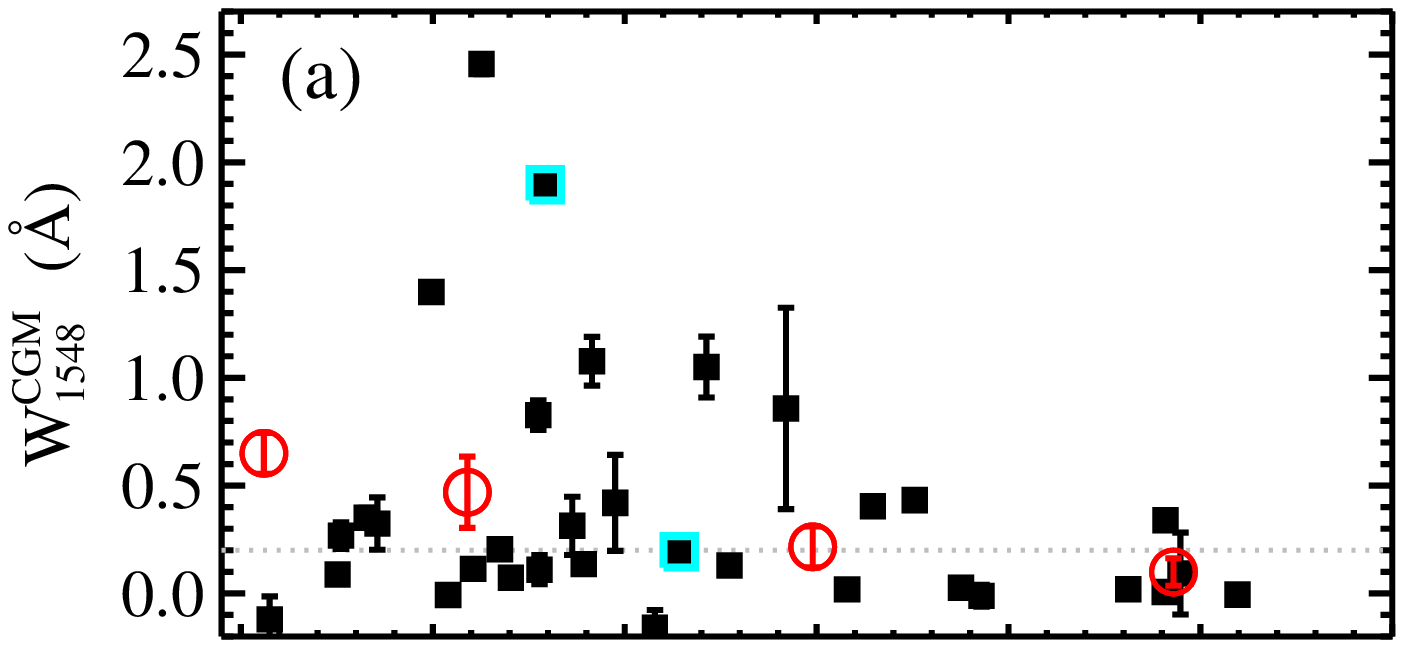}
\includegraphics[angle=0,width=\columnwidth]{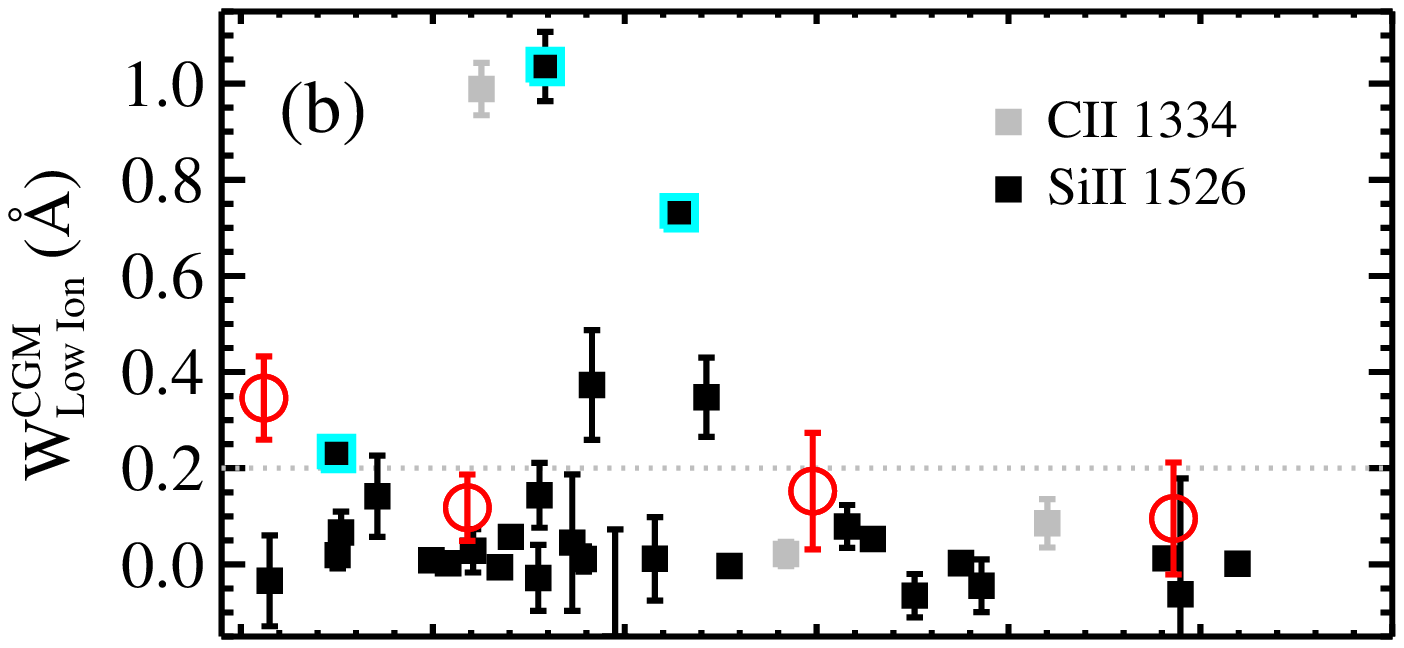}
\includegraphics[angle=0,width=\columnwidth]{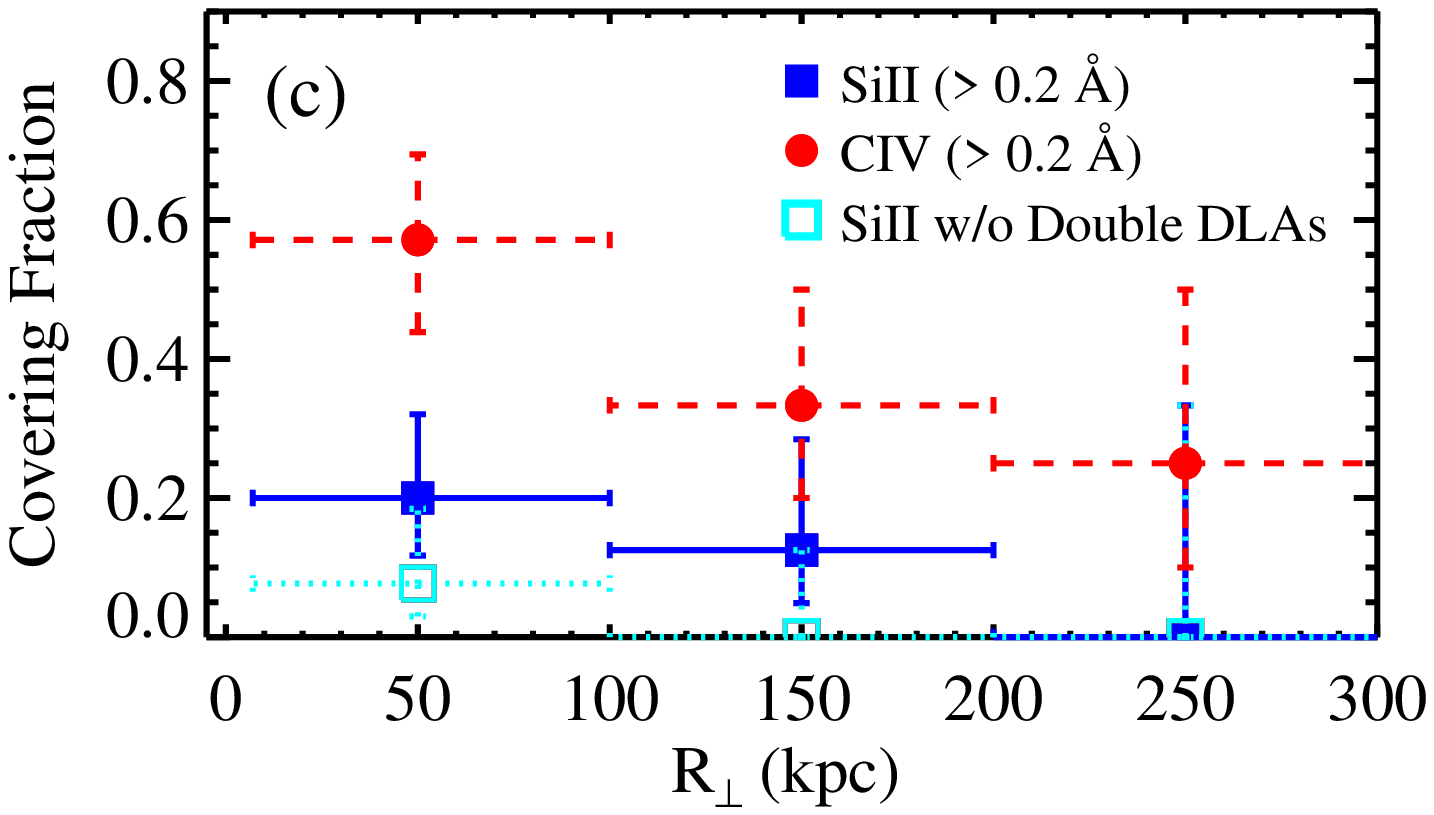}
\caption[]{\emph{(a)}  $W_{1548}^{\rm CGM}$ vs.\ \rperp measured in the CGM around DLAs (filled black squares).   %The points outlined in cyan show $W_{1548}$ measured in 
%sightlines distributed evenly across a 1 Mpc box centered on the Eris simulation \citep{Shen2012}.  
Points outlined in cyan indicate CGM sightlines with $N_\mathrm{HI} \ge 10^{20.1}\rm cm^{-2}$.
Red open circles show \avgWCIV\ measured in coadded DLA sightlines (near $\mrperp = 0$ kpc) and CGM sightlines divided into 3 subsamples according to $R_{\perp}$.
The horizontal dotted gray line indicates our `strong line' limit of 0.2 \AA.
%The solid and dashed 
%red lines show the median and $\pm25$th-percentile EW values measured in sightlines distributed evenly across a 1 Mpc box centered on the Eris %simulation \citep{Shen2012}.
\emph{(b)} $W_{1526}^{\rm CGM}$ vs.\ \rperp (black).  For CGM sightlines which lack spectroscopic coverage of \ion{Si}{2}, 
%, or for 
%which \ion{C}{2} falls within the \lya\ forest, 
$W_{1334}^{\rm CGM}$ is instead shown in gray (after rescaling as described in the text).  %Cyan points show $W_{1334}$ measured 
Red open circles, cyan outlines, and the dotted gray line are as described for panel (\emph{a}).
\emph{(c)} The fraction of CGM systems having $W^{\rm CGM} > 0.2$ \AA\ within 3 subsamples divided according to $R_{\perp}$ for \ion{Si}{2} 1526 (blue) and \ion{C}{4} (red).  The \rperp\ range for each subsample is shown with horizontal error bars, and the vertical error bars  indicate the Wilson score $68\%$ confidence intervals.  Cyan squares show the covering fraction of $W^{\rm CGM}_{1526} > 0.2$ \AA\ systems after excluding CGM sightlines with $N_\mathrm{HI} \ge 10^{20.1}\rm cm^{-2}$ (that is, excluding `double DLAs').
\label{fig.metals}}
\end{center}
\end{figure}

\begin{figure}%[ht]
\begin{center}
\vskip 0.2in
\includegraphics[angle=0,width=\columnwidth]{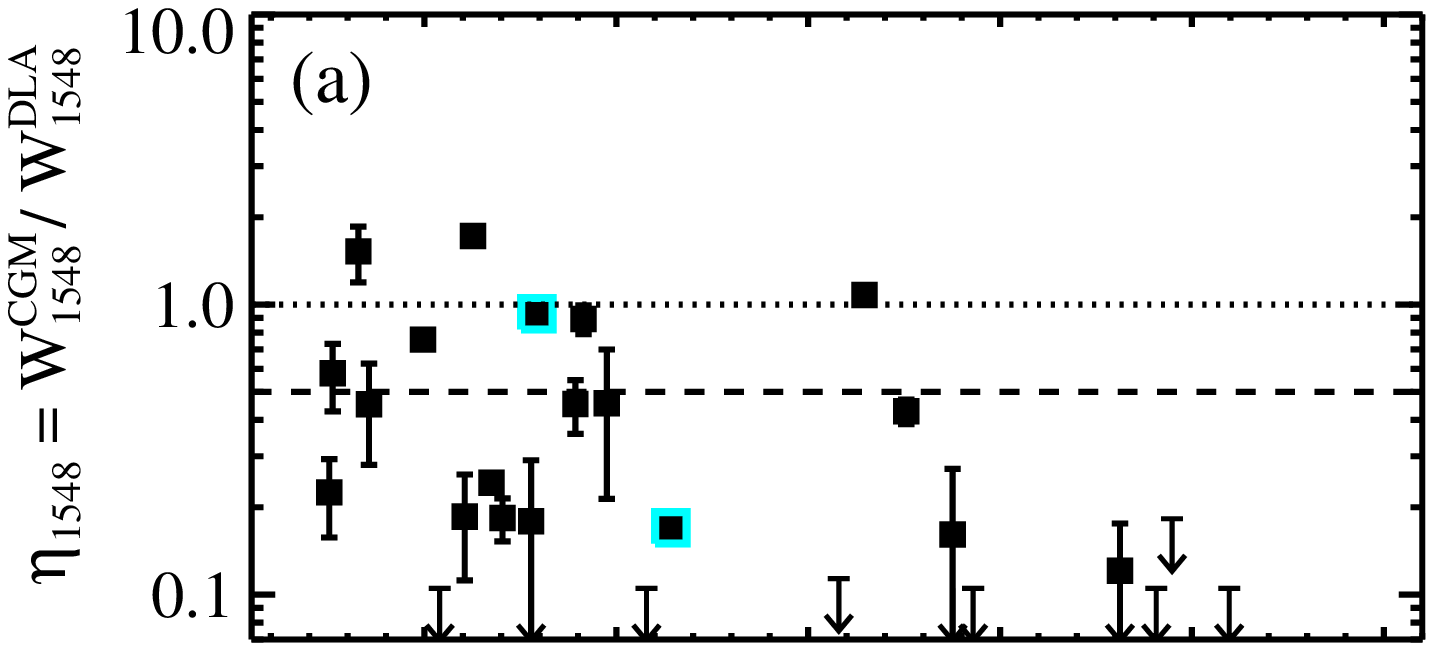}
\includegraphics[angle=0,width=\columnwidth]{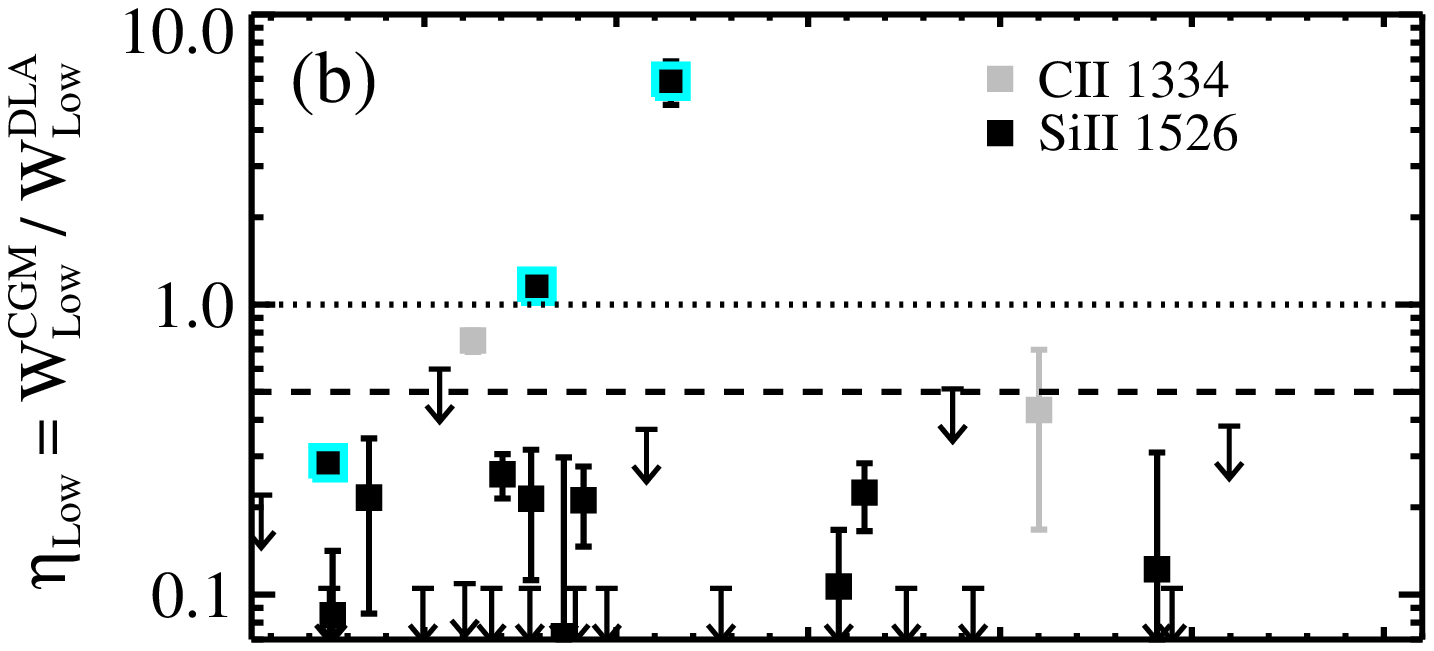}
\includegraphics[angle=0,width=\columnwidth]{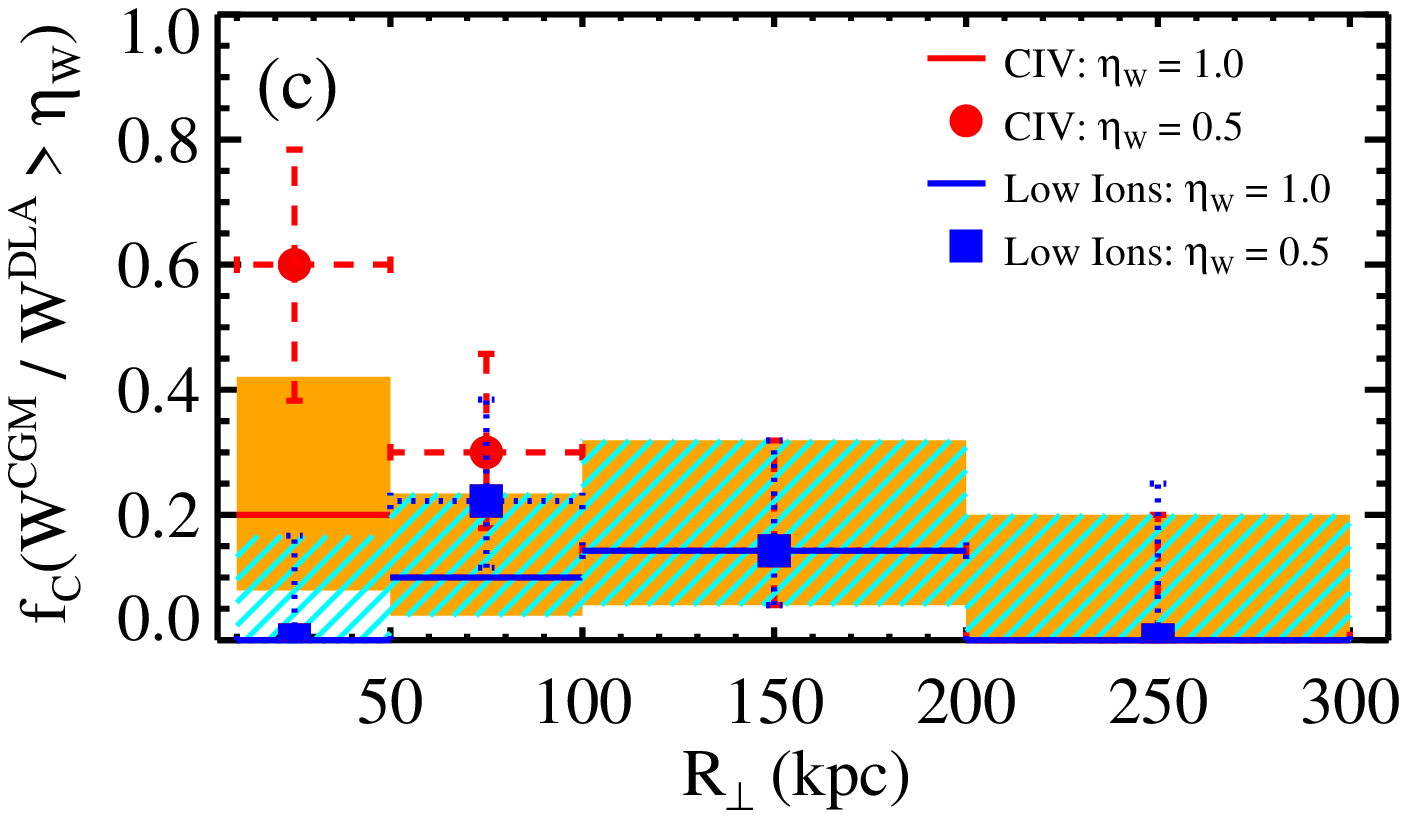}
\caption[]{\emph{(a)}  The ratio ($\eta_{1548}$) of $W_\mathrm{1548}$ measured in each CGM sightline ($W^\mathrm{CGM}_{1548}$) to $W_\mathrm{1548}$ in the associated DLA ($W^\mathrm{DLA}_{1548}$), plotted versus \rperp.  Only systems with $W^\mathrm{DLA}_{1548}$ measurements which are unaffected by line blending are included.  %CGM sightlines with securely-detected \ion{C}{4} are shown with solid black squares, and limits are indicated with red arrows.
Ratios falling below $\eta_{1548} = 0.07$ are indicated with downward arrows placed at $\eta_{1548} \sim 0.1$.
Points outlined in cyan indicate CGM sightlines with $N_\mathrm{HI} \ge 10^{20.1}\rm cm^{-2}$.
The horizontal lines are added to guide the eye at ratios of 1.0 and 0.5.
\emph{(b)} Same as panel \emph{(a)}, for low-ionization transitions.  $W^\mathrm{CGM}_{1526} / W^\mathrm{DLA}_{1526}$ values are shown with black squares.  $W^\mathrm{CGM}_{1334} / W^\mathrm{DLA}_{1334}$ values are shown in gray in cases for which we lack coverage of \ion{Si}{2} 1526 in both the CGM and DLA sightlines.
\emph{(c)} The fraction of systems having $W^\mathrm{CGM} / W^\mathrm{DLA} > \eta_{W}$ in four subsamples divided by $R_{\perp}$
(where $\eta_W$ refers to either $\eta_{1548}$ or $\eta_{\rm Low}$) for $\eta_W = 1.0$ (solid horizontal lines) and $\eta_W = 0.5$ (squares and circles).  The Wilson score 68\% confidence intervals are shown with colored boxes and error bars, respectively.  $f_C (W^\mathrm{CGM}_{1548} / W^\mathrm{DLA}_{1548} > \eta_{1548})$ values are shown in red and orange, 
and $f_C (W^\mathrm{CGM}_\mathrm{Low} / W^\mathrm{DLA}_\mathrm{Low} > \eta_{\rm Low})$ values are shown in blue and cyan.
\label{fig.dtb_cgm}}
\end{center}
\end{figure}

\subsection{Metal-Line Absorption in DLA Environments}\label{sec.results_metals}

Figure~\ref{fig.metals}a shows our measurement of $W_{1548}$ in each CGM sightline, $W^\mathrm{CGM}_{1548}$.  %Absorption detected at greater than $2\sigma$ significance is indicated with solid filled squares, and non-detections are shown at our $2\sigma$ upper limit values with red arrows.  
%% DONE: JFH Again, I think that this EW Is always deefined, and so you confuse the issue by showing these
%% limits. Just show the measurements themselves, and if they are < 2\sigma, just highlight them 
%% somehow. Point with error bars are more informative then limits. 
For reference, the CGM sightlines having $N_\mathrm{HI} \ge 10^{20.1}\rm cm^{-2}$ are marked with open cyan squares.
We detect very large $W^\mathrm{CGM}_{1548} > 0.6$ \AA\ to projected distances as large as 121 kpc.  However,  within this distance there is significant scatter in $W^\mathrm{CGM}_{1548}$, with many sightlines exhibiting only $W^\mathrm{CGM}_{1548} \sim 0.1$ \AA.  Beyond 150 kpc, we measure $W^\mathrm{CGM}_{1548}$ as large as $\sim0.3$ \AA, but are more likely to find $W^\mathrm{CGM}_{1548} < 0.2$ \AA\ (see also Figure~\ref{fig.metals}c).  
A Kendall's $\tau$ rank correlation test does not rule out a lack of correlation between
$W^\mathrm{CGM}_{1548}$  and \rperp\ (yielding a two-sided probability $P=0.24$), reflecting the overall large scatter in these values at a given \rperp.
%suggestive of 
%a weak anti-correlation between the \ion{C}{4} absorption strength and projected distance from DLAs
%of marginal statistical significance.

The red open circles show $\langle W_{1548} \rangle$ measured in the coadded spectra discussed in \S\ref{sec.stacks}, with \avgWCIV\ in the coadded DLA sightlines marked at $\mrperp = 6$ kpc.  The corresponding error bars are determined using our bootstrapping method, and thus reflect both the measurement uncertainty and the scatter in $W^\mathrm{CGM}_{1548}$ values for each subsample.  We find that while $\langle W_{1548} \rangle$ appears to decrease with increasing $R_{\perp}$,  the trend is a weak one:
even the \avgWCIV \ measured in the DLA sightlines is consistent with \avgWCIV \ in the CGM at $\mrperp \sim50$ kpc within the measured $1\sigma$ scatter, and the \avgWCIV \ measured at $100~\mathrm{kpc} < R_{\perp} < 200~\mathrm{kpc}$ differs from the latter by $<2\sigma$.  
%the values measured in the two outer $R_{\perp}$ bins are consistent within their $1\sigma$ error bars, and the
%\avgWCIV \ measured at $\sim50$ kpc differs from \avgWCIV \ measured at $100~\mathrm{kpc} < R_{\perp} < 200~\mathrm{kpc}$ by only %$2\sigma$.  
Kendall's $\tau$ test for a lack of correlation between \avgWCIV \ and \rperp\ yields a low two-sided probability ($P=0.04$) only if the value measured in DLA sightlines is included.  Without this `down-the-barrel' measurement, 
the probability is $P=0.12$, suggestive of a weak anti-correlation of marginal statistical significance.
%there is 
%a relatively high probability ($P=0.12$) that these measurements are uncorrelated.
%no significant anti-correlation with \rperp\ ($P=0.11$).

 %{\bf Continue from here.}

Figure~\ref{fig.metals}b shows our measurements of CGM absorption in low-ionization metal lines ($W^\mathrm{CGM}_\mathrm{Low Ion}$): in most cases we plot the equivalent width of \ion{Si}{2} 1526 ($W^\mathrm{CGM}_{1526}$; black), but we add measurements of the \ion{C}{2} 1334 equivalent width ($W^\mathrm{CGM}_{1334}$; gray) for CGM sightlines which lack coverage of \ion{Si}{2} 1526.  These latter values have been multiplied by the ratio of the rest wavelengths of the \ion{Si}{2} and \ion{C}{2} transitions (i.e., by 1526.7066 \AA/1334.5323 \AA).
%% DONE: JFH Explain how you converted one to the other? You should be doing that right?
The largest $W^\mathrm{CGM}_\mathrm{Low Ion}$ values are $\sim1$ \AA, with 2 of these 3 strong systems arising in ``double'' DLAs (having $N^\mathrm{CGM}_\mathrm{HI} \ge 10^{20.1}\rm cm^{-2}$).  Otherwise, the vast majority ($90\%$) of the remaining 29 (non-DLA) CGM sightlines yield weak absorption ($W^\mathrm{CGM}_\mathrm{Low Ion} < 0.2$ \AA), including the  lensed QSO sightline with $R_{\perp} \sim 7.5$ kpc.  Our measurements of \avgWSiII \ (red open circles) are consistent within the $1\sigma$ uncertainty intervals across the three CGM subsamples, with \avgWSiII\ in the coadded DLA sightlines exceeding that in the CGM by only $(1.3-2)\sigma$.
%% DONE: JFH Again there should be no upper limits in the stacks!!! 
Here, we find no statistically significant anti-correlation between either \avgWSiII \ or $W^\mathrm{CGM}_{1526}$ and \rperp.
% we achieve a $\sim2\sigma$-significant detection at $R_{\perp} < 100$ kpc of \avgWSiII \ $= 0.22\pm0.11$ \AA, and place $2\sigma$ upper limits on the absorption strength for the subsamples at larger $R_{\perp}$ of \avgWSiII \ $< 0.24$ \AA.

We next compute covering fractions for strong metal-line absorption.  We consider a system to be ``strong'' if the equivalent width measurement satisfies $W^{\rm CGM}/\sigma_W^{\rm CGM} > 3$ (where $\sigma_W^{\rm CGM}$ is the uncertainty in $W^{\rm CGM}$) and $W^{\rm CGM} > 0.2$ \AA.  All systems with securely-detected lines (having $W^{\rm CGM}/\sigma_W^{\rm CGM} > 3$) with $W^{\rm CGM}$ below 0.2 \AA\ and all systems having $W^{\rm CGM}/\sigma_W^{\rm CGM} < 3$ with $3\sigma$ upper limits on $W^{\rm CGM}$ less than 0.2 \AA\  are treated as sightlines without strong absorption.  We consider constraints from sightlines with $3\sigma$ upper limits on $W^{\rm CGM}$ larger than 0.2 \AA\ to be ambiguous in this context, and do not include them in covering fraction estimates.
%Turning to the estimates of metal-line covering fractions shown in Figure~\ref{fig.metals}c, 
As shown in Figure~\ref{fig.metals}c, 
we detect strong ($> 0.2$ \AA) \ion{C}{4} absorption in $57^{+12}_{-13}$\% of our sightlines within 100 kpc of a DLA.  Beyond this \rperp, we estimate a lower $f_C (W^{\rm CGM}_{1548} > 0.2~\rm \AA )\sim 25-33\%$, but find that the $f_C$ measurements in all $R_{\perp}$ bins are consistent within their $1\sigma$ uncertainties.  We measure $f_C (W^{\rm CGM}_\mathrm{1526} > 0.2~\rm \AA)  = 20^{+12}_{-8}$\% within 100 kpc, an incidence $2.1\sigma$ lower than that measured for \ion{C}{4}, and estimate similarly low $f_C (W^{\rm CGM}_\mathrm{1526} > 0.2~\rm \AA)$ values at larger $R_{\perp}$ (also consistent with our results for $f_C (W^{\rm CGM}_{1548} > 0.2~\rm \AA)$ at these distances).  
The $f_C (W^{\rm CGM}_\mathrm{1526} > 0.2~\rm \AA)$ values are slightly lower than, although statistically consistent with, our measurement of a $\sim30-40\%$ incidence of optically thick
\ion{H}{1} described in \S\ref{sec.results_hi}; this is in spite of our use 
of the presence of strong low-ionization metal absorption as a criterion for the detection of optically thick material.  
This slight discrepancy is in part due to two systems (J0201+0032 and J1153+3530) which exhibit clear damping wings
but for which we measure weak ($\lesssim0.2$ \AA) \ion{Si}{2} absorption, and in part due to 
our incomplete spectroscopic coverage of \ion{Si}{2} or \lya\ forest contamination of this transition in a few sightlines.
We additionally note that if ``double DLA'' systems are excluded, 
the estimated covering fraction of low-ionization material within 100 kpc falls below 10\% ($f_C (W^{\rm CGM}_\mathrm{1526} > 0.2~\rm \AA)\sim8\%$, from 
1 strong system among 13 total sightlines).
%% DONE: JFH Isn't the covering factor of low-ions by construction the covering factor of optically thick gas
%% Is there any case of optically thick in HI, where SiII is not > 0.2A? That is deserving of comment. 

Finally, we find that the \ion{Si}{4} 1393 absorption around DLAs is intermediate in strength between that of \ion{C}{4} and low-ionization absorption.  We measure $W_{1393}^{\rm CGM} > 0.2$ \AA\ in  6 out of 14 sightlines within 100 kpc of DLAs, yielding $f_C (W^{\rm CGM}_{1393}> 0.2 ) = 0.43_{-0.12}^{+0.13}$; i.e., a value slightly higher than 
$f_C (W^{\rm CGM}_{1526}> 0.2)$ but lower than $f_C (W_{1548}^{\rm CGM} > 0.2)$.
At larger impact parameters $100~\mathrm{kpc} < \mrperp < 300$ kpc, we measure only $f_C (W^{\rm CGM}_{1393}> 0.2 ) = 0.10_{-0.06}^{+0.13}$.
Similarly, our coadded spectra covering \ion{Si}{4} at $\mrperp>100$ kpc exhibit negligible absorption, with the coadd of sightlines within 
$\mrperp < 100$ kpc showing a modest \avgWSiIV $\sim 0.15$ \AA\ (Table~\ref{tab.stacks}).

In Figure~\ref{fig.dtb_cgm}, we compare our measurements of the CGM metal-line absorption strength with the strength of metal absorption measured along the associated DLA sightline, or `down the barrel'.  Panel \emph{(a)} shows the ratio of $W_{1548}$ measured in the CGM, $W^\mathrm{CGM}_{1548}$, to that measured in the DLA, $W^\mathrm{DLA}_{1548}$, as a function of sightline separation.  Here we only include systems for which we have  unblended coverage of the \ion{C}{4} transition along the DLA sightline:
%yield a $2\sigma$-significant detection of \ion{C}{4} in order to be included; 
of the 33 systems shown in Figure~\ref{fig.metals}a, 28 meet this criterion.  %CGM sightlines exhibiting statistically-significant \ion{C}{4} absorption are shown with solid black squares, and sightlines for which we obtain an upper limit on the absorption strength are indicated with red arrows. 
In cases for which the ratio is $<0.07$, a symbol is shown either at the $2\sigma$ upper limit on the ratio, 
or at $\sim0.1$ if the upper limit is below the range of the y-axis.
 Double DLA systems are highlighted with cyan open squares.  Particularly within $R_{\perp} < 100$ kpc, $W^\mathrm{CGM}_{1548}$ is frequently at least half as large as $W^\mathrm{DLA}_{1548}$.  This finding is quantified in panel \emph{(c)}, in which we show the fraction of pairs ($f_C$) exhibiting $W^\mathrm{CGM}_{1548} / W^\mathrm{DLA}_{1548}$ larger than a fiducial ratio, $\eta_\mathrm{W}$, calculated by dividing the number of pairs satisfying 
$W^\mathrm{CGM}_{1548} / W^\mathrm{DLA}_{1548} > \eta_\mathrm{W}$ by the total number of pairs in a given range in $R_{\perp}$.  
We choose values of $\eta_\mathrm{W} = 1$ (horizontal red bars with confidence intervals in orange) and $\eta_\mathrm{W} = 0.5$ (filled red circles).  Sightlines at $R_{\perp} \sim 7-50$ kpc have a $\sim20\%$ probability of exhibiting $W^\mathrm{CGM}_{1548}$ as high as that measured in the associated DLA, and have a $\sim60\%$ probability of exhibiting $W^\mathrm{CGM}_{1548}$ which is at least half as strong as $W^\mathrm{DLA}_{1548}$.  
The incidence of similarly high $W$ ratios decreases at $R_{\perp} \sim 50-100$ kpc but remains significant
($f_C(W^\mathrm{CGM}_{1548} / W^\mathrm{DLA}_{1548} > 0.5)\sim30\%$).  
These results imply that the bulk of the \ion{C}{4} equivalent width observed `down the barrel' along DLA sightlines traces the motions of gas extending well beyond the cold neutral material giving rise to the \ion{H}{1} absorption in the systems.  This high-ionization absorption may instead be tracing halo gas kinematics dominated by virial motions and/or galactic winds out to distances $\gtrsim50$ kpc.  
This scenario was first suggested by the finding that the velocity structure of unsaturated, low-ionization metal absorption 
(tracing neutral material) differs significantly from the velocity structure of \ion{C}{4} in DLAs \citep{WP2000a}.  However, our measurements offer the first direct constraints on the three-dimensional geometry of this high-ionization absorption.

Figure~\ref{fig.dtb_cgm}b shows the same $W$ ratios for the \ion{Si}{2} (black) and \ion{C}{2} (gray) transitions.  
Here, it is unusual for $W^\mathrm{CGM}_{1526} / W^\mathrm{DLA}_{1526}$ to exceed $\eta_\mathrm{Low} = 0.5$:
if `double DLAs' are excluded, \emph{no} CGM sightlines exhibit low-ionization equivalent widths greater than those measured toward the associated DLA.
Overall, Figure~\ref{fig.dtb_cgm}c shows that only $\sim20\%$ of sightlines at 50 kpc $< R_{\perp} < 100$ kpc exhibit $W^\mathrm{CGM}_{1526} / W^\mathrm{DLA}_{1526} > 0.5$.  
There is, however,  a low incidence ($\sim15\%$) of systems having $\eta_{\rm Low} > 0.5$
% $W^\mathrm{CGM}_{1526} / W^\mathrm{DLA}_{1526} > 0.2$ are relatively common (with an incidence $\sim40\%$) 
out to $R_{\perp} = 200$ kpc.  Thus, while \ion{Si}{2} absorption in DLAs appears to arise predominantly from gas within $\lesssim10$ kpc of the neutral material, there is a sub-dominant contribution from a gaseous component extending over  $>100$ kpc scales.  
%% DONE: JFH Why are you referring to cold ISM and disks, when we have no idea what the heck DLAs are???
%% Rework this language discussion to be model-independent. 
We discuss this point further in the context of previous results \citep[e.g.,][]{Prochaska2008} in \S\ref{sec.dlas_2d}.

%% DONE: JFH I can't figure out what the order of the panels is here, and often cannot read the impact
%% parameter labels. I suggest you order these by impact parameter. 
\begin{figure*}%[ht]
\begin{center}
\includegraphics[angle=90,width=0.9\textwidth]{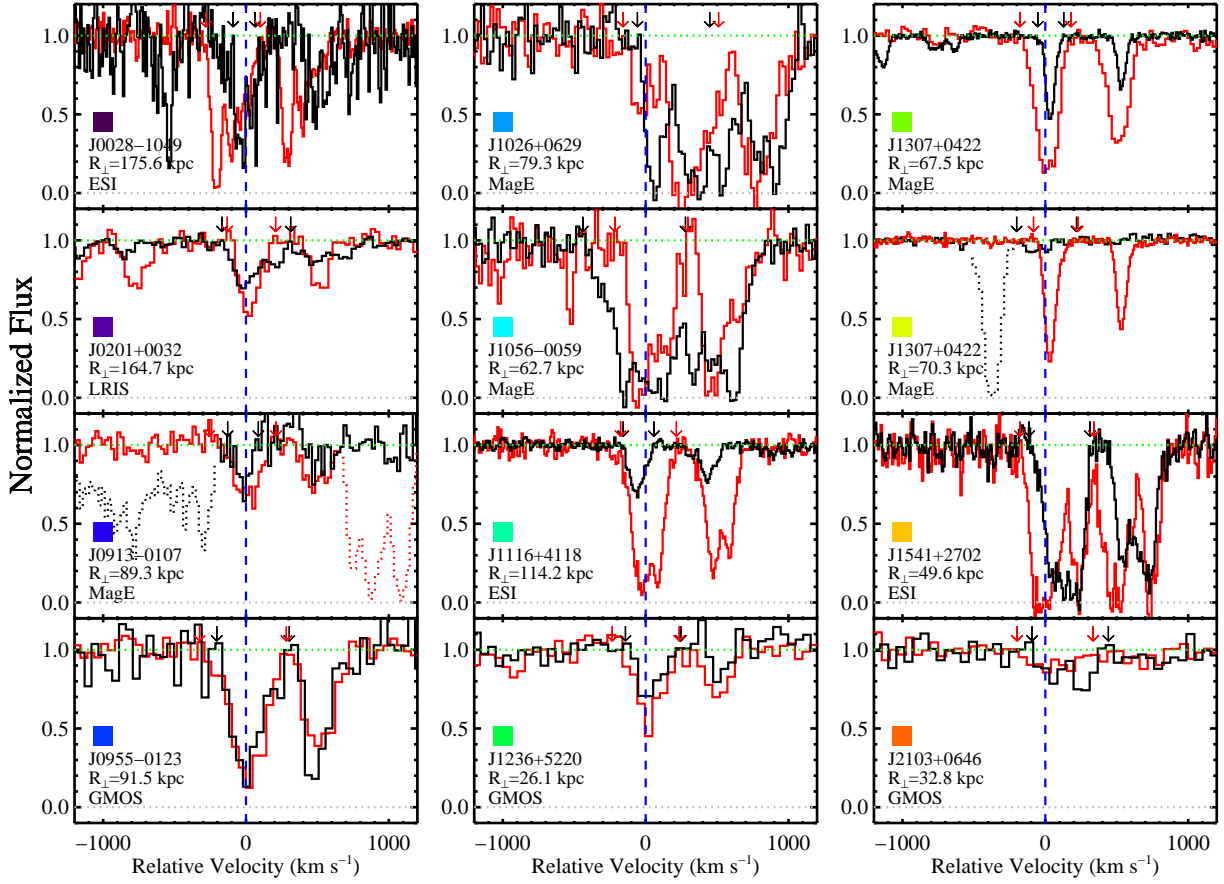}
\caption[]{\ion{C}{4} absorption profiles for systems having $W_{1548}/\sigma_{W_{1548}} > 4$ in both the CGM (black) and DLA (red) sightlines.  Systems are ordered by the QSO pair ID,  listed at the bottom of each panel along with the projected separation of the pair and the instrument used to obtain the spectra. 
 The relative velocity is $0\mkms$ at \zdla\ (indicated with a blue vertical dashed line), determined from an approximate centroid of the low-ionization absorption in the DLA sightline.  The small downward arrows show the velocity range over which we measure $W_{1548}$, $\delta v^{1548}$, and $\Delta v_{0.75}$ for each system.  The large colored squares match each pair to the corresponding points in the left- and right-hand panels of  Figure~\ref{fig.kinematics}.  Strong absorption which is physically unrelated to the DLA-CGM systems is shown with dotted histograms.
%{\bf [Mask unrelated absorption with dotted histograms]}
\label{fig.showciv}}
\end{center}
\end{figure*}

\begin{figure*}%[ht]
\begin{center}
\vskip 0.2in
\includegraphics[angle=0,width=0.6\columnwidth]{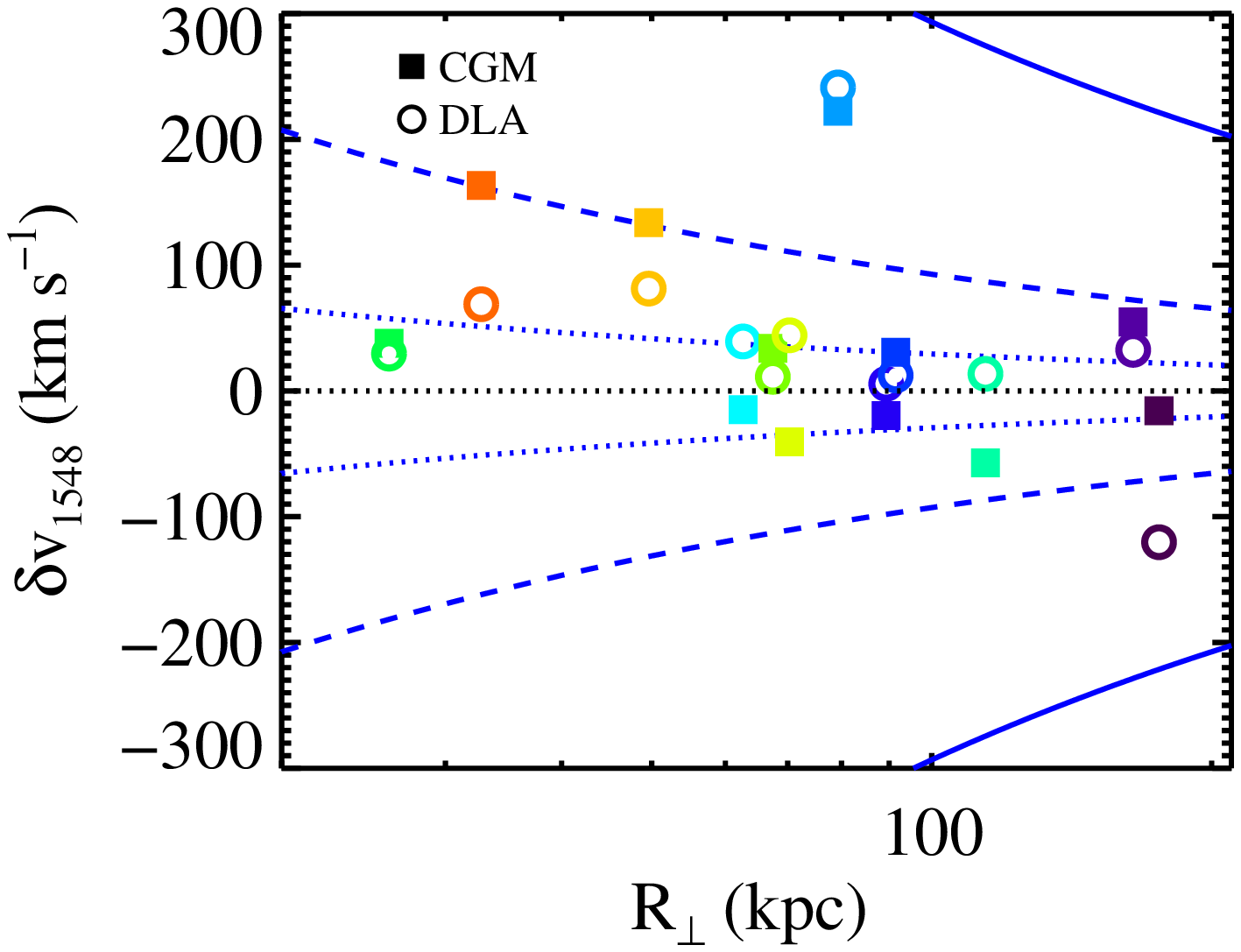}
\includegraphics[angle=0,width=0.6\columnwidth]{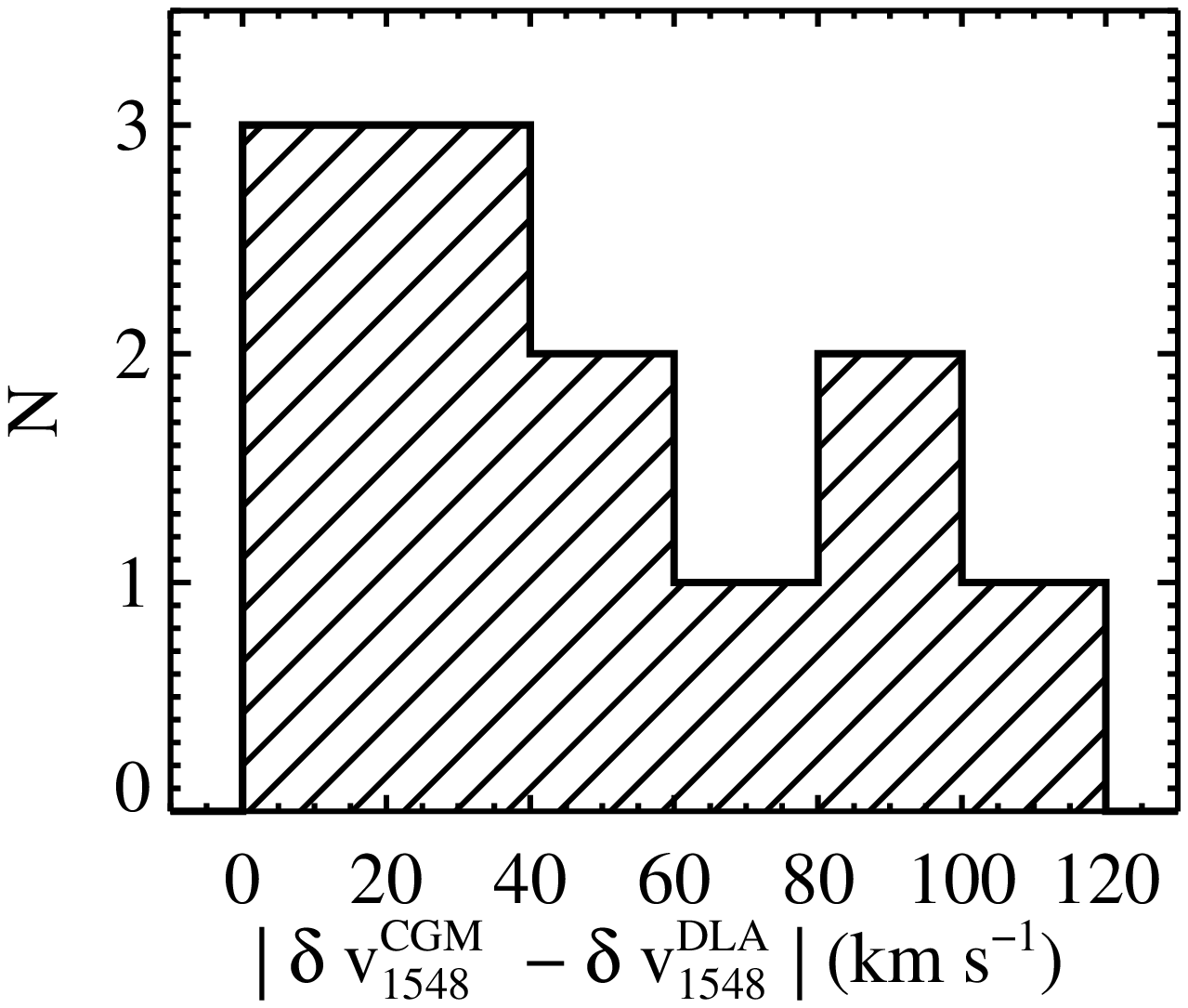}
\includegraphics[angle=0,width=0.6\columnwidth]{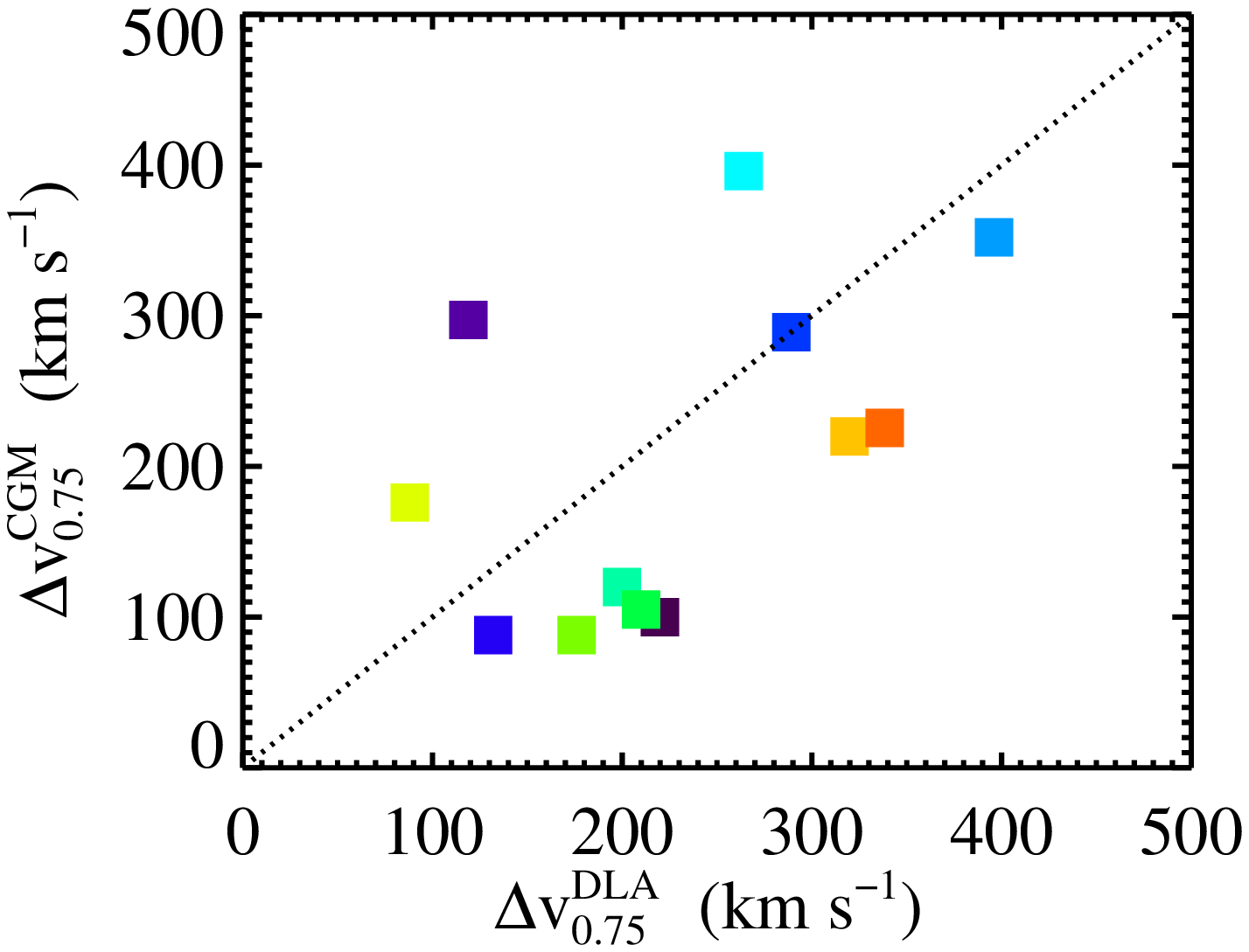}
\caption[]{\emph{Left:} Velocity offset between \zdla\ and the flux-weighted wavelength centroid of \ion{C}{4} absorption ($\delta v_{1548}$) in the CGM (solid squares) and DLA (open circles) sightlines in systems for which $W_{1548}/\sigma_{W_{1548}} > 4$.  The color of the points marking each pair indicates the corresponding panel in Figure~\ref{fig.showciv}.  The blue curves show the escape velocity in the radial direction (\vesc) as a function of total (not projected) distance from the center of a dark matter halo with mass $10^{10}\msun$ (dotted), $10^{11}\msun$ (dashed), and $10^{12}\msun$ (solid).  
\emph{Middle:} The distribution of offsets between $\delta v^\mathrm{CGM}_{1548}$ and $\delta v^\mathrm{DLA}_{1548}$.  These quantities do not differ by more than $105\mkms$ for any pair of sightlines.
\emph{Right:} The velocity width over which \ion{C}{4} absorption expresses $75\%$ of its total $W_{1548}$ ($\Delta v_{0.75}$) in DLA vs.\ CGM sightlines.  The point color indicates the corresponding system in the left-most panel and in Figure 8.  The dotted line shows a 1:1 relation.  These widths are within $100\mkms$ of each other in 7 of 12 cases, and never differ by more than $200\mkms$.
\label{fig.kinematics}}
\end{center}
\end{figure*}

\subsection{\ion{C}{4} Absorption Kinematics in DLA Environments}\label{sec.results_kinematics}

%% DONE: JFH Figure 7 and Figure 8 references in text appear to be out of order. You are referencing
%% Figure 8 here, but have not yet discussed Figure 7??
As discussed in the previous subsection, the strong similarity in the values of $W^\mathrm{CGM}_{1548}$ and $W^\mathrm{DLA}_{1548}$, particularly between sightlines separated by $\lesssim 100$ kpc, suggests that these quantities are dominated by absorbing gas extending over large distances from the DLA ($\mrperp \sim 7-100$ kpc).  Motivated by the high quality and high spectral resolution of the data available for many of these sightlines, here we perform a more detailed comparison of the properties of these line profiles.  
Figure~\ref{fig.showciv} shows \ion{C}{4} profiles for the 12 systems with $W_{1548} / \sigma_{W_{1548}} > 4$ in both the CGM and DLA sightlines.  Eight of these pairs were observed at echellette resolution, such that our spectra reveal the detailed velocity structure of the profiles.  The separations of these sightlines range from $\mrperp = 26$ to 176 kpc.    

Several points become evident from examination of this figure.  First, we remind the reader that the systemic velocity (\zdla) is determined from the centroid of the low-ionization metal absorption arising in the DLAs.  This velocity is often very similar to the velocity centroid of higher-ionization absorption, although we see that the DLA \ion{C}{4} profile is significantly offset from \zdla\ (by a few hundred $\mkms$) in a handful of cases (most notably for J1026+0629).  These offsets notwithstanding, the central velocities, velocity widths, and even the detailed shapes of the DLA and CGM \ion{C}{4} profiles are remarkably similar.  
%% DONE: JFH Black arrows looks messed up for J1307+0422 second panel. 

To quantify these similarities, we calculate the flux-weighted wavelength centroid of each profile, 
$\delta \lambda_{1548} = \sum\limits_i (1 - f_i) \lambda_i / \sum\limits_i (1-f_i)$, 
where $f_i$ and $\lambda_i$ are the continuum-normalized flux and wavelength of individual pixels comprising the profile of each system.
We show the
velocities of these centroids relative to \zdla\ 
%% DONE: JFH Clarify that z_dla is the low-ion centroid. 
($\delta v_{1548}$) in Figure~\ref{fig.kinematics} (left).  
Measurements for DLA and CGM sightlines are shown with open circles and filled squares, respectively, and the symbols for each pair are given a unique color to indicate the corresponding profiles in Figure~\ref{fig.showciv}.  In general, the values of $\delta v_{1548}$ for the  sightlines in  each pair are close, and where they are offset from $\delta v_{1548} = 0 \mkms$ they are mostly offset in the same sense.
%% NOT DONE: JFH perhaps connect these with a line. 
%% KHRR: Ignoring this for now
We show the distribution of the offsets between $\delta v_{1548}^{\rm CGM}$ and $\delta v_{1548}^{\rm DLA}$ in the middle panel of  Figure~\ref{fig.kinematics}.  These differences are never larger than $105\mkms$, even for the pairs with the largest sightline separations (up to $\mrperp = 176$ kpc), and are $< 60\mkms$ for 8 of 12 pairs.  
We note that a number of these systems have 
$\delta v_{1548}^\mathrm{DLA}$ exceeding $100-200\mkms$, such that there is a higher degree of coherence between the \ion{C}{4} absorption covering $\gtrsim100$ kpc scales in these systems than that exhibited by low- and high-ionization absorption along the same QSO sightline.  
Figure~\ref{fig.kinematics} (left) also indicates the radial velocity required for escape from the potential well of dark matter halos over a range of masses ($M_h = 10^{10} - 10^{12}\msun$), or $v_\mathrm{esc} = \sqrt{2GM_h / R}$, with $R=\mrperp$.
There are few instances in which the CGM \ion{C}{4} absorption has a central velocity surpassing these values, even for quite low $M_h \lesssim 10^{11}\msun$.  However, this material may have an additional component to its velocity vector in the plane of the sky to which our measurements are not sensitive. Furthermore, material with kinematics at the extremes of these quite broad profiles may indeed have the energy to escape from halos with $M_h \lesssim 10^{11}\msun$, even if motions transverse to the line of sight are neglected.
 %% JFH Perhaps it is worth trying to make Figure 9b but normalized by the \Delta_v_0.75

To quantify the velocity width of these profiles,
% we first calculate a column density for each profile pixel via the pixel optical depth method \citep{Cowie1998,Songaila1998,Ellison2000,Schaye2000}.  We find the maximum column of the profile $N_\mathrm{pix}^\mathrm{max}$, and identify all pixels having $N_\mathrm{pix} > 0.25 N_\mathrm{pix}^\mathrm{max}$.   From among these high-$N_\mathrm{pix}$ pixels, we locate those closest to the blue and red profile edges, and calculate the velocity difference between them ($\Delta v_{0.25}$). 
we identify the set of pixels encompassing $75\%$ of the total profile $W$, defining the central pixel in this set be the pixel whose relative velocity is closest to 
$\delta v_{1548}$ (i.e., the flux-weighted \ion{C}{4} 1548 velocity centroid measured as described above).
%% DONE: JFH Unclear why this is a delta and what this is a delta in reference to. Clarify. Also, you now
%% appear to have two defintiions of delta_1548. I've gotten quite confused by this discussion. 
From this set of pixels, we locate those closest to the blue and red profile edges, and calculate the velocity difference between them ($\Delta v_{0.75}$). 
 We compare our measurements of this quantity for each pair of sightlines in the right-most panel of Figure~\ref{fig.kinematics}.  Although there is a large range in the  $\Delta v_{0.75}$ values ($100 - 450\mkms$), $\Delta v^\mathrm{CGM}_{0.75}$ and $\Delta v^\mathrm{DLA}_{0.75}$ differ by more than $100\mkms$ in only 5 of 12 pairs and are weakly correlated (at a $\sim90\%$ confidence level).  Furthermore, $\Delta v_{0.75}$ is almost always larger than the value 
$| \delta v_{1548}^\mathrm{CGM} -  \delta v_{1548}^\mathrm{DLA}|$, and exceeds the latter by $>100\mkms$ in 6 pairs.  
As anticipated above, the velocities of the pixels at the blue and red edges of these quite broad line profiles (identified in the process of estimating $\Delta v_{0.75}$) frequently lie outside of the envelop defined by the halo escape velocity if $M_h \lesssim 10^{11}\msun$.

These comparisons evoke a scenario in which \ion{C}{4} absorption around DLAs arises from gaseous structures having a large velocity dispersion from structure to structure (yielding large velocity widths), but which extend over many tens of kpc with a high degree of kinematic coherence on these scales.  Indeed, such coherence among \ion{C}{4} systems detected along paired QSO sightlines has been noted previously \citep[e.g.,][]{Rauch2001,Martin2010}, but our study is the first to measure this in the vicinity of DLAs.
We discuss the implications of these results and their potential to constrain the physical drivers of \ion{C}{4} gas kinematics in \S\ref{sec.civ}.

%\begin{figure}
%\begin{center}
%\includegraphics[angle=0,width=\columnwidth,trim=0 120 160 0,clip=]{../Figures/pro/fig_dlacgm_recenter_eris.ps}
%\caption[]{\avgWLya\ (top), \avgWCII\ (middle), and \avgWCIV\ (bottom) measured in coadded spectra of DLA (at $\mrperp = 0$ kpc)
%and CGM sightlines (black squares).  
%The solid and dashed 
%blue lines show the median and $\pm25$th-percentile EW values measured in sightlines distributed evenly across a 1 Mpc box %centered on the Eris simulation \citep{Shen2012}.  Blue diamonds and error bars show the average EW and dispersion measured in %1000 randomly drawn sets of  sightlines having the same \rperp\ distribution as our DLA-CGM dataset.  The purple, red, orange, %and yellow curves and diamonds show the same measurements, assuming the center of the system ($\mrperp = 0$ kpc) is located %20, 40, 60, and 80 kpc from the center of the Eris box, respectively.
%\label{fig.eris}}
%\end{center}
%\end{figure}

\section{Discussion}\label{sec.discussion}

\subsection{A `Two-Dimensional' View of DLAs}\label{sec.dlas_2d}

\subsubsection{The Spatial Extent of DLAs}
Much of our understanding of DLAs relies heavily on 
studies of the absorption along single, pencil-beam sightlines piercing 
neutral gas in the host galaxy along with any more diffuse material associated with the galaxy's halo in the same beam.  
However, as noted in \S\ref{sec.intro}, studies of DLAs toward lensed QSOs \citep[e.g.,][]{Cooke2010} have recently begun to augment these single-sightline analyses, constraining the spatial extent of damped absorption and the coherence of metal-line kinematics over relatively small scales ($\lesssim10$ kpc).  
\citet{Ellison2007} presented the first exploration of the extent of DLAs over the scales of galaxy halos,  identifying a $z=2.66$ absorption system having \nhi\ $> 10^{20.1}\cmsq$ in spectroscopy of both sightlines toward the $z\sim3$ binary QSO J1116+4118 (see also Figure~\ref{fig.nhi}, top panel).

The present work adds considerable fidelity to this latter, `two-dimensional' approach to the study of DLAs and their environment.
First, the measurements shown in Figure~\ref{fig.nhi} offer direct constraints on the spatial extent of the high column density  material giving rise to DLAs on scales larger than $\sim10$ kpc.  Of the 30 CGM sightlines in our sample with sufficient S/N to assess \nhi, only three exhibit \nhi\ $\ge 10^{20.1}\cmsq$. %over a range in \rperp\ from 30 to 114 kpc. 
%% NOT DONE: JFH Perhaps you should make a DLA covering factor plot, i.e. f_C(N_HI > 20.3)
%% KHRR -- just skipping this for now; I feel like the text is enough (but not strongly!) 
Within $\mrperp <120$ kpc (the maximum \rperp\ among these double DLA pairs), absorption with \nhi\ $\ge 10^{20.1}\cmsq$ is \emph{absent} from 13 CGM sightlines, most notably from $\sim75\%$ of sightlines having $\mrperp < 30$ kpc.  
%% DONE: JFH These sentences are contradictory. You say only 3 systems have double DLAs, and then you 
%% say 14 systems of 30 don't have damped absorption. I'm very confused, by what you mean by damped
%% absorption then? Usually that means DLA. 
This is strong confirmation of the conclusion of \citet{Cooke2010} that the physical extent of \nhi\ $\ge10^{20.3}\cmsq$ absorption in a `typical' DLA must be $< 10$ kpc.  Taking our measurements at face value, they indicate either (1) that all DLAs have radii $< 10$ kpc, with $\sim10-20\%$ occurring in overdense environments hosting multiple damped systems;  (2) that DLA gas is distributed on scales $> 10$ kpc with a low covering factor; or (3) 
that $\sim10-20\%$ of DLAs have physical extents $\gtrsim30-120$ kpc, with all others having much smaller sizes. 
 As we expect these systems to occupy halos having virial radii $\lesssim100$ kpc, the latter scenario would require high-density gas disks to extend over at least $15-60\%$ of their halo virial diameter (as proposed in, e.g., \citealt{Maller2001}).
%Given that such high \nhi\ material is never found to extend over these scales in analyses of cosmological `zoom-in' simulations of halos over a wide mass range ($10^{11.2} - 10^{12.6}\msun$; \citealt{Fumagalli2013}), we favor the former picture.  {\bf [I actually don't know if this is true -- confirm with Michele!]}.  

Moreover, we note that the covering fraction of \nhi\ $\ge 10^{20.1}\cmsq$ material within $\mrperp < 100$ kpc of DLAs is $f_C^\mathrm{DLA} (\mrperp < 100~{\rm kpc}) = 0.13^{+0.11}_{-0.07}$.  
This measurement may be compared with constraints on the DLA cross section offered by the clustering analysis of 
\citet{Font-Ribera2012} as follows.  The expectation value %$\langle f_C^\mathrm{DLA} (\mrperp < 100~{\rm kpc})\rangle$ 
of the DLA covering fraction within $\mrperp = 100$ kpc measured from a statistical sampling of \nhi\ in dark matter halos with masses ranging down to a minimum mass $M_0$ is
\begin{equation} \label{equ.fc} 
  \langle f_C^\mathrm{DLA} \rangle = \frac{\int\limits_{M_0}^{\infty} P(M_h)~f_C^{\rm DLA}(M_h; \mrperp < 100~{\rm kpc})~dM_h}{\int \limits_{M_0}^{\infty} P(M_h) dM_h}. 
\end{equation}
Here, we assume that 
our experimental setup is sensitive to halos with $M_h > M_0$; i.e., that all of these halos host a DLA and hence 
may fall into our `primary' DLA sample.  We further assume that 
\begin{equation*}
P(M_h) = (c/H_0)\Sigma_{\rm DLA}(M_h) n(M_h)
\end{equation*}
describes the typical incidence of DLAs as a function of halo mass $M_h$, with 
$\Sigma_{\rm DLA} (M_h)$ equal to the DLA cross section (in $\rm kpc^2$) and $n(M_h)$ 
equal to the comoving number density of halos with mass in the interval ($M_h$, $M_h+dM_h$).

If we consider all halos to be isolated such that their DLA cross sections do not overlap on the sky, 
we can 
additionally state that the DLA covering fraction measured within $\mrperp < 100$ kpc  for a halo of mass $M_h$ is
\begin{equation*}
  f_C^{\rm DLA} (M_h; \mrperp < 100~{\rm kpc}) = \frac{\Sigma_{\rm DLA}(M_h)}{\pi (100~{\rm kpc})^2}. 
\end{equation*}
Here we are assuming that the full DLA cross section arises within $\mrperp < 100$ kpc, and that 
$\Sigma_{\rm DLA}$ cannot exceed $\pi (100~\rm kpc)^2$.
The value $\langle f_C^\mathrm{DLA} \rangle$ in Equ.~\ref{equ.fc} is then fully specified given a functional form for 
$\Sigma_{\rm DLA}(M_h)$ and a minimum DLA halo mass $M_0$.

Motivated by trends in the distribution of neutral material over a range in halo masses exhibited in cosmological hydrodynamical simulations, \citet{Font-Ribera2012} explored two parameterizations of $\Sigma_{\rm DLA}(M_h)$.
They first adopted a form
\begin{equation}
  \Sigma_{\rm DLA} (M_h) = \Sigma_0 (M_h/M_0)^{\alpha}, 
\end{equation}
with $\Sigma_0$ a constant.   Their estimate of the bias factor of DLAs, in combination with the observed DLA 
incidence rate, place simultaneous constraints on $\Sigma_0$, $M_0$, and $\alpha$.  
For instance, they found that $\alpha = 1$ with $M_0 = 10^{10}\msun$ requires $\Sigma_{\rm DLA} (10^{12}\msun)  = 1400~\rm kpc^2$.
This particular model yields a low value of $\langle f_C^\mathrm{DLA} \rangle = 0.08$, 
%{\bf (should this be 0.12?)}, 
consistent with our measurement.  
On the other hand, a model 
of the form
\begin{equation}
  \Sigma_{\rm DLA} = \Sigma_0 (M_h/M_0)^2 (1 + M_h/M_0)^{\alpha-2}
\end{equation}
with $\alpha = 1$ and $M_0 = 10^{10}\msun$ satisfies 
the DLA bias and incidence rate with $\Sigma_{\rm DLA} (10^{12}\msun) = 2400~\rm kpc^2$, but yields a much higher
$\langle f_C^\mathrm{DLA} \rangle = 0.26$.  This value likewise falls nearly within the $\pm1\sigma$ uncertainties in our estimate of $f_C^\mathrm{DLA} (\mrperp < 100~\rm kpc)$.  
Thus, our current constraints on $f_C^{\rm DLA}$ are 
 in accord with a large neutral gas cross section (with a characteristic length scale 
$R_{\rm char} \sim \sqrt{1400~\rm kpc/\pi} \sim 20$ kpc) arising in high-mass 
dark matter halos ($M_h \sim 10^{12}\msun$).  As noted above, however, given 
the high incidence of non-detections within $\mrperp < 30$ kpc and in the \citet{Cooke2010}
study, this material is most likely distributed with a covering fraction less than unity.
Moreover, the factor of $>3$ variation in $\langle f_C^\mathrm{DLA} \rangle$ between the two models described above suggests that a larger sample of QSO sightlines within $\mrperp < 100$ kpc of DLAs may eventually aid in breaking the degeneracies in these model parameters, further elucidating the relationship between the morphology of DLA absorption and dark matter halo mass.

%% DONE: JFH Let's talk through the assumptions that you are making in the Font-Ribera comparison. 

%If we assume that every DLA in our sample occupies a $10^{12}\msun$ dark matter halo with virial radius $100$ kpc, this implies that the cross section for DLA absorption is 
%$\Sigma_\mathrm{DLA} \sim f_C^\mathrm{DLA} \pi R^2_\mathrm{vir} \sim 4100$ kpc$^2$.
%This large $\Sigma_\mathrm{DLA}$ value is more than sufficient to satisfy both the DLA rate of incidence and bias factor as estimated by \citet{Font-Ribera2012}.  These authors, adopting a functional form $\Sigma_\mathrm{DLA} \propto (M/M_0)^2 (1+ M/M_0)^{\alpha-2}$ for the dependence of $\Sigma_\mathrm{DLA}$ on halo mass ($M$) (with $M_0$ the ``minimum'' DLA halo mass), find that for $10^{12}\msun$ halos $\Sigma_\mathrm{DLA} \sim 2400 - 4400$ kpc$^2$ if $\alpha = 0.5-1$ and $M_0 = 10^{10} - 10^{11.5}\msun$.
%Future studies using a larger sample of CGM sightlines within $\mrperp < 100$ kpc of DLAs have the potential to break the 
%strong degeneracy between $\alpha$ and $M_0$ in the context of this model.  
%{\bf[Am I thinking about this correctly?  No I am not!  X points out that one must integrate $\Sigma(M) * $ the assumed DLA incidence as a function of halo mass to come up with a predicted $f_C^\mathrm{DLA}$. ]}
 
\subsubsection{The Origin of Low-Ionization Absorption Associated with DLAs}
%In the absence of a second DLA along the CGM sightline, 
Among CGM sightlines which do not exhibit a second DLA, 
we detect optically thick \ion{H}{1} within $\mrperp < 100$ kpc with an incidence of $\sim23\%$.  We emphasize that we cannot rule out the presence of optically thick material in \emph{any} of the CGM sightlines within this projected distance, and that the true incidence of such absorption may be significantly higher.  We can, however, 
place stringent limits on the incidence of low-ionization metal absorption in many of our CGM sightlines.  As shown in Figure~\ref{fig.metals}b, we detect strong \ion{Si}{2} with $W_{1526}^{\rm CGM} > 0.2$ \AA\ in only one sightline which does not also probe a DLA (and which has $\mrperp < 100$ kpc),
with measurements of $W_{1526}^{\rm CGM}$ falling well below 0.2 \AA\ in the vast majority of the remaining sightlines.
The resulting covering fraction for strong low-ionization
absorption in environments outside the high-density neutral material giving 
rise to DLAs is $\sim 0.08$.  
%% JFH I see no reason to exclude the other DLAs here and compute a separate covering factor for the
%% non-DLAs. 
%% KHRR: Ignoring this for now.

This finding has implications for the interpretation of low-ionization absorption kinematics and equivalent widths measured `down the barrel' toward 
DLAs themselves.  Such low $W^\mathrm{CGM}_{1526}$ limits at $\mrperp \sim 30-100$ kpc suggest that $W_{1526}^{\rm DLA}$
%(measured toward DLAs 
measurements must be dominated by material within a physical distance $R_\mathrm{3D} \lesssim 30$ kpc.  Furthermore, Figure~\ref{fig.dtb_cgm} shows that 
$W^\mathrm{CGM}_{1526}$ is nearly always $< 50\%$ of $W^\mathrm{DLA}_{1526}$, likewise indicating that the low-ion kinematics are typically driven by gas motions close to the DLA.  
If the tight correlation between DLA metallicity and $W^\mathrm{DLA}_{1526}$ \citep{Prochaska2008,Neeleman2013} is indeed 
driven by a galaxy mass--metallicity relation, this suggests that $W^\mathrm{DLA}_{1526}$ must preferentially trace galaxy dynamics on small scales, 
%% DONE: JFH Again, what evidence do we have that DLAs are disks besides pure speculation?
analogous to emission-line tracers of \ion{H}{2} region kinematics \citep[e.g.][]{Weiner2006}.
In contrast to these results, we have found that the kinematics of higher-ionization material (e.g., \ion{C}{4}) must arise predominantly from
the motions of gas extending over much larger scales.  We discuss the processes which may be most relevant to these motions in \S\ref{sec.civ}.

\begin{figure}
\begin{center}
\includegraphics[angle=0,width=\columnwidth]{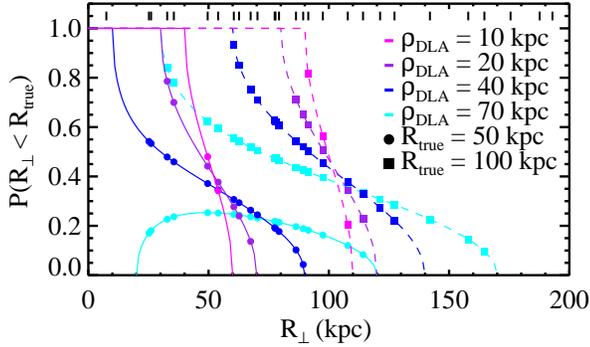}
\caption[]{Probability that a sightline at \rperp\ falls within a `true' projected distance from 
the associate halo center ($R_{\rm true}$), $P(\mrperp < R_{\rm true})$, as a function of \rperp\ 
for values of $R_{\rm true} = 50$ kpc (solid curves) and $R_{\rm true} = 100$ kpc (dashed curves).  Colors correspond to different values of 
$\rho_{\rm DLA}$ as indicated in the legend.  The set of \rperp\ values for the QSO pairs with coverage of \ion{C}{4} in the CGM sightline is shown with black vertical hashes toward the top of the figure.  The points show the value of $P(\mrperp < R_{\rm true})$ corresponding to each  sightline for 
$R_{\rm true} = 50$ kpc (circles) and $R_{\rm true} = 100$ kpc (squares), excluding points at $P(\mrperp < R_{\rm true})=1$.
\label{fig.cartoon_recenter}}
\end{center}
\end{figure}

\subsection{The DLA-CGM and Magnitude-Selected Galaxy Environments}

\subsubsection{Geometrical Considerations}\label{sec.geometry}
As one of our primary goals is to understand DLAs in the context of their host dark matter halos and their relation to star formation at high redshift, we wish to draw comparisons between the absorption strength of material in and around DLAs with that around optically-selected
 galaxies at $z\sim2$.  Ideally, we would directly compare the CGM absorption strength as a function of projected distance from the centers (or density peaks) of DLA host halos with that of halos of known mass scale.  However, because the precise location of DLAs within their surrounding dark matter distribution is not well understood, we must first consider how our experimental design affects our ability to constrain the projected radial absorption profile of the halos selected via our chosen technique.

As noted in \S\ref{sec.intro}, in the few cases for which an emission counterpart to a previously-known DLA has been recovered, they are typically located within $\lesssim20$ kpc of the QSO sightline \citep[e.g.,][]{Peroux2011,Krogager2012}, suggestive of a scenario in which DLAs arise 
close to the peak halo density locus.  On the other hand, cosmological `zoom-in' simulations predict that DLAs can trace inflowing streams or cool outflows extending to the host halo virial radius \citep{Fumagalli2011,Bird2014a}, leaving open the possibility that a significant portion of the DLA cross section is contributed by systems many tens of kpc from the nearest halo center.  In this case, the `true' projected distance ($R_{\rm true}$) from the center of a DLA-selected halo for a given CGM sightline in our sample may likewise be many tens of kpc larger or smaller than the QSO pair sightline separation (\rperp).  
In particular, given a projected distance from the halo center for a DLA, $\rho_{\rm DLA}$, $R_{\rm true}$ must fall in the range 
$\mrperp - \rho_{\rm DLA} \le R_{\rm true} \le \mrperp + \rho_{\rm DLA}$.  

To determine the probability distribution of $R_{\rm true}$ within this range of values, we consider a circle with radius defined by the vector ${\bf \mrperp}$  
and centered on the DLA.
The DLA is located at 
$\boldsymbol{\rho_{\rm DLA}}$, with the origin of the coordinate system defined to be at the halo center.  
We refer to the angle between the vectors ${\bf \mrperp}$ and $\boldsymbol{\rho_{\rm DLA}}$ as $\theta$.
The vector connecting the origin to the CGM sightline, ${\bf R_{\rm true}}$, forms the third side of a triangle 
with ${\bf \mrperp}$ and $\boldsymbol{\rho_{\rm DLA}}$, 
and its length may therefore be written $|{\bf R_{\rm true}}| = R_{\rm true} = (\mrperp^2 + \rho_{\rm DLA}^2 - 2\mrperp \rho_{\rm DLA} \cos \theta)^{1/2}$.  

Under the assumption that there is no preferred direction for ${\bf \mrperp}$, i.e., that $\theta$ has a uniform probability distribution in the range 
$0 \le \theta \le 2\pi$, we draw $\theta$ values at random to 
 estimate the resultant probability distribution for $R_{\rm true}$, $P(R_{\rm true})$.
We find that $P(R_{\rm true})$ is sharply peaked toward both $\mrperp - \rho_{\rm DLA}$ and $\mrperp + \rho_{\rm DLA}$, meaning that $R_{\rm true}$ is significantly 
more likely to have a value close to these extremes than close to \rperp.  
For example, if $\rho_{\rm DLA} = 20$ kpc and $\mrperp = 100$ kpc, the total probability that $R_{\rm true} < 85$ kpc or $R_{\rm true} > 115$ kpc is 
$46\%$, whereas the probability that $95~{\rm kpc} < R_{\rm true} < 105$ kpc is only $16\%$.  The form of this distribution must be considered 
when interpreting the results presented in Figures~\ref{fig.wlya}, \ref{fig.nhi}, \ref{fig.metals}, and \ref{fig.dtb_cgm}: each sightline shown 
has a non-negligible probability of probing an $R_{\rm true}$ offset from the indicated \rperp\  by an amount $\approx \rho_{\rm DLA}$.  If $\rho_{\rm DLA}$ 
is indeed small ($\lesssim 20$ kpc), this offset will be $\lesssim7\%$ of the x-axis coverage of these figures.  If $\rho_{\rm DLA}$ is instead on the order 
of $\sim100$ kpc, the systematic uncertainty in $R_{\rm true}$ will span much of the \rperp\ range shown.

%Given a particular projected distance from the halo center for a DLA, $\rho_{\rm DLA}$, 
In preparation for comparing \avgW\ measured in coadded DLA-CGM sightlines to that measured around magnitude-selected samples, we also wish to
 calculate the probability of a sightline at \rperp\ falling within a projected distance $R_{\rm true}$. 
Here, we consider the intersection of two circles: (1) one of radius $R_{\rm true}$ and centered at the origin (i.e., the halo center), and (2) one of radius \rperp\ and centered at $\rho_{\rm DLA}$.  The probability that a sightline at \rperp\ falls within $R_{\rm true}$ is then simply the fraction of the circumference of circle (2) which falls within circle (1).  This probability can be written:
%\[
\begin{multline*}
P(\mrperp < R_{\rm true}) = \\
\begin{cases}
  1, & \text{if } \rho_{\rm DLA} \leq R_{\rm true} - \mrperp\\
  1 - \frac{1}{\pi} \sin^{-1} \frac{\rm A_{int}}{2\mrperp}, & \text{if }  R_{\rm true} - \mrperp < \rho_{\rm DLA} \leq\\
 & \frac{\rho_{\rm DLA} - R_{\rm true}^2 + \mrperp^2}{2\rho_{\rm DLA}}\\
\frac{1}{\pi} \sin^{-1} \frac{\rm A_{int}}{2\mrperp}, & \text{if } \rho_{\rm DLA} > \frac{\rho_{\rm DLA} - R_{\rm true}^2 + \mrperp^2}{2\rho_{\rm DLA}}.
\end{cases}
\end{multline*}
%\]
\noindent Here, $\rm A_{int}$ is the length of the chord defined by the intersection points of the two circles. 
When comparing DLA-CGM absorption measurements against those measured around magnitude-selected samples to a given $R_{\rm true}$, 
higher values of $P(\mrperp < R_{\rm true})$ indicate higher probabilities that our DLA-CGM measurements with $\mrperp < R_{\rm true}$ actually fall within this $R_{\rm true}$, and hence that we are more likely to be comparing physically analogous regions.

We show the distribution of $P(\mrperp < R_{\rm true})$ expected for our sample adopting representative values of $\rho_{\rm DLA}$ and $R_{\rm true}$ in 
Figure~\ref{fig.cartoon_recenter}.  
The set of \rperp\ values for the QSO pairs with coverage of \ion{C}{4} in the CGM sightline is shown with black vertical hashes toward the top of the figure.  The colored curves show the probability $P(\mrperp < R_{\rm true})$ as a function of \rperp\ 
for values of $R_{\rm true} = 50$ kpc (solid) and $R_{\rm true} = 100$ kpc (dashed), with different colors corresponding to different values of 
$\rho_{\rm DLA}$ as indicated in the legend.  The points show the value of $P(\mrperp < R_{\rm true})$ corresponding to each sample sightline (although values of $P(\mrperp < R_{\rm true})=1$ are not plotted).

This figure demonstrates that for $\rho_{\rm DLA} \leq 20$ kpc, most sightlines having $\mrperp \leq 50$ kpc  or $\leq 100$ kpc 
have a high probability of lying within $R_{\rm true} \leq 50$ kpc or $\leq 100$ kpc, respectively.  For $\rho_{\rm DLA} = 20$ kpc, only 4 of 18 sightlines within $\mrperp < 100$ kpc have $P(\mrperp < 100 ~\rm kpc) < 1$, and in 3 of these 4 cases $P(\mrperp < 100 ~\rm kpc) \gtrsim 0.6$.  
Moreover, there are only two sightlines at $\mrperp > 100$ kpc with a non-zero $P(\mrperp < 100 ~\rm kpc)$, and these probability values are low ($\lesssim0.35$).  Therefore, in coadded spectra of all CGM sightlines having $\mrperp \leq 100$ kpc, 4 of the sightlines will have a $\sim30-50\%$ probability of lying at $R_{\rm true} > 100$ kpc.  Under the assumption that the CGM absorption strength declines with $R_{\rm true}$, the presumably weaker absorption in these few sightlines will tend to dilute the absorption signal measured in the coadded spectrum.  At the same time, a coadded spectrum of sightlines with $\mrperp > 100$ kpc may include a few sightlines with $R_{\rm true} < 100$ kpc: specifically, two sightlines with $\mrperp > 100$ kpc have a $\sim25-35\%$ probability of having $R_{\rm true} < 100$ kpc.  These sightlines therefore may tend to enhance the absorption signal measured at larger impact parameters.

These effects become more pronounced for larger values of $\rho_{\rm DLA}$.  For instance, the average value of $P(\mrperp <100~\rm kpc)$ for all sightlines with $\mrperp < 100$ kpc is 0.97 for $\rho_{\rm DLA} = 10$ kpc, 0.92 for $\rho_{\rm DLA} = 20$ kpc, 0.78 for $\rho_{\rm DLA} = 40$ kpc, 
and 0.61 for $\rho_{\rm DLA} = 70$ kpc.  Similarly, the likelihood of spurious enhancement of the absorption signal at large \rperp\ increases with $\rho_{\rm DLA}$: the average value of $P(\mrperp <100~\rm kpc)$ for all sightlines having $\mrperp > 100$ kpc is 0.04 for $\rho_{\rm DLA} = 20$ kpc and 0.12 for $\rho_{\rm DLA} = 70$ kpc. 

In more qualitative terms, our uncertainty in the value of $R_{\rm true}$ for our sample sightlines can be considered an additional source of 
systematic uncertainty in our assessment of the average CGM absorption strength as a function of projected distance from the centers of DLA host halos.  Under the assumption that this absorption declines in strength with $R_{\rm true}$, we expect that the primary repercussion of this uncertainty is a `dilution' or underestimation of the CGM absorption signal at small impact parameters.  The foregoing analysis suggests that the enhancement of CGM absorption at large \rperp\ due to the inclusion of sightlines at small $R_{\rm true}$ occurs with a relatively low probability.  
Because $\rho_{\rm DLA}$ is not well constrained and may span a wide range of values, we do not attempt to correct for these effects here.  However, they will be considered as we proceed with our comparison to previous results on the CGM absorption strength around optically-selected samples.

%Returning to our goal of understanding DLAs in the context of their host dark matter halo mass and their relation to star formation at high redshift, 
% we now compare the absorption strength of material in and around DLAs with that around optically-selected
%host galaxies at $z\sim2$.  

\begin{figure}
\begin{center}
\includegraphics[angle=0,width=\columnwidth,trim=0 120 160 0,clip=]{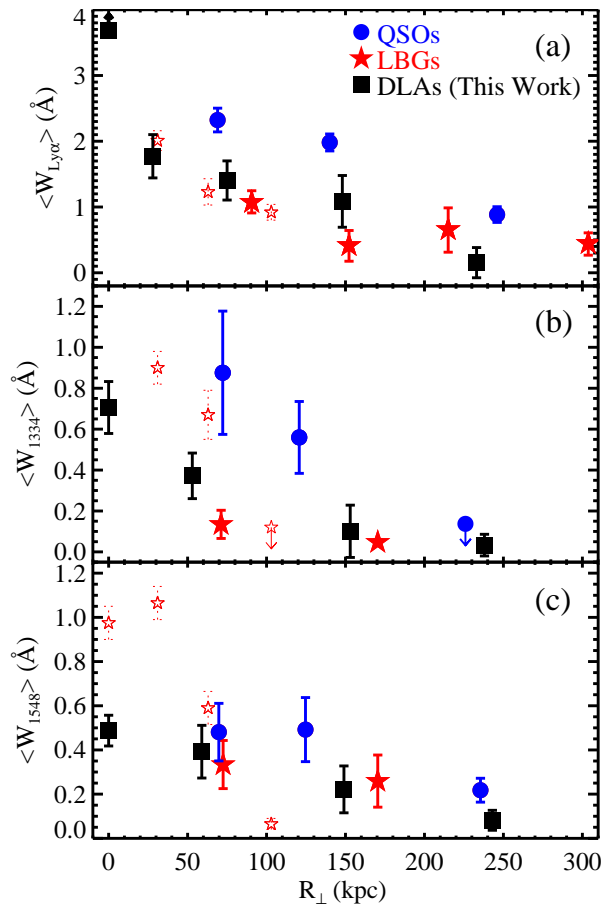}
\caption[]{\emph{(a)}  \avgWLya \ measured in the coadded spectra generated as described in \S\ref{sec.stacks} of DLA (at \rperp $= 0$ kpc) and CGM sightlines (black).
 Measurements of CGM absorption in coadded spectra of sightlines probing foreground QSO host halos are shown in blue (QPQ5).  The CGM absorption strength around LBGs measured along background LBG sightlines is shown with open red stars \citep{Steidel2010}, and LBG-CGM absorption measured toward background QSOs is indicated with solid red stars \citep{Adelberger2005,Simcoe2006,Rakic2011,Rudie2012,Crighton2014}.  
%The cyan solid and dashed lines indicate the median and $\pm25$th-percentile EWs measured in a 1 Mpc box centered on Eris \citep{Shen2012}.  
The CGM around DLAs exhibits \avgWLya \ similar to the material surrounding LBGs.
\emph{(b)} Same as panel \emph{(a)}, for \avgWCII.  The CGM around DLAs generally yields \avgWCII \ consistent with the CGM absorption strength measured around LBGs, although the DLA-CGM \avgWCII \ at $R_{\perp} \sim 50$ kpc is 
marginally discrepant with both measurements of the LBG-CGM at $\mrperp \sim 60-70$ kpc shown.
%the value reported in \citet{Steidel2010}
\emph{(c)} Same as panel \emph{(a)}, for \avgWCIV.  
\label{fig.comparison}}
\end{center}
\end{figure}

\subsubsection{Comparison with the CGM around Bright Galaxies and QSOs}\label{sec.comparison}
Figure~\ref{fig.comparison} shows \avgWLya \ (a), 
\avgWCII \ (b), and \avgWCIV \ (c) measured from the coadded spectra in Figure~\ref{fig.stacks} 
(black squares).  Symbols at $\mrperp = 0$ kpc show equivalent widths measured in the coadded spectra of DLA sightlines.
The absorption strength of CGM material around QSO host galaxies 
measured using a similar dataset is shown with filled blue circles (QPQ5).
The CGM absorption strength around LBGs measured along coadded background galaxy sightlines is shown with 
red open stars \citep{Steidel2010}, and measurements of the LBG-CGM absorption strength 
toward background QSOs assembled from the literature 
\citep{Adelberger2005,Simcoe2006,Rakic2011,Rudie2012,Crighton2014} are plotted with red filled stars.  
%% DONE: JFH I think this is now Crighton 2014:)

First,  regarding average equivalent widths measured along DLA sightlines, we find that they are significantly lower than equivalent widths measured down the barrel in coadded spectra of LBGs.  In particular, \citet{Steidel2010} measure $\langle W^\mathrm{LBG}_{1334} \rangle = 1.6 - 1.8$ \AA\ and $\langle W^\mathrm{LBG}_{1526} \rangle = 1.3 - 1.5$ \AA, values more than twice as large as $\langle W^\mathrm{DLA}_{1334} \rangle$ and $\langle W^\mathrm{DLA}_{1526} \rangle$ (Figure~\ref{fig.comparison}; Table~\ref{tab.stacks}).  
%% JFH Why don't you put the Steidel down the barrel numbers on the plot at rperp = 0 as well for the
%% the metals??
The relatively low absorption strength in DLAs suggests that their low-ion absorption profiles are tracing 
material with less extreme kinematics on average.  This may be a consequence of DLAs having lower host halo masses, particularly 
given the established correlation between $W^\mathrm{DLA}_{1526}$ and metallicity. 
%% JFH Again lost here. You say low-ions are tracing the 30 kpc environment, which may not be large
%% enough to sample halo kniematics. 
However, \citet{Steidel2010} argued that the large LBG \ion{C}{2} equivalent widths are due to large-scale outflows driven by star formation in the galaxies based on the overall blueshift of the transition (by $\sim 100-800\mkms$).  The low $\langle W^\mathrm{DLA}_{1334} \rangle$ may therefore instead indicate either that DLA host galaxies drive less extreme outflows, or that these outflows are not traced by the low-ion absorption because, e.g., the DLAs are not co-spatial with galactic star formation \citep{Fumagalli2014DLA}.  
%The identification of bright, emission counterparts of a few high-metallicity DLAs at impact parameters of up to $\sim20-30$ kpc leave open this latter possibility \citep{Fynbo2013} {\bf [although maybe not, since these DLAs probably have super high SiII EW?]}.   %as does the analysis presented in \S\ref{sec.eris}.
%% DONE: JFH Clarify what you mean by ``this latter possibility''. The writing is currently ambigous. 
%% JFH Discuss the CGM similarity

Turning to the CGM, we find that the equivalent widths of CGM absorption around DLAs and LBGs
are consistent within the measurement errors.  With the exception of the $\langle W^\mathrm{CGM}_{1334} \rangle$
measurement at $\sim 50$ kpc, which falls between the
\citet{Steidel2010} value of \avgWCII\ at \rperp $\sim60$ kpc
and the independent LBG-CGM measurement toward background QSOs at $\sim70$ kpc, 
%which at \rperp $\sim60$
%kpc is $1.8\sigma$ above $\langle W^\mathrm{CGM}_{1334} \rangle$ (and similarly discrepant
%with the independent LBG-CGM measured toward background QSOs; see
%solid red stars), 
every  DLA-CGM $\langle W \rangle$ reported is within
$\lesssim1\sigma$ of the neighboring LBG-CGM values.  This
suggests that on average, both LBGs and DLAs are surrounded by similar
gaseous environments, in spite of any differences in the distribution
of halo masses and/or star formation histories among the two populations.  
We additionally note that the cosmological `zoom-in' simulations of \citet{CAFG2014} 
predict this overall similarity, under the assumption that DLAs do indeed occupy 
smaller halos than bright LBGs.  In particular, they report that  
the covering fraction of optically thick \ion{H}{1} within $\mrperp < 100$ kpc
remains approximately constant over a range in halo mass $M_h \sim 10^{11-12}\msun$.  They do not 
report \avgW\ of \lya\ or metal-line absorption in the simulated CGM; however, 
we expect comparisons with more detailed predictions from such work to yield useful constraints on 
the physics adopted by the simulations.

Moreover, the
significant decline in both \avgWLya\ and \wlya\ with \rperp\ measured in our DLA-CGM sightlines is similar to
the trend exhibited in nearly all studies of CGM absorption centered
around magnitude-selected systems.
%This decline suggests that DLAs are
%located primarily near local maxima in CGM absorption strength, especially 
%considering that 
%% JFH This is misleading. The down the barrel bin should not be included in this 
%% decline with impact parameter, as it is damped and has a huge EW. If you only consider
%% the CGM bins, you have a pretty mild decreasing trend with R_perp. That said, I believe
%% that there is a correlation, but I don't think it is so convincing that you can argue
%% that DLAs reside at the centers of their DM halos. It certainly suggests that DLAs are 
%% centered in gaseious structures which have a correlation of ~ 100 kpc, but I think that is about
%% it. 
This finding conflicts with a picture in which DLAs are dominated by
absorption on the outskirts ($\mrperp \sim 100$ kpc) of the halos hosting bright LBGs, and
instead suggests that DLAs tend to arise close to the centers of their
halos.  Moreover, as discussed in \S\ref{sec.geometry}, 
if there is indeed a small offset between the DLA locations and their halo centers, 
the CGM absorption signal at a given $R_{\rm true}$ will likely be underestimated.  
This suggests that the consistency between DLA-CGM and LBG-CGM absorption measurements 
cannot be due to spurious sampling of regions with small $R_{\rm true}$, and is 
robust to the systematic scatter introduced by our experimental design.
%Thus, `correcting' for such an 
%effect would not likely lower our \avgW\ measurements significantly below the LBG-CGM.

%{\bf Does this need to get more quantitative?  Can it?  I
%  could ``reobserve'' the LBG-CGM from varying impact parameters to
%  see where we start to get inconsistency with the DLA-CGM.  But I
%  think the results will be suggestive at best.  Plus this is kind of
%  hokey.}

%However, spectroscopy 
%of higher fidelity and with improved S/N may yet reveal significant variations in absorption strength and/or  
%kinematics around these two types of host galaxies.

%%%%%%%%% Figure out issue of centering
%% LBG Wlya measurements from Rakic:
%% R=90kpc,   EW=1.07, up to 1.25, down to 0.91 Ang
%% R=152kpc, EW=0.42, up to 0.64, down to 0.18 Ang
%% R=215kpc, EW=0.66, up to 0.99, down to 0.31 Ang
%% Comparing the EW_DLA=1.79Ang bin to the EW_LBG=1.07Ang bin, we get a 2.2sigma diff
%% Comparing the EW_DLA=1.36Ang bin to the EW_LBG=0.42Ang bin, we get a 2.6sigma diff
%% But these differences will get washed out when you average over EWs on either side of the 
%% appropriate LBG rperp.

In contrast to the LBG-CGM, the CGM around QSOs yields marginally stronger low-ionization absorption than that around 
DLAs.  The QSO-CGM \avgWLya \ values are $\gtrsim2.2\sigma$ higher than the measured DLA-CGM 
absorption at comparable \rperp\ within 200 kpc.  
%{\bf [Probably want to use a DLA-CGM coadd with an \rperp range 0-100 kpc.]}
The QSO-CGM \avgWCII \ is likewise $\sim2\sigma$ stronger than our measurements of \avgWCII \ at \rperp $\lesssim200$ kpc
from DLAs.  
The general finding that the CGM around 
QSOs gives rise to the strongest absorption in low-ionization transitions (i.e., \lya, \ion{C}{2}) of any galaxy 
environment probed to date was discussed in detail in QPQ7, and the absorption in DLA environments assessed here 
offers no exception.  Indeed, it is noted in QPQ7  that QSO host halos exhibit
the strongest low-ionization CGM both at a given \rperp\ and at a given $\mrperp / R_{\rm vir}$
(with the fiducial QSO host halo virial radius $R_{\rm vir}^{\rm QSO} \sim 160$ kpc).
Furthermore, it was argued that this strong, cool gas absorption must result primarily from the 
relatively high masses of the halos hosting QSOs \citep{White2012}.

However, the QSO-CGM and DLA-CGM \avgWCIV \ values are discrepant only at $100~\mathrm{kpc} < R_{\perp} < 200$ kpc, and 
are very close within 100 kpc.  
Indeed, if we rescale the QSO-CGM measurements to account for the larger virial radii of the host halos
(and assume $R_{\rm vir}^{\rm DLA} \sim R_{\rm vir}^{\rm LBG} \sim 100$ kpc), we find that the blue, red, and black points in Figure~\ref{fig.comparison}c
lie nearly on top of each other.
This similarity is particularly noteworthy given it has been explicitly demonstrated 
that DLAs are very rarely detected within 200 kpc of QSOs (QPQ6). Considering this
concordance of the CGM \ion{C}{4} absorption in the context of host halo mass, we must conclude either that the absorption 
strength of this higher-ionization material has a weak mass dependence (if any), or that
DLAs and QSOs occupy halos which give rise to similar \ion{C}{4} kinematic widths, e.g., 
because DLA- and QSO-hosts in fact have similar virial masses.

Our finding that \ion{C}{4}-absorbing material is distributed over large scales (\S\ref{sec.results_kinematics}) and is therefore 
likely tracing virial halo motions tends to support the latter scenario over the former.
Furthermore,  the large cross-correlation amplitude measured in a 
clustering analysis of strong \ion{C}{4} systems and QSO host galaxies (QPQ7)
implies that these \ion{C}{4} absorbers do indeed occupy the same dark matter overdensities as bright QSOs, 
and additionally provides strong evidence against a scenario in which \ion{C}{4} is insensitive to halo mass.
On the other hand, it is difficult to reconcile the idea of a close association between QSOs and DLAs and 
the relative weakness of low-ionization absorption in the DLA-CGM.

These tensions notwithstanding, the comparisons described above offer new insight into the origin of 
metals extending many tens to hundreds of kpc from galaxies at $z\sim2$.  Metal absorption
in LBG environments has been attributed in the literature to powerful, metal-rich gas outflows driven to $> 100$ kpc distances by strong star formation activity in the central galaxy \citep{Steidel2010}.  
Bright LBGs, with typical SFRs $\sim20-50\msunyr$ \citep[e.g.,][]{Erb2006b}, 
do indeed exhibit strong outflows when observed `down the barrel', with metal-line absorption 
extending blueward of systemic velocity by up to $\sim800\mkms$ \citep{Steidel2010}.  However, 
the spatial extent and ultimate fate of this high-velocity material has remained unconstrained: to give 
rise to the observed blueshifted absorption, it need only cover the young stars in the LBGs extending over scales of 
a few kpc \citep{Rubin2013}.  

Adding a new layer to this picture, 
the present study has revealed a strong similarily between the CGM metal absorption strength around both LBGs and DLAs; that is, in the environments surrounding galaxies with SFRs which differ by at least an order of magnitude.  
In particular, \citet{Fumagalli2014DLAb} place a limit on the in-situ SFR of typical DLAs within $\sim6$ kpc of the QSO sightlines of $\lesssim0.65\msunyr$, and further determine that only a small 
minority of DLAs ($<13\%$) have SFRs $>2\msunyr$ within $\sim10$ kpc.  
We consider it implausible that systems with such low SFRs could give rise to  
powerful gas outflows similar to those attributed to bright LBGs, and yet the material in their surroundings 
exhibits very similar \ion{C}{2} and \ion{C}{4} absorption strengths.  
This suggests an alternative origin for the metals in \emph{both} the LBG- and DLA-CGM.  
Of course, some fraction of the DLA population is certainly in the vicinity of LBGs, 
allowing the possibility that the DLA-CGM is on occasion enriched by LBG winds.  
The probability of such enrichment may be estimated by invoking the cross-correlation function between DLAs and LBGs measured in \citet{Cooke2006} and the LBG luminosity function of \citet{Reddy2008}.
We find that a spherical volume extending 100 proper kpc from a DLA at $z=2.5$ has only a $\sim10\%$ 
probability of containing a bright ($R < 25.5$) LBG.  The vast majority of DLAs, therefore, appear to 
lie well beyond a plausible enrichment `radius' from ongoing, LBG-driven galactic outflows.
Alternative enrichment mechanisms for LBG and DLA environments could 
include tidal stripping or the accretion of gas which has been enriched and expelled from dwarf galaxies at an earlier epoch \citep{Shen2012}.  
We discuss these scenarios in more detail along with additional supporting evidence in the next subsection. %\S\ref{sec.civ}.

\begin{figure}%[ht]
\begin{center}
\includegraphics[angle=90,width=1.1\columnwidth]{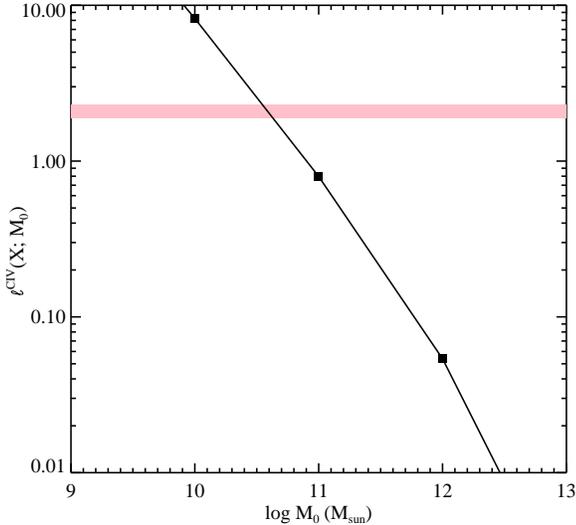}
\caption[]{The predicted incidence of strong \ion{C}{4} assuming every dark matter halo with mass $>M_\mathrm{0}$ hosts a DLA and is surrounded by $W_{1548} > 0.3$ \AA\ absorption  with a covering fraction $f_C = 50\%$ to $R_\mathrm{CIV} = 100$ kpc.  The pink band shows the observed random incidence of \ion{C}{4} having $W_{1548} > 0.3$ \AA\ (QPQ7).  If every DLA contributes to the random incidence of \ion{C}{4} absorption in this manner, the DLA population must predominantly arise in halos with masses $\gtrsim 10^{10.5}\msun$.
\label{fig.lox_civ}}
\end{center}
\end{figure}

\subsection{The CIV Halos of DLAs}\label{sec.civ}

In \S\ref{sec.results_metals}, we emphasized the high covering fraction of
strong \ion{C}{4} around DLAs: $\approx 57\%$ for $\mrperp < 100$ kpc.
We showed that the \ion{C}{4} equivalent widths along the CGM
sightlines frequently exceed half that measured along the corresponding DLA
sightline.  These properties imply a ubiquitous, highly ionized, and
enriched medium tracing the environments surrounding DLAs.
%with the enriched volume extending beyond the virial radius of these halos in some cases.
 Investigating further, in \S\ref{sec.results_kinematics}
we demonstrated a high degree of kinematic coherence between 
the \ion{C}{4} absorption along each pair of sightlines.  This coherence is exemplified by 
the small offsets in the flux-weighted velocity centroids of these profiles, which are 
$< 60\mkms$ for 8 of 12 pairs and never exceed $105\mkms$ over projected distances as large as $\mrperp = 176$ kpc.
Finally, inspired by the similarity between the LBG- and DLA-CGM absorption discussed in 
\S\ref{sec.comparison}, we note that \citet{Turner2014} detected enhanced \ion{C}{4} absorption
out to 2 Mpc from their LBG sample.  While this absorption is weak
 (0.01 \AA\ $ < $\avgWCIV\ $ < $ 0.1 \AA\ at $200~\rm kpc < \mrperp < 2$ Mpc), 
this finding nevertheless suggests that the \ion{C}{4} absorption around 
DLAs may in fact extend to much larger scales than are probed in this study.

%There is an additional aspect of the \ion{C}{4} absorption to
%emphasize: an apparent kinematic coherence between the absorption
%along each pair of sightlines.  

%In Figure~\ref{fig:cohere}, we present the XX~pairs that exhibit $5
%\sigma$ detections for \ion{C}{4}~1548 absorption in each sightline.
%[restrcti to less than 150kpc??]
%It is readiliy apparent that the profiles strongly overlap, both in
%velocity and component structure.  Quantitatively, we observed that
%XX/XX of these pairs are co-aligned to within XX~\kms.  [Describe the
%actual statistic used]
%Indeed, there is not a single sightlie that exhibits a greater than
%XX~\kms\ offset.

%{\bf [EVOKE: large gaseous structures (extending over large scales) with a large kinematic dispersion between them.]}

Together, these results offer unique constraints on the formation and
evolution of \ion{C}{4} `halos'.  On the one hand, the high incidence
of \ion{C}{4} requires wide-spread enrichment in the highly
ionized gas phase.    Previous works have invoked strong galactic winds to
enrich halo gas and drive the material to large scales \citep[e.g.,][]{Aguirre2001,Oppenheimer2006}.  
Some have further argued that the \ion{C}{4} observed along sightlines probing DLAs
directly traces the wind \citep{Fox2007}.
However, such a scenario lies in apparent conflict with the kinematics of the \ion{C}{4}-absorbing gas
associated with DLAs.  First, the coherence in kinematics between
sightlines of a given pair implies modest motions within the \ion{C}{4}-absorbing medium.  Wind speeds exceeding $100 \mkms$ are likely ruled
out by the observations, unless one invokes a fine-tuned geometry to
minimize velocity differences at $\sim 100$ kpc separations.
Second, the \ion{C}{4} absorption
appears to be dominated by gas extending over large distances.
The large covering fraction of 
$W^\mathrm{CGM}_{1548} > 0.2$ \AA\ absorption evident in Figure~\ref{fig.metals}c, 
along with the $\sim60\%$ frequency with which $\eta_{1548} > 0.5$
(Figure~\ref{fig.dtb_cgm}c), imply that the properties are largely
determined by gas at $R_\mathrm{3D} \sim 30-100$ kpc.
Such a geometry in turn implies a limited contribution from galactic wind
kinematics, which we expect to dominate on smaller scales (i.e., close to
the star-forming regions).
Considering these points together, we regard a wind-dominated scenario to be
implausible and encourage comparison of these observations to models
of galaxy formation which invoke high mass-loading factors in lower
mass galaxies (e.g., \citealt[][]{Vogelsberger2013,Vogelsberger2014}, Crain et al.\ 2014, in prep).

In lieu of winds, what may the \ion{C}{4} gas be tracing?
Reviewing the properties, one requires a medium that was previously
enriched, that is distributed over scales of $\sim 100$ kpc,
and has relatively quiescient kinematics.  
Perhaps the simplest picture to invoke is a 
highly-ionized medium 
that pervades the dark matter halos hosting
DLAs and in many cases extends beyond the virial radii of these halos.
%% JFH Quasi-static makes absolutely no physical sense here. The
%% pressure of this gas is way to low to support itself against gravity,
%% i.e. it would need to be 10^6 K for that. THe picture I think you want to invoke
%% is of cool photoionized (or possibly collisionally ionized clouds) in virial 
%% motion in the halo of the DLA. Even if the strcutures are filamentary, there is no
%% way for them to remain at rest, i.e. they will be moving in the halo potential
%% because there is no pressure support. They could be pressure confined by a hotter
%% phase, but even then you'll get motions chracteristic of the virial velocity. 
One may envision a filamentary structure of diffuse, enriched
gas that expresses \ion{C}{4} and has kinematics dominated by gravitational motions,
similar to the scenario first proposed
by \citet{Rauch1997}.  
%This filamentary gas may be slowly
%contracting/expanding and/or the gas may be slowly accreting onto the
%underlying dark matter halo.  %{\bf [Maybe add that these structures can, depending on their 
%geometry, give rise to large velocity widths along the line of sight.]}  
The early enrichment of filamentary infalling material may also
 explain the high metallicities of weak ($0.3~{\rm \AA} < W_{2796} < 1~{\rm \AA}$)
\ion{Mg}{2} absorbers at high redshifts ($z>2$; \citealt{Matejek2013}).
\citet{Rauch1997} additionally showed that such structures 
can naturally give rise to a broad range in \ion{C}{4} velocity widths, 
depending on their geometry and the orientation of the line of sight. 
%Most importantly, we contend that the 
%gas motions must not be strongly affected by the potential well of the DLA halo:
%the observed coherence scale is larger than the expected extent of these halos ($R_\mathrm{vir} \lesssim 90$ kpc), 
%particularly if we assume the DLA is close to the halo center.
%% JFH It makes no physical sense for the gas to not by affected by the potential well of the halo.
%% There is nothing to support it against gravity. Its own pressure is too low to support it. If it is
%% mixed in with a hot phase, then it could be pressure confined, but then it will be moving
%% through the hot medium at near virial velocity. The coherence you see could imply the kinematics
%% of this gas is correlated, but there is no way for it to be at rest. 
%(i.e.\ the gas does not have virialized motions). 
We do note at least one potential conflict with this simple scenario.
Observations along DLA sightlines have shown that the kinematics of
the \ion{C}{4} gas are more complex than those traced by the low-ion gas.  
In the cases of DLAs exhibiting both very broad low and high-ion absorption,
one may need to invoke additional motions on small scales (e.g.\ winds) to
account for the measurements.  The potential impact of winds on the DLA \ion{C}{4} absorption in 
such cases could in principle be tested by comparing the coherence of the \ion{C}{4} between 
paired sightlines with the same measurement in less extreme systems.
%% JFH I don't think this paragarph makes much physical sense. 

Irrespective of the origin of the \ion{C}{4} gas surrounding DLAs, we
may leverage the high covering fraction to provide a constraint on
the minimum mass of halos hosting DLAs, as follows.
We adopt two assumptions motivated by our observations and current
theories on neutral gas in high-$z$ galaxies:
 (1) we assume that all halos above a minimum mass $M_{0}$ contain
 sufficient \ion{H}{1} gas to satisfy our DLA criterion (\nhi\ $\ge 
  10^{20.1}\cmsq$);
 and (2) we assume every halo hosting DLAs exhibits extended, strong \ion{C}{4} absorption with
 covering fraction $f_C$  as observed in
 this study.
Under these assumptions, we may calculate the incidence of strong \ion
{C}{4} absorption as a function of $M_{0}$:  
\begin{equation}
\ell^{\rm CIV}(X;M_{0}) = n(M_{0}) f_C \pi R_{\rm CIV}^2,
\end{equation}
where $R_{\rm CIV}$ is the extent of the \ion{C}{4} gas, $f_C$ is  the
covering fraction for $\mrperp < R_{\rm CIV}$, and where $n(M_{0})$ is the
comoving number density of halos with $M_h>M_{0}$. 

Figure~\ref{fig.lox_civ} presents $\ell^{\rm CIV}(X; M_{0})$ assuming a
$\Lambda$CDM cosmology at $z=2.2$ with
%{\bf [should we use a different redshift? -- the median z of the sample is 2.2]}
$f_C = 0.5$ and $R_{\rm CIV} = 100$ kpc.
The pink band overplotted on the figure shows the incidence of strong ($W_{1548}>0.3$ \AA) \ion{C}{4} systems
at $z=2.1$ along random quasar sightlines (QPQ7).  We find that 
$\ell^{\rm CIV}(X; M_{0})$ exceeds the random value for $M_{0} <
10^{10.5} \msun$.  
This constraint is conservative in the sense that 
%We further note that 
strong \ion{C}{4} absorption
may also occur in astrophysical sightlines located far from DLAs. %in
%this respect our constraint on $M_{\rm min}$ is conservative.
However, our assumption that DLAs occupy all halos with masses $> M_{0}$
may not hold in practice, and relaxing this assumption would indeed allow for a contribution to the DLA population from 
halos with $M_h<M_{0}$.

%[finish up]

%[We might comment somewhere that the mass in metals may
%easily exceed that related to the neutral gas in DLAs.  Fox et
%al. have likely made this argument previously.  Or we can take their
%estimate for the ionized phase but now extend it well beyond the DLA]

The high incidence of strong \ion{C}{4} absorption and the large scales 
over which it is distributed point to a substantial reservoir of metals 
in the diffuse material surrounding DLAs at $z\sim2$.  We may roughly estimate the total 
mass in carbon in this CGM reservoir as follows:
\begin{equation*}
  M^{\rm CGM}_{\rm C} = \pi f_{\rm C} R_{\rm CIV}^2 \langle N_{\rm CIV} \rangle m_{\rm C} \frac{1}{x_{\rm CIV}}, 
\end{equation*}
where $x_{\rm CIV} = N_{\rm CIV} / N_{\rm C}$ and $m_{\rm C}$ is the mass of the carbon atom.
Here we conservatively set
$x_{\rm CIV} = 0.3$, as \citet{Fox2007OVI} demonstrated this to be the maximum possible ionization fraction
in models assuming either photo- or collisional ionization.
Adopting a typical column density of 
$\langle N_{\rm CIV}\rangle = 10^{14}\cmsq$ for \ion{C}{4} in DLAs  
\citep{Fox2007}, with $f_{\rm C} = 0.5$ and $R_{\rm CIV} = 100$ kpc
as above, we find $ M^{\rm CGM}_{\rm C} \approx 5\times 10^{5} \msun$.

We compare this value to an estimate of the total mass in carbon in the stars and neutral ISM of DLAs themselves.
We first assume that DLAs occupy dark matter halos having masses $M_h = 10^{11.5}\msun$ (only slightly lower in 
mass than the halos hosting LBGs; e.g., \citealt{Rakic2013}).
Such halos host galaxies having stellar masses $M^{\rm DLA}_* \approx 10^{9.3}\msun$ at $z\sim2$ \citep{Moster2013}, 
and which are likely to have high ISM gas fractions $f_{\rm gas} \gtrsim 0.5$ \citep{Tacconi2013}.  The total 
baryonic mass of these systems is then $M^{\rm DLA}_{\rm bar} = \frac{1}{1 - f_{\rm gas}} M^{\rm DLA}_*$, and the total mass
in carbon is
\begin{equation*}
M^{\rm DLA}_{\rm C} = \frac{Z_{\rm DLA}}{Z_{\odot}} {\rm (C/H)_{\odot}} \frac{1}{1 - f_{\rm gas}}  M^{\rm DLA}_*.  
\end{equation*}
If we assume a typical DLA metallicity of 
 $\log \frac{Z_{\rm DLA}}{Z_{\odot}} = -1.5$ \citep{Neeleman2013} and adopt a value for the solar abundance of carbon $\rm (C/H)_{\odot} = 10^{-3.57}$ \citep{Asplund2009}, 
we find $M^{\rm DLA}_{\rm C} \approx 3\times 10^{4}\msun$, an order of magnitude lower than $M^{\rm CGM}_{\rm C}$.
DLAs in less massive halos will likely have yet lower metallicities, and hence may
be deficient in carbon relative to the solar abundance pattern \citep{Cooke2011}.  We expect that the discrepancy 
between the carbon mass in the stars and ISM of such systems and the mass of carbon in their CGM 
will therefore be even more extreme.

The preceding estimates are crude, and suffer from uncertainties comparable to the differential 
between the two values obtained.  However, the exercise illustrates that diffuse CGM 
material must make up an important fraction of the universal metal budget as early as 
10 Gyr ago (see also \citealt{Lehner2014}), and may contain 
more metals than the stellar material and star-forming regions it surrounds.  This widespread distribution of 
metals, in combination with the apparent quiescence of the \ion{C}{4} kinematics,
 point to efficient enrichment processes occurring at much earlier times; e.g., powerful stellar feedback with 
high mass-loading factors from galaxies at $z\gtrsim3$. 
% {\bf[It would 
%take gas going $300\mkms$ a Gyr to travel 300 kpc -- and there's $\sim1$ Gyr between $z=3$ and 2.  
%Is this too silly?]}  
We await deep, rest-frame UV spectroscopy of star-forming systems at 
$z>3$, aided by gravitational lensing  \citep{Bayliss2013} or with next-generation 30m-class telescopes,
to confirm this picture.

\section{Conclusions}
With the goal of understanding the relationship between reservoirs of neutral hydrogen 
and star formation at early times, we have searched spectroscopy of paired QSO sightlines 
with projected separations $\mrperp < 300$ kpc for 
instances of damped \lya\ absorption in the foreground.  We use the second 
QSO sightline in each pair to characterize the \lya\ and metal-line absorption 
as a function of \rperp\ around 40 such systems.  
Our primary findings are as follows:

\begin{itemize}
\item Damped absorption rarely extends over scales $\gtrsim10$ kpc: 
the measured incidence of paired sightlines both exhibiting DLAs within $\mrperp < 100$ kpc
is $f_C^{\rm DLA} \sim 0.13$.  This incidence, although low, is consistent with 
a model in which the cross section of DLAs increases with halo mass and extends over $1400~\rm kpc^{2}$ in halos 
with $M_h = 10^{12}\msun$ \citep{Font-Ribera2012}.

\item We place a lower limit on the incidence of optically thick \ion{H}{1} absorption 
within 200 kpc of DLAs of $\gtrsim20-40\%$.  However,  
strong absorption from low-ionization metal transitions is rare in these environments, 
such that the covering fraction of $W^{\rm CGM}_{1526}> 0.2$ \AA\ absorption is $20^{+12}_{-8}\%$ %of sightlines
within $\mrperp < 100 $ kpc.  
%Furthermore, the majority of this absorption occurs in CGM sightlines having 
%\nhi\ $ > 10^{20.1}\cmsq$ (i.e., in `double DLAs').  
%This low incidence suggests that the true covering fraction of optically thick, metal-enriched
% \ion{H}{1} is not significantly higher than the limit listed above.  
These measurements 
 suggest that the low-ionization metal absorption observed toward DLAs themselves
is dominated by material within $\lesssim30$ kpc of the DLA, rather than by an extended 
gaseous halo.

\item We measure a high incidence ($57^{+12}_{-13}\%$) of strong \ion{C}{4} absorption 
($W^{\rm CGM}_{1548} > 0.2$ \AA) within 100 kpc of DLAs, with the absorption strength in the CGM 
frequently at least half that in the nearby DLA.  This absorption exhibits a high degree of 
kinematic coherence on scales of $\sim26-176$ kpc, with flux-weighted velocity centroids 
in the sightline pairs falling within $< 105\mkms$ in all cases.  These profiles must be
dominated by the motions of material at large physical separations from the DLAs; i.e., at $R_{\rm 3D}\sim 30-100$ kpc.
%This finding in turn implies that the \ion{C}{4} absorption cannot be 
%dominated by active galactic winds, and may instead arise from extended, metal-enriched gas filaments.
Under the assumption that all dark matter halos with masses above $M_{\rm 0}$ host DLAs,  
the ubiquity of \ion{C}{4} absorption in DLA environments requires that $M_{\rm 0} \gtrsim 10^{10.5}\msun$.  

%CIV has a high cf and coherent kinematics, constraint on min DLA halo mass
\item The equivalent width of \lya\ absorption in the DLA-CGM is anticorrelated with \rperp\ at $>98\%$ confidence.  This finding suggests that DLAs cannot predominantly arise near the outskirts of 
bright LBG host halos, and instead are likely located 
close to their halo centers.

\item  The average \lya\ and metal 
absorption strength in the environments extending to
$300$ kpc from DLAs %is significantly
% weaker than that observed in massive QSO host halos ($M_h\sim10^{12.5}\msun$).
%, and 
%thus DLAs do not likely arise in such dense environments.
%However, this absorption, along with absorption traced by \ion{C}{4}, 
is of similar strength to that exhibited by the CGM around LBGs.
This implies either (1) that the DLA population is dominated by systems 
hosted by halos similar in mass to those hosting LBGs ($M_h\sim10^{11.6-12}\msun$), or
(2) CGM absorption at $z\sim2$ does not change with halo mass over 
the range $10^{10.5}\msun \lesssim M_h \lesssim 10^{12}\msun$.

\item The close conformity between the DLA- and LBG-CGM, in combination with 
recently-reported limits on the SFRs of DLA hosts \citep{Fumagalli2014DLAb}, 
suggest that the distribution of metals in the outer regions of DLA- and LBG-host halos 
occurs via the dynamical accretion of previously-enriched material 
rather than via ongoing cool gas outflows.
%We further conclude that while DLAs cannot be dominated by absorption that is 
%precisely co-spatial with LBGs \citep{Fumagalli2014DLA}, the trend of declining 
%\avgWLya\ with \rperp\ suggests DLAs arise close to the centers of their host halos.
  
\end{itemize}

The foregoing discussion reports our initial efforts to constrain the properties of 
the diffuse DLA-CGM and the physical processes relevant to its origin.  Future directions will include 
analysis of detailed metal abundances and kinematics in our echellette-quality
spectroscopy for constraints on the degree of CGM metal enrichment relative to 
that of DLA material.  
Direct comparisons of these results with 
the metal abundances and kinematics of CGM material around DLAs in cosmological `zoom-in' simulations 
\citep[e.g.,][]{Shen2012,CAFG2014} and simulations of large cosmological volumes \citep[e.g.,][]{Bird2014a,Vogelsberger2014}
will inform future interpretation and provide crucial leverage on feedback prescriptions.
In combination with future efforts to characterize the emission from DLA hosts to deeper limits
than have yet been achieved \citep[e.g.,][]{Fumagalli2014DLAb}, these experiments 
will ultimately link the early reservoirs of neutral material with the formation of luminous structures
on every mass scale.

%So, what are DLAs after all?  Seems like we have a range $10^{10.5}\msun < M_h < 10^{12}\msun$.
%But does the claim that DLAs must be centered in their halos conflict with the need for a low 
%$f_C^{\rm DLA}$ within $\lesssim30$ kpc?  Probably not, as ``close to the center'' is not very quantitative.  
%Should we try to strengthen this point?
%Discuss how can we improve these constraints.

\acknowledgements

The authors wish to thank Sara Ellison, Crystal Martin, and 
George Djorgovski for aiding in collection and reduction of 
the Keck/ESI spectroscopy used in this study.
We thank Neil Crighton, Simeon Bird, Jeff Cooke, Claude-Andr{\'e} Faucher-Gigu{\`e}re, 
Michele Fumagalli, John O'Meara, Celine P{\'e}roux, and Sijing Shen for valuable discussion of this work, 
and we additionally acknowledge Sijing Shen for sharing the results of her analysis of the CGM properties 
of the Eris2 simulation.  Finally, we are grateful to Art Wolfe for sharing his insight on these results, 
and for his inspiring and pioneering work in this field.

K.H.R.R. acknowledges the generous support of the Clay Postdoctoral Fellowship, and 
J.F.H. acknowledges support from the Alexander von Humboldt foundation 
in the context of the Sofja Kovalevskaja Award. The Humboldt foundation is funded by the 
German Federal Ministry for Education and Research.
J.X.P. and M.L. acknowledge support from National Science Foundation (NSF) grants AST-1010004, 
AST-1109452, AST-1109447, and AST-1412981.

This paper includes data gathered with the 6.5 meter Magellan Telescopes located at Las Campanas Observatory, Chile.
Some of the data presented herein were obtained at the W. M. Keck Observatory, 
which is operated as a scientific partnership among the California Institute of Technology, 
the University of California, and the National Aeronautics and Space Administration. 
The Observatory was made possible by the generous financial support of the W. M. Keck Foundation. 
%Some of the Keck data were obtained through the NSF Telescope System Instrumentation Program (TSIP), 
%supported by AURA through the NSF under AURA Cooperative Agreement AST 01-32798 as amended.

%\clearpage
The authors wish to recognize and acknowledge the very significant
cultural role and reverence that the summit of Mauna Kea has always
had within the indigenous Hawaiian community.  We are most fortunate
to have the opportunity to conduct observations from this mountain.

%\bibliography{adssample}

\clearpage
{%\footnotesize
\centering
\begin{deluxetable*}{lcccccccc}
\tablecolumns{9}
\tablewidth{0pc}
\tablecaption{QSO Pair Observations and DLA Sample\label{tab.dlas}}\\
%\hline \hline  \\[-2ex]
\tablehead{
\colhead{QSO Pair Name \tablenotemark{a}} & \colhead{$z^\mathrm{QSO}_\mathrm{b/g}$} & \colhead{$z^\mathrm{QSO}_\mathrm{f/g}$} & \colhead{$R_{\perp}$} & \colhead{$z_\mathrm{DLA}$} & \colhead{$N^\mathrm{DLA}_\mathrm{HI}$} & \colhead{Instrument} & \colhead{Resolution} & \colhead{Date\tablenotemark{b}} \\ [0.5ex]
\colhead{}  & \colhead{}  &  \colhead{}  & \colhead{kpc} & \colhead{}  & \colhead{$\rm log [cm^{-2}]$}  & \colhead{} &\colhead{}  &\colhead{}}  %\\
%\hline \hline
%\endfirsthead
%\multicolumn{9}{c}{{\tablename} \thetable{} -- Continued} \\ [0.5ex]
%\hline \hline
%QSO Pair Name &  $z^{QSO}_{b/g}$ & $z^{QSO}_{f/g}$ & $R_{\perp}$ & $z_\mathrm{DLA}$ & $N^\mathrm{DLA}_\mathrm{HI}$ & Instrument  & Resolution & Date \\ [0.5ex]
%  &  &   & kpc &  & $\rm log ~[cm^{-2}]$ & & & \\
%\hline \\[-1.8ex]
%\endhead
%\\[-1.8ex] \hline \hline
%\multicolumn{9}{c}{{Continued on Next Page\ldots}} \\
%\endfoot
%\\[-1.8ex] \hline \hline
%\endlastfoot
 J0004-0844  &     3.00 &     3.00 &  35.6 &  2.75877 &  20.7 & Magellan/MagE &  4000 & 2008-06-28\\
 J0028-1049  &     3.10 &     2.62 & 175.6 &  2.58793 &  20.9 & Keck/ESI &  5000 & 2008-07-04\\
 J0040+0035 &     2.76 &     2.75 &  97.5 &  2.12990 &  20.6 & BOSS &  2000 & 2010-09-05\\
 J0201+0032 &     2.30 &     2.29 & 164.7 &  2.07593 &  20.1 & Gemini/GMOS &  1700 & 2004-11-19\\
 J0800+3542 &     2.07 &     1.98 & 202.1 &  1.78820 &  20.9 & Keck/LRISb &  2400 & 2007-04-13\\
 \nodata & \nodata & \nodata & \nodata & \nodata & \nodata & SDSS &  2000 & 2001-11-25\\
 J0833+3535 &     2.34 &     2.34 & 244.8 &  2.19860 &  20.2 & BOSS &  2000 & 2010-12-02\\
 J0913-0107 &     2.92 &     2.75 &  89.3 &  2.68874 &  20.3 & Magellan/MagE &  4000 & 2014-02-01\\
 J0920+1311  &     2.43 &     2.42 &  54.0 &  2.03572 &  20.1 & Magellan/MagE &  4000 & 2008-04-06\\
 J0920+1311  &     2.43 &     2.42 &  54.5 &  1.60723 &  20.2 & Magellan/MagE &  4000 & 2008-04-06\\
 J0932+0925 &     2.60 &     2.41 & 240.7 &  2.25198 &  20.6 & Magellan/MagE &  4000 & 2009-03-26\\
 J0955-0123 &     2.84 &     2.83 &  91.5 &  2.72677 &  20.8 & Gemini/GMOS &  1700 & 2013-09-02\\
 J1010+4037 &     2.51 &     2.18 & 193.0 &  2.04454 &  20.7 & BOSS &  2000 & 2011-01-09\\
 J1026+0629 &     3.12 &     3.12 &  79.3 &  2.56408 &  21.5 & Magellan/MagE &  4000 & 2009-03-23\\
 J1026+0629  &     3.12 &     3.12 &  77.8 &  2.78217 &  20.4 & Magellan/MagE &  4000 & 2009-03-23\\
 J1029+2623  &     2.21 &     2.20 &   7.5 &  1.97830 &  20.5 & Keck/LRISb &  2400 & 2007-01-17\\
 \nodata & \nodata & \nodata & \nodata & \nodata & \nodata & Keck/LRISr &  -999 & 2007-01-17\\
 \nodata & \nodata & \nodata & \nodata & \nodata & \nodata & SDSS &  2000 & 2006-02-28\\
 J1038+5027 &     3.24 &     3.13 & 240.9 &  2.79502 &  20.2 & Keck/ESI &  5000 & 2008-01-04\\
 J1045+4351 &     3.01 &     2.44 & 232.3 &  2.19625 &  21.3 & BOSS &  2000 & 2011-04-02\\
 \nodata & \nodata & \nodata & \nodata & \nodata & \nodata & Keck/LRISb &  2400 & 2008-05-08\\
 J1056-0059  &     2.13 &     2.12 &  62.7 &  1.96682 &  20.3 & Magellan/MagE &  4000 & 2014-02-01\\
 J1116+4118 &     3.00 &     2.94 & 114.2 &  2.66270 &  20.2 & Keck/ESI &  5000 & 2006-03-04\\
 J1150+0453  &     2.52 &     2.52 &  60.5 &  2.00063 &  20.6 & Gemini/GMOS &  1700 & 2013-08-30\\
 J1153+3530 &     3.05 &     2.43 &  77.5 &  2.34610 &  20.7 & BOSS &  2000 & 2011-03-28\\
 \nodata & \nodata & \nodata & \nodata & \nodata & \nodata & Keck/LRISb &  2400 & 2008-05-08\\
 J1236+5220  &     2.58 &     2.57 &  26.1 &  2.39644 &  21.1 & Gemini/GMOS &  1700 & 2013-07-11\\
 J1240+4329  &     3.26 &     3.25 &  24.9 &  3.09694 &  20.5 & Keck/ESI &  5000 & 2014-02-05\\
 J1240+4329  &     3.26 &     3.25 &  25.2 &  2.97887 &  21.1 & Keck/ESI &  5000 & 2014-02-05\\
 J1306+6158  &     2.17 &     2.10 & 142.0 &  1.88163 &  20.4 & Keck/LRISb &  2400 & 2005-03-09\\
 \nodata & \nodata & \nodata & \nodata & \nodata & \nodata & SDSS &  2000 & 2002-02-20\\
 J1307+0422 &     3.04 &     3.01 &  67.5 &  2.76528 &  20.1 & Magellan/MIKE & 22000 & 2004-05-09\\
 \nodata & \nodata & \nodata & \nodata & \nodata & \nodata & Magellan/MagE &  4000 & 2008-06-28\\
 J1307+0422 &     3.04 &     3.01 &  70.3 &  2.24969 &  20.1 & Magellan/MIKE & 22000 & 2004-05-09\\
 \nodata & \nodata & \nodata & \nodata & \nodata & \nodata & Magellan/MagE &  4000 & 2008-06-28\\
 J1416+3510 &     2.92 &     2.48 & 134.4 &  2.08596 &  20.4 & BOSS &  2000 & 2010-03-13\\
 J1428+0232 &     3.02 &     3.01 & 158.0 &  2.62613 &  21.1 & Gemini/GMOS &  1700 & 2013-09-02\\
 J1529+2314 &     2.64 &     2.49 &  86.3 &  2.07736 &  20.4 & BOSS &  2000 & 2011-04-29\\
 J1541+2702 &     3.63 &     3.62 &  49.6 &  3.32992 &  20.1 & Keck/ESI &  5000 & 2011-04-29\\
 J1542+1733  &     3.26 &     2.78 & 231.2 &  2.42660 &  21.1 & Keck/ESI &  5000 & 2008-06-05\\
 J1559+4943  &     1.95 &     1.86 & 210.1 &  1.78449 &  20.8 & Keck/LRISb &  2400 & 2008-05-08\\
 J1613+0808 &     2.39 &     2.38 &  84.4 &  1.61703 &  20.5 & Magellan/MagE &  4000 & 2008-06-29\\
 J1627+4606  &     4.11 &     3.81 & 259.7 &  3.54960 &  20.5 & Keck/ESI &  5000 & 2007-04-11\\
 J1630+1152 &     3.28 &     3.26 & 187.6 &  3.18047 &  20.3 & Gemini/GMOS &  1700 & 2012-09-06\\
 J1719+2549  &     2.17 &     2.17 & 127.3 &  2.01900 &  20.8 & Gemini/GMOS &  1700 & 2004-04-23\\
 J2103+0646  &     2.57 &     2.55 &  32.8 &  2.13902 &  20.5 & Gemini/GMOS &  1700 & 2013-07-11\\
 J2141-0229 &     2.71 &     2.70 & 107.9 &  2.10624 &  20.4 & Gemini/GMOS &  1700 & 2013-07-11\\
 J2146-0753  &     2.58 &     2.11 & 121.3 &  1.85306 &  20.5 & Keck/LRISb &  2400 & 2007-08-17\\
 \nodata & \nodata & \nodata & \nodata & \nodata & \nodata & SDSS & 2000 & 2001-09-21%\\
\enddata
\vspace{-0.4cm}
\tablenotetext{a}{Pair names followed by ``..." indicate objects for which more than one instrument was used. }\\
\tablenotetext{b}{UT date (YYYY-MM-DD) of the first night this object was observed by the instrument in column (7). }\\
\end{deluxetable*}
%\end{center}
\clearpage

{%\tiny
\begin{turnpage}
\begin{deluxetable}{lcccccccccc}
\caption{DLA and CGM Absorption Line Measurements\label{tab.dlaews}}\\
\hline \hline  \\[-2ex]
QSO Pair Name & $R_{\perp}$ & $z_\mathrm{DLA}$ & $W^\mathrm{DLA}_{1334}$ & $W^\mathrm{DLA}_{1526}$ & $W^\mathrm{DLA}_\mathrm{1548}$ & $N^\mathrm{CGM,a}_\mathrm{HI}$ & $W^\mathrm{CGM}_{\rm Ly\alpha}$ & $W^\mathrm{CGM}_{1334}$ & $W^\mathrm{CGM}_{1526}$ & $W^\mathrm{CGM}_{1548}$ \\ [0.5ex]
  & kpc &  & \AA & \AA & \AA & $\rm log [cm^{-2}]$ & \AA & \AA & \AA & \AA \\
\hline \hline
\endfirsthead
\multicolumn{11}{c}{{\tablename} \thetable{} -- Continued} \\ [0.5ex]
\hline \hline
QSO Pair Name & $R_{\perp}$ & $z_\mathrm{DLA}$ & $W^\mathrm{DLA}_{1334}$ & $W^\mathrm{DLA}_{1526}$ & $W^\mathrm{DLA}_\mathrm{1548}$ & $N^\mathrm{CGM,a}_\mathrm{HI}$ & $W^\mathrm{CGM}_{\rm Ly\alpha}$ & $W^\mathrm{CGM}_{1334}$ & $W^\mathrm{CGM}_{1526}$ & $W^\mathrm{CGM}_{1548}$ \\ [0.5ex]
  & kpc &  & \AA & \AA & \AA & $\rm log [cm^{-2}]$ & \AA & \AA & \AA & \AA \\
\hline \\[-1.8ex]
\endhead
\\[-1.8ex] \hline \hline
\multicolumn{11}{c}{{Continued on Next Page\ldots}} \\
\endfoot
\\[-1.8ex] \hline \hline
\multicolumn{11}{l}{Note 1: Transitions which fall in the Ly$\alpha$ forest of the corresponding QSO are indicated with ``...".}\\
\multicolumn{11}{l}{Note 2: Transitions which are affected by blending with unassociated systems are marked with $^b$.}\\
\multicolumn{11}{l}{$^a$ $N_{\mathrm HI}$ in the CGM sightline.  Values are not listed for spectra which have insufficient S/N to constrain $N_{\rm HI}$ (S/N $< 9.5~\rm \AA^{-1}$ at $\lambda_{\rm obs}^{\rm DLA}$.) }\\
\endlastfoot
 J0004-0844  &  35.6 &  2.75877 & $ 0.564 \pm  0.037$ & $ 0.656 \pm  0.055$ & $ 0.715 \pm  0.052$ & \nodata & $ 2.564 \pm  0.229$ & $ 0.011 \pm  0.130$ & $ 0.142 \pm  0.084$ & $ 0.324 \pm  0.121$\\
 J0028-1049  & 175.6 &  2.58793 & \nodata & $ 0.199 \pm  0.020$ & $ 1.009 \pm  0.023$ & \nodata & $ 0.688 \pm  0.096$ & $-0.063 \pm  0.085$ & $-0.065 \pm  0.045$ & $ 0.433 \pm  0.041$\\
 J0040+0035 &  97.5 &  2.12990 & \nodata & $ 0.333 \pm  0.027$ & $ 0.919 \pm  0.037$ & \nodata & $ 1.855 \pm  0.436$ & \nodata & $-0.171 \pm  0.243$ & $ 0.420 \pm  0.223$\\
 J0201+0032 & 164.7 &  2.07593 & $ 0.616 \pm  0.028$ & $ 0.235 \pm  0.026$ & $ 0.375 \pm  0.025$ & $< 19.3$ & $ 2.670 \pm  0.055$ & $ 0.118 \pm  0.009$ & $ 0.053 \pm  0.013$ & $ 0.405 \pm  0.013$\\
 J0800+3542 & 202.1 &  1.78820 & $ 0.918 \pm  0.013$ & $ 0.754 \pm  0.165$ & $ 0.984 \pm  0.164$ & $< 17.2$ & $ 0.600 \pm  0.046$ & \nodata & \nodata & \nodata\\
 J0833+3535 & 244.8 &  2.19860 & $ 2.017 \pm  0.069$ & $ 1.258 \pm  0.065$ & $ 2.081 \pm  0.063$ & \nodata & $ 0.633 \pm  0.315$ & $ 0.312 \pm  0.220$ & $-0.062 \pm  0.241$ & $ 0.092 \pm  0.190$\\
 J0913-0107 &  89.3 &  2.68874 & $ 1.272 \pm  0.033$ & $ 0.788 \pm  0.028$ & $ 0.299 \pm  0.031$ & $< 18.7$ & $ 1.287 \pm  0.033$ & $-0.062 \pm  0.058$ & $ 0.011 \pm  0.026$ & $ 0.136 \pm  0.025$\\
 J0920+1311  &  54.0 &  2.03572 & \nodata & $ 0.049 \pm  0.014$ & $ 0.219 \pm  0.009$ & $< 18.0$ & $ 0.854 \pm  0.027$ & \nodata & $ 0.003 \pm  0.015$ & $-0.008 \pm  0.011$\\
 J0920+1311  &  54.5 &  1.60723 & \nodata & \nodata & \nodata & \nodata & $ 1.132 \pm  0.158$ & \nodata & \nodata & \nodata\\
 J0932+0925 & 240.7 &  2.25198 & $ 0.652 \pm  0.049$ & $ 0.277 \pm  0.057$ & $ 0.265 \pm  0.034$ & $< 15.0$ & $ 0.262 \pm  0.023$ & \nodata & $ 0.049 \pm  0.014^b$ & $ 0.007 \pm  0.008$\\
 J0955-0123 &  91.5 &  2.72677 & $ 1.957 \pm  0.068$ & $ 1.765 \pm  0.065$ & $ 1.207 \pm  0.051$ & $< 19.6$ & $ 2.811 \pm  0.192$ & $ 0.532 \pm  0.113$ & $ 0.373 \pm  0.114$ & $ 1.077 \pm  0.113$\\
 J1010+4037 & 193.0 &  2.04454 & \nodata & $ 1.511 \pm  0.074$ & $ 1.446 \pm  0.064$ & $< 18.4$ & $ 0.897 \pm  0.073$ & $ 0.042 \pm  0.051$ & $-0.044 \pm  0.055$ & $-0.011 \pm  0.051$\\
 J1026+0629 &  79.3 &  2.56408 & \nodata & $ 0.896 \pm  0.055$ & $ 2.040 \pm  0.064$ & $ 20.1$ & $ 7.783 \pm  0.172$ & \nodata & $ 1.036 \pm  0.072$ & $ 1.896 \pm  0.050$\\
 J1026+0629  &  77.8 &  2.78217 & $ 0.785 \pm  0.040$ & $ 0.672 \pm  0.052$ & $ 0.614 \pm  0.069$ & $< 18.6$ & $ 1.290 \pm  0.109$ & $ 0.034 \pm  0.028$ & $ 0.144 \pm  0.067$ & $ 0.110 \pm  0.067$\\
 J1029+2623  &   7.5 &  1.97830 & \nodata & $-0.882 \pm  0.577$ & $ 0.019 \pm  0.214$ & $< 18.7$ & $ 1.568 \pm  0.026$ & $ 1.421 \pm  0.089^b$ & $-0.034 \pm  0.094$ & $-0.121 \pm  0.106$\\
 J1038+5027 & 240.9 &  2.79502 & $ 0.227 \pm  0.006$ & $ 0.103 \pm  0.007$ & $ 0.022 \pm  0.006^b$ & $< 18.9$ & $ 1.578 \pm  0.037$ & \nodata & $ 0.013 \pm  0.019$ & $ 0.340 \pm  0.021$\\
 J1045+4351 & 232.3 &  2.19625 & $ 0.311 \pm  0.022$ & $ 0.177 \pm  0.088$ & $ 0.131 \pm  0.141$ & $< 19.6$ & $ 3.057 \pm  0.018$ & \nodata & \nodata & \nodata\\
 J1056-0059  &  62.7 &  1.96682 & $ 1.150 \pm  0.075$ & $ 1.003 \pm  0.062$ & $ 1.425 \pm  0.075$ & $< 19.4$ & $ 3.062 \pm  0.102$ & $ 0.864 \pm  0.048$ & $ 0.726 \pm  0.027^b$ & $ 2.456 \pm  0.046$\\
 J1116+4118 & 114.2 &  2.66270 & $ 0.492 \pm  0.009$ & $ 0.124 \pm  0.021$ & $ 1.136 \pm  0.012$ & $ 20.4$ & $10.078 \pm  0.023$ & $ 1.004 \pm  0.006$ & $ 0.731 \pm  0.016$ & $ 0.193 \pm  0.006$\\
 J1150+0453  &  60.5 &  2.00063 & \nodata & $ 0.828 \pm  0.044$ & $ 0.613 \pm  0.056$ & \nodata & $ 0.954 \pm  0.202$ & \nodata & $ 0.028 \pm  0.045$ & $ 0.114 \pm  0.044$\\
 J1153+3530 &  77.5 &  2.34610 & \nodata & $ 0.330 \pm  0.266$ & $ 1.035 \pm  0.303^b$ & $< 19.4$ & $ 3.413 \pm  0.011$ & $ 0.411 \pm  0.067$ & $-0.028 \pm  0.069$ & $ 0.827 \pm  0.068$\\
 J1236+5220  &  26.1 &  2.39644 & $ 0.975 \pm  0.060$ & $ 0.779 \pm  0.048$ & $ 0.462 \pm  0.058$ & $< 19.1$ & $ 1.637 \pm  0.103$ & $ 0.173 \pm  0.072$ & $ 0.066 \pm  0.044$ & $ 0.268 \pm  0.062$\\
 J1240+4329  &  24.9 &  3.09694 & $ 0.879 \pm  0.018$ & $ 0.810 \pm  0.018$ & $ 0.259 \pm  0.018$ & $ 20.1$ & $ 5.868 \pm  0.108$ & $ 0.646 \pm  0.040$ & $ 0.231 \pm  0.027$ & $ 0.128 \pm  0.028^b$\\
 J1240+4329  &  25.2 &  2.97887 & $ 1.319 \pm  0.017$ & $ 0.997 \pm  0.021$ & $ 0.391 \pm  0.019$ & $< 19.2$ & $ 1.924 \pm  0.055$ & $ 0.032 \pm  0.022$ & $ 0.019 \pm  0.028$ & $ 0.088 \pm  0.026$\\
 J1306+6158  & 142.0 &  1.88163 & \nodata & \nodata & \nodata & $< 17.2$ & $ 0.041 \pm  0.069$ & $ 0.019 \pm  0.022$ & $ 1.355 \pm  0.390^b$ & $ 0.858 \pm  0.467$\\
 J1307+0422 &  67.5 &  2.76528 & $ 0.670 \pm  0.008$ & $ 0.536 \pm  0.013$ & $ 0.842 \pm  0.016$ & $< 18.1$ & $ 0.766 \pm  0.014$ & $-0.005 \pm  0.005$ & $-0.006 \pm  0.007$ & $ 0.204 \pm  0.007$\\
 J1307+0422 &  70.3 &  2.24969 & \nodata & $ 0.220 \pm  0.006$ & $ 0.393 \pm  0.006$ & $< 18.7$ & $ 1.140 \pm  0.020$ & \nodata & $ 0.057 \pm  0.010$ & $ 0.072 \pm  0.012$\\
 J1416+3510 & 134.4 &  2.08596 & \nodata & $ 0.606 \pm  0.159$ & $ 0.642 \pm  0.129$ & \nodata & $ 2.311 \pm  0.200$ & \nodata & \nodata & \nodata\\
 J1428+0232 & 158.0 &  2.62613 & \nodata & $ 0.740 \pm  0.035$ & $ 0.632 \pm  0.028$ & $< 19.2$ & $ 2.015 \pm  0.065$ & \nodata & $ 0.079 \pm  0.045$ & $ 0.019 \pm  0.036$\\
 J1529+2314 &  86.3 &  2.07736 & \nodata & $ 0.630 \pm  0.086$ & $ 0.057 \pm  0.096$ & \nodata & $ 0.190 \pm  0.224$ & \nodata & $ 0.045 \pm  0.142$ & $ 0.313 \pm  0.135$\\
 J1541+2702 &  49.6 &  3.32992 & $ 0.999 \pm  0.019$ & $ 0.546 \pm  0.028$ & $ 1.847 \pm  0.034$ & $< 19.0$ & $ 1.779 \pm  0.024$ & $ 0.237 \pm  0.017$ & $ 0.008 \pm  0.018$ & $ 1.398 \pm  0.023$\\
 J1542+1733  & 231.2 &  2.42660 & \nodata & $ 0.751 \pm  0.012$ & $ 0.160 \pm  0.014$ & $< 18.5$ & $ 0.882 \pm  0.046$ & \nodata & $ 0.360 \pm  0.007^b$ & $ 0.019 \pm  0.009$\\
 J1559+4943  & 210.1 &  1.78449 & $ 0.173 \pm  0.029$ & \nodata & \nodata & $< 17.2$ & $ 0.281 \pm  0.104$ & $ 0.075 \pm  0.044$ & \nodata & \nodata\\
 J1613+0808 &  84.4 &  1.61703 & \nodata & \nodata & \nodata & \nodata & $ 1.971 \pm  0.139$ & \nodata & \nodata & \nodata\\
 J1627+4606  & 259.7 &  3.54960 & \nodata & $ 0.032 \pm  0.008$ & $ 0.060 \pm  0.009$ & $< 14.8$ & $ 0.324 \pm  0.008$ & $-0.003 \pm  0.006$ & $ 0.001 \pm  0.006$ & $-0.006 \pm  0.007$\\
 J1630+1152 & 187.6 &  3.18047 & $ 0.192 \pm  0.073$ & $ 0.069 \pm  0.036$ & $ 0.161 \pm  0.032$ & $< 18.6$ & $ 1.030 \pm  0.010$ & $ 0.002 \pm  0.029$ & $ 0.003 \pm  0.018$ & $ 0.026 \pm  0.017$\\
 J1719+2549  & 127.3 &  2.01900 & $ 1.178 \pm  0.024$ & $ 0.694 \pm  0.017$ & $ 0.647 \pm  0.013^b$ & $< 17.8$ & $ 0.854 \pm  0.056$ & $-0.029 \pm  0.021$ & $-0.003 \pm  0.015$ & $ 0.129 \pm  0.020$\\
 J2103+0646  &  32.8 &  2.13902 & \nodata & $ 0.429 \pm  0.036$ & $ 0.230 \pm  0.040$ & $< 19.3$ & $ 2.231 \pm  0.212$ & \nodata & $ 0.298 \pm  0.041^b$ & $ 0.351 \pm  0.046$\\
 J2141-0229 & 107.9 &  2.10624 & \nodata & $ 0.468 \pm  0.066$ & $ 0.763 \pm  0.064$ & \nodata & $ 1.055 \pm  0.310$ & \nodata & $ 0.012 \pm  0.087$ & $-0.159 \pm  0.082$\\
 J2146-0753  & 121.3 &  1.85306 & \nodata & \nodata & \nodata & $< 18.8$ & $ 1.292 \pm  0.013$ & $ 0.418 \pm  0.007^b$ & $ 0.347 \pm  0.082$ & $ 1.050 \pm  0.141$\\
\end{deluxetable}
\clearpage
\end{turnpage}}

\begin{deluxetable}{lcccccccccc}
\caption{$\langle W \rangle$ measured in coadded DLA and CGM sightlines\label{tab.stacks}}\\
\hline \hline  \\[-2ex]
\rperp Range & \multicolumn{2}{c}{HI $\rm Ly\alpha$}  & \multicolumn{2}{c}{CII 1334} & \multicolumn{2}{c}{SiIV 1393} & \multicolumn{2}{c}{SiII 1526} & \multicolumn{2}{c}{CIV 1548} \\ [0.5ex]
     &  $\langle \mrperp \rangle$   & $\langle W_\mathrm{Ly\alpha} \rangle$ & $\langle \mrperp \rangle$ & $\langle W_\mathrm{1334} \rangle$ & $\langle \mrperp \rangle$ & $\langle W_\mathrm{1393} \rangle$ & $\langle \mrperp \rangle$ & $\langle W_\mathrm{1526} \rangle$ & $\langle \mrperp \rangle$ & $\langle W_\mathrm{1548} \rangle$ \\
 kpc & kpc & \AA & kpc & \AA & kpc & \AA & kpc & \AA & kpc & \AA \\
\hline \hline
\endfirsthead
\multicolumn{11}{c}{{\tablename} \thetable{} -- Continued} \\ [0.5ex]
\hline \hline
\hline \\[-1.8ex]
\endhead
\\[-1.8ex] \hline \hline
\multicolumn{11}{c}{{Continued on Next Page\ldots}} \\
\endfoot
\\[-1.8ex] \hline \hline
\multicolumn{11}{l}{Note 1: All $\langle W \rangle$ values are measured in a relative velocity window $-500\mkms < \delta v < 500\mkms$, with the exception of $\langle W_{1548} \rangle$,}\\
\multicolumn{11}{l}{which is measured in a window  $-500\mkms < \delta v < 249\mkms$.  }\\
\multicolumn{11}{l}{Note 2: We have excluded spectra from these coadds for which the transition of interest lies in the Ly$\alpha$ forest; i.e., if}\\
\multicolumn{11}{l}{$\lambda < (1215.6701~\mathrm{\AA}) * (1 + z_{\rm QSO}) + 20~\rm \AA$.}\\
\endlastfoot
 $ \rm DLAs$ &   0 & $ 3.69\pm 0.05$ &   0 & $ 0.71\pm 0.13$ &   0 & $ 0.32\pm 0.06$ &   0 & $ 0.35\pm 0.09$ &   0 & $ 0.49\pm 0.07$\\
 $0 < \mrperp < 50$ &  28 & $ 1.77\pm 0.33$ & \nodata & \nodata & \nodata & \nodata & \nodata & \nodata & \nodata & \nodata\\
 $50 < \mrperp < 100$ &  75 & $ 1.40\pm 0.30$ & \nodata & \nodata & \nodata & \nodata & \nodata & \nodata & \nodata & \nodata\\
 $0 < \mrperp < 100$ &  60 & $ 1.53\pm 0.25$ &  53 & $ 0.37\pm 0.11$ &  54 & $ 0.15\pm 0.12$ &  59 & $ 0.12\pm 0.07$ &  59 & $ 0.39\pm 0.12$\\
 $100 < \mrperp < 200$ & 148 & $ 1.09\pm 0.39$ & 153 & $ 0.10\pm 0.13$ & 154 & $-0.08\pm 0.04$ & 149 & $ 0.15\pm 0.12$ & 149 & $ 0.22\pm 0.11$\\
 $200 < \mrperp < 300$ & 233 & $ 0.15\pm 0.23$ & 238 & $ 0.03\pm 0.05$ & 238 & $ 0.04\pm 0.06$ & 243 & $ 0.10\pm 0.12$ & 243 & $ 0.08\pm 0.05$\\
 $0 < \mrperp < 300$ & 119 & $ 1.10\pm 0.18$ & 112 & $ 0.23\pm 0.08$ & 117 & $ 0.06\pm 0.07$ & 112 & $ 0.12\pm 0.06$ & 112 & $ 0.30\pm 0.07$\\
\end{deluxetable}

\appendix

Figure~\ref{fig.showspecs1} shows our spectroscopy of all DLAs and the corresponding CGM systems  included in this study.

%% Use psselect for this
%% psselect -p1 fig_dlacgm_showallspecs.ps fig_dlacgm_showallspecs1.ps
\begin{figure*}[ht]
\begin{center}
\includegraphics[angle=90,width=0.8\textwidth]{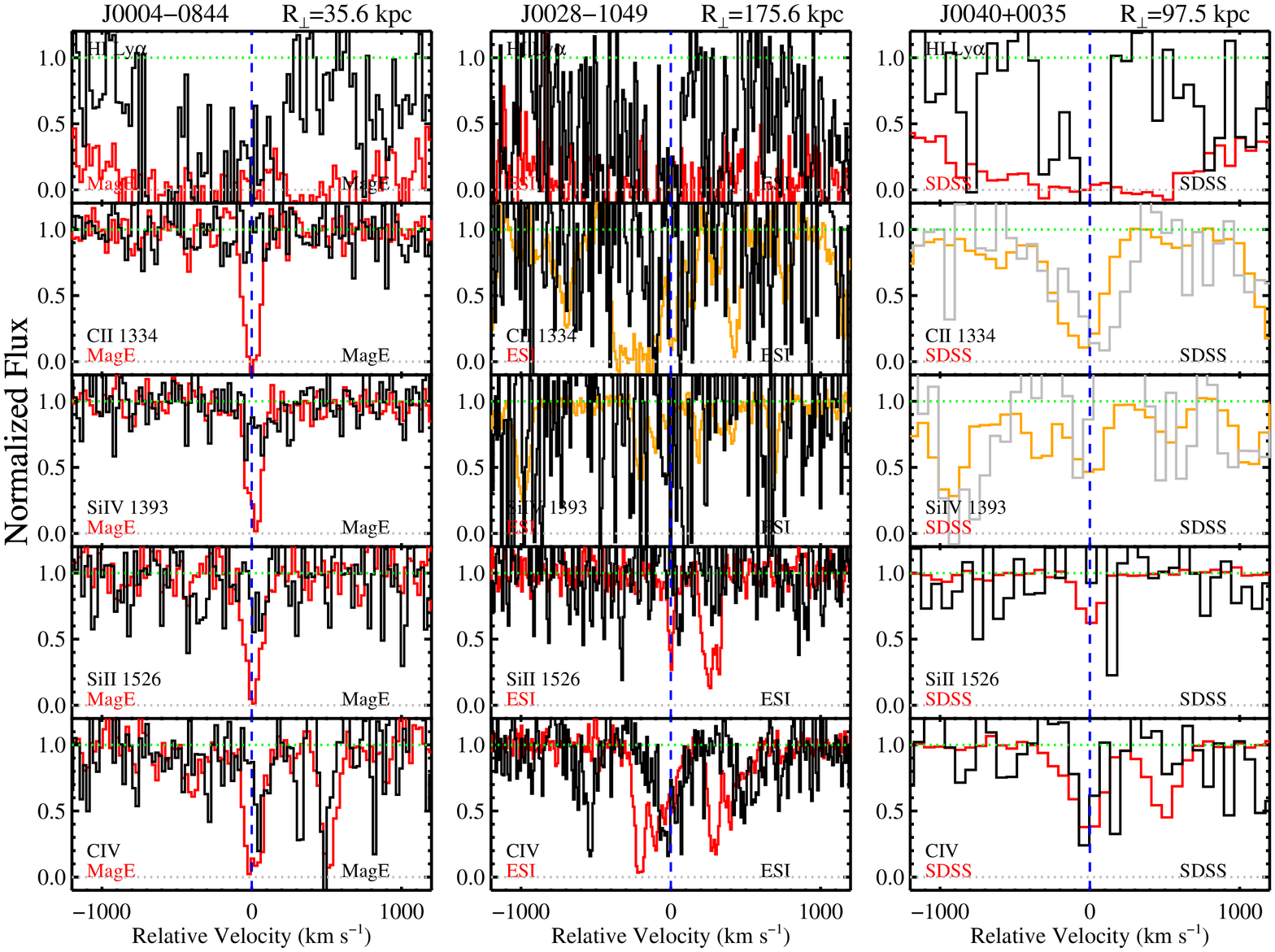}
\caption[]{QSO pair spectroscopy.  
Each column shows the \lya, \ion{C}{2}, \ion{Si}{4}, \ion{Si}{2}, and \ion{C}{4} absorption transitions due to a DLA (red histogram).  The black histogram shows the CGM absorption 
probed by the secondary QSO in each pair at the same redshift as the DLA.  The physical separation 
of each pair and the QSO pair ID are given at the top of each column, and the instrument used to obtain each spectrum is marked in red and black, respectively.  Spectral regions falling in the \lya\ forest are shown with orange (for DLAs) or gray (for CGM sightlines) histograms (for metal lines only).  Metal-line transitions affected by blending with unassociated systems are shown with dotted lines.
A subset of our 
sample was observed at high spectral resolution (FWHM $\lesssim50\mkms$) 
with, e.g., Magellan/MagE or Keck/ESI.  
The majority of our pairs, however, were observed with medium-resolution setups (FWHM $\sim 125-180\mkms$; with Keck/LRIS, Gemini/GMOS, etc).  
\label{fig.showspecs1}}
\end{center}
\end{figure*}

\setcounter{figure}{0}
\begin{figure*}%[ht]
\begin{center}
\includegraphics[angle=90,width=0.8\textwidth]{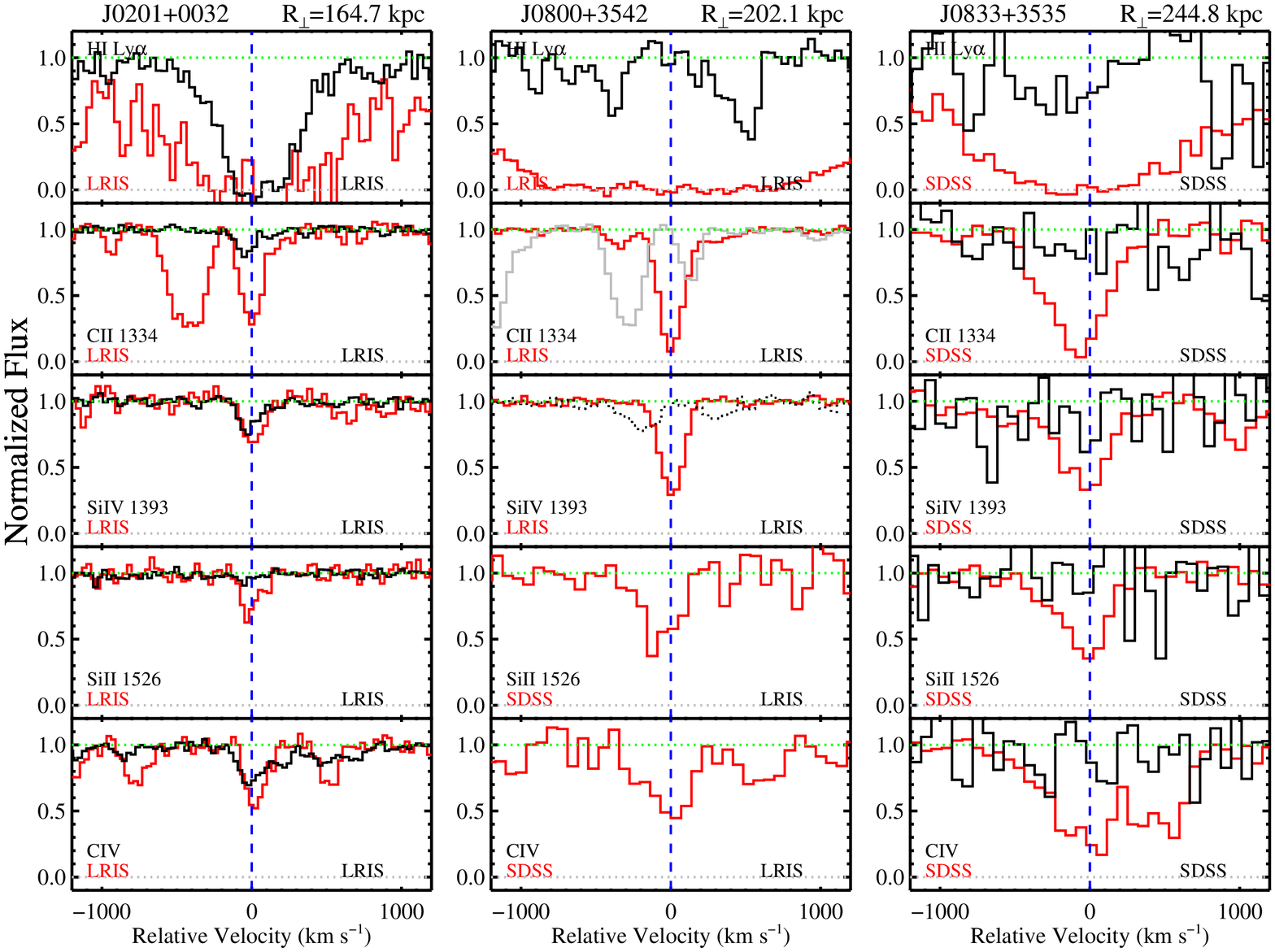}
\includegraphics[angle=90,width=0.8\textwidth]{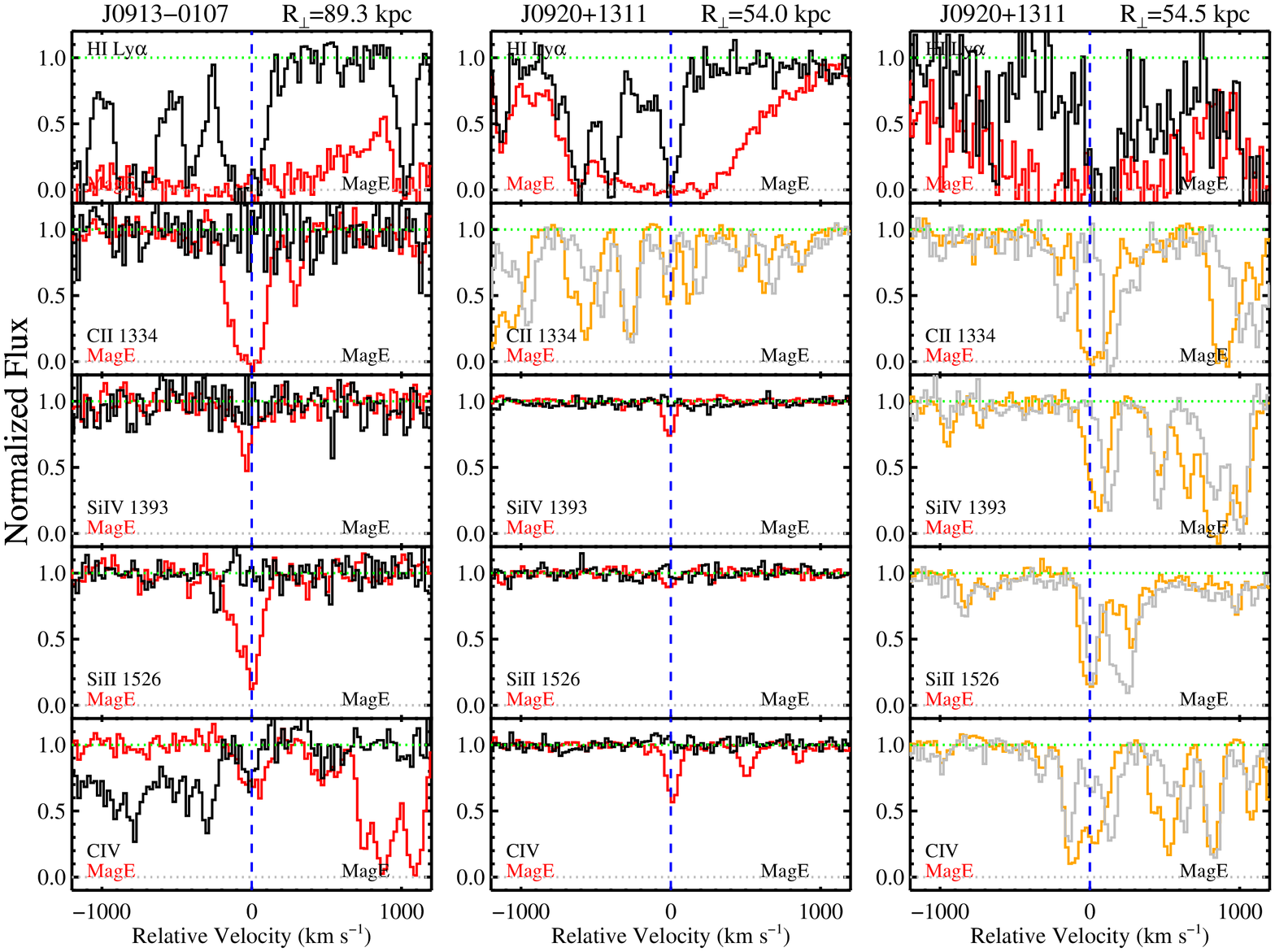}
\caption[]{ -- continued%Figure~\ref{fig.showspecs1} (continued)
\label{fig.showspecs2}}
\end{center}
\end{figure*}

\setcounter{figure}{0}
\begin{figure*}%[ht]
\begin{center}
\includegraphics[angle=90,width=0.8\textwidth]{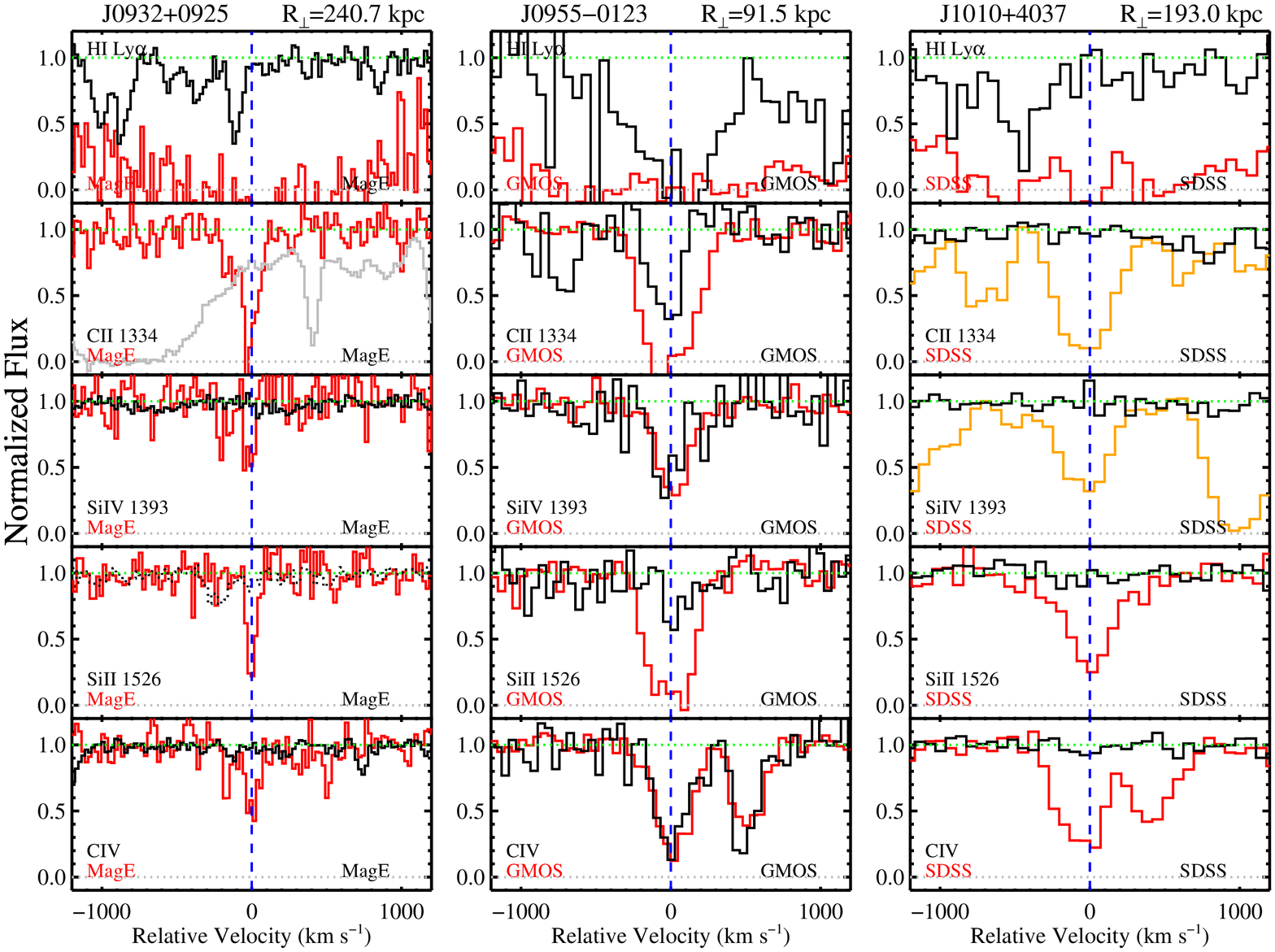}
\includegraphics[angle=90,width=0.8\textwidth]{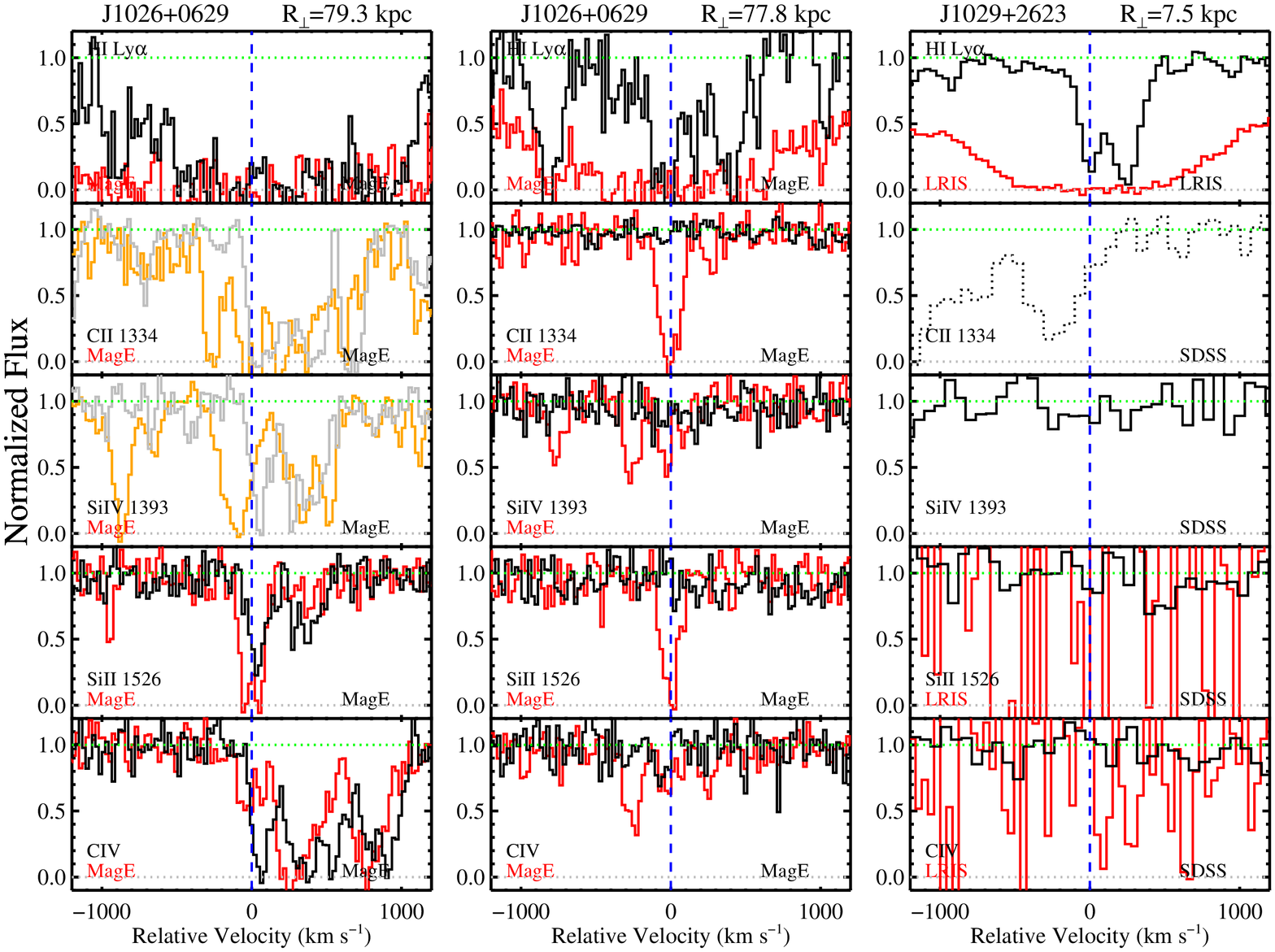}
\caption[]{ -- continued
\label{fig.showspecs3}}
\end{center}
\end{figure*}

\setcounter{figure}{0}
\begin{figure*}%[ht]
\begin{center}
\includegraphics[angle=90,width=0.8\textwidth]{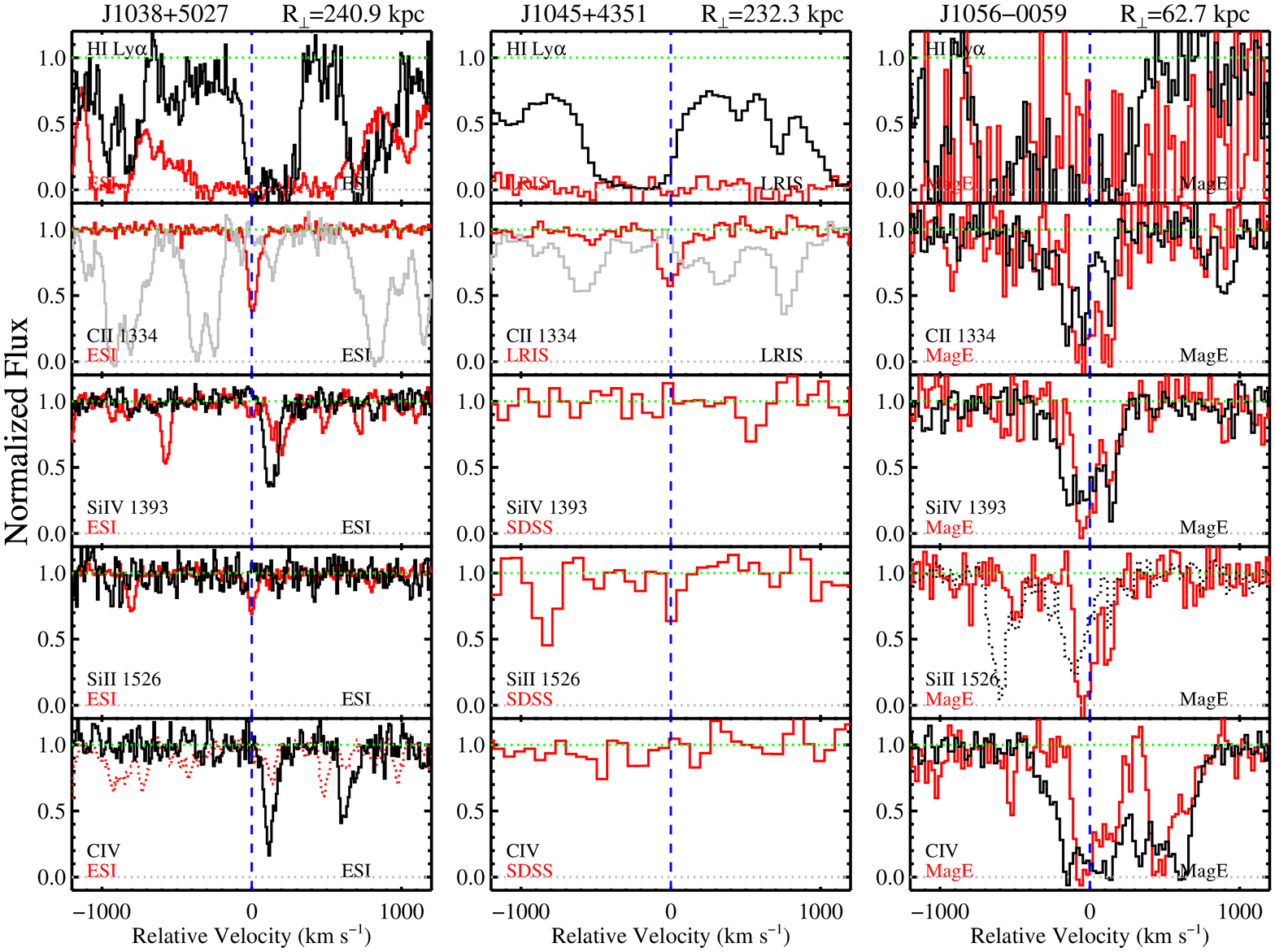}
\includegraphics[angle=90,width=0.8\textwidth]{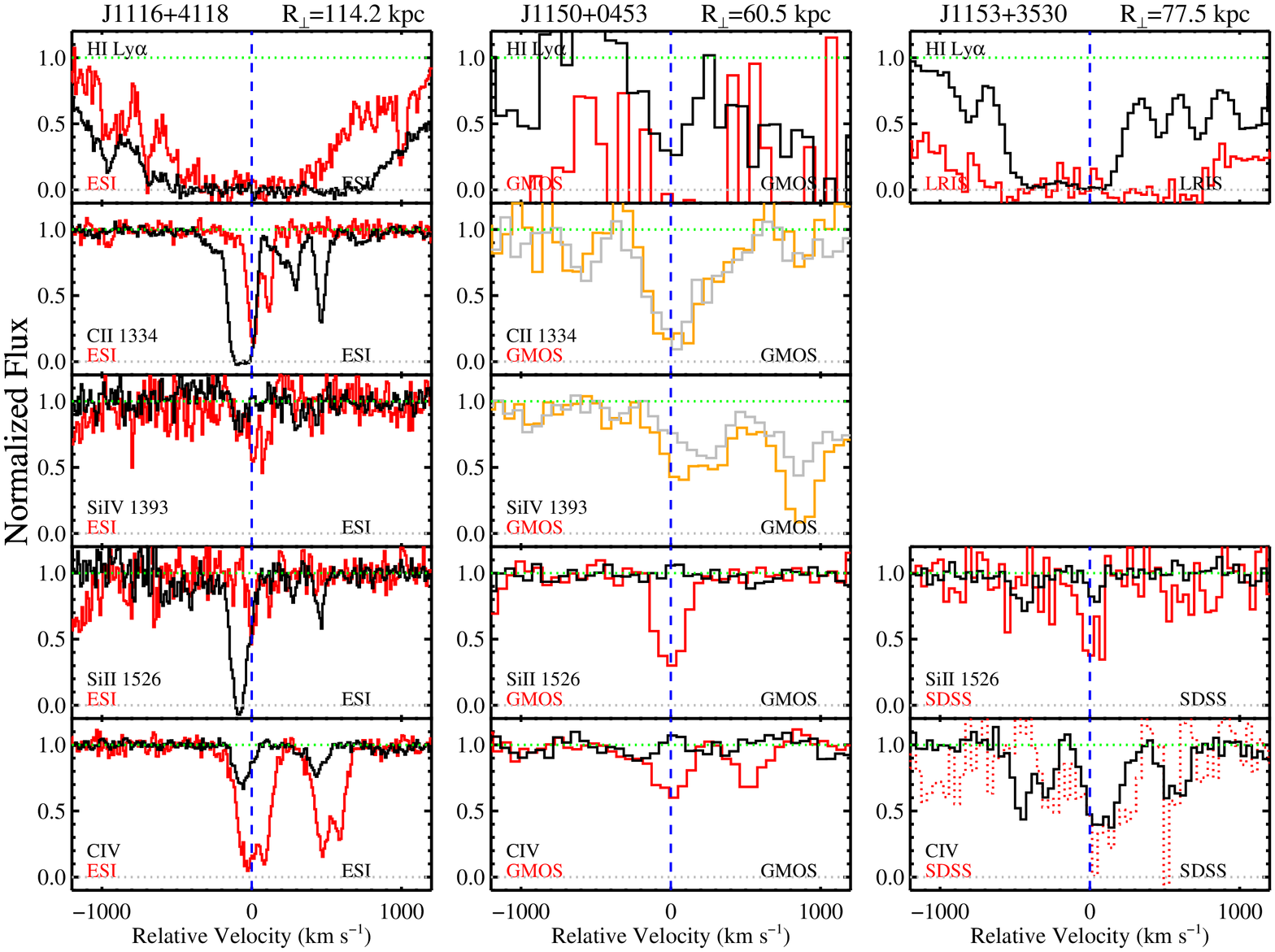}
\caption[]{ -- continued
\label{fig.showspecs4}}
\end{center}
\end{figure*}

\setcounter{figure}{0}
\begin{figure*}%[ht]
\begin{center}
\includegraphics[angle=90,width=0.8\textwidth]{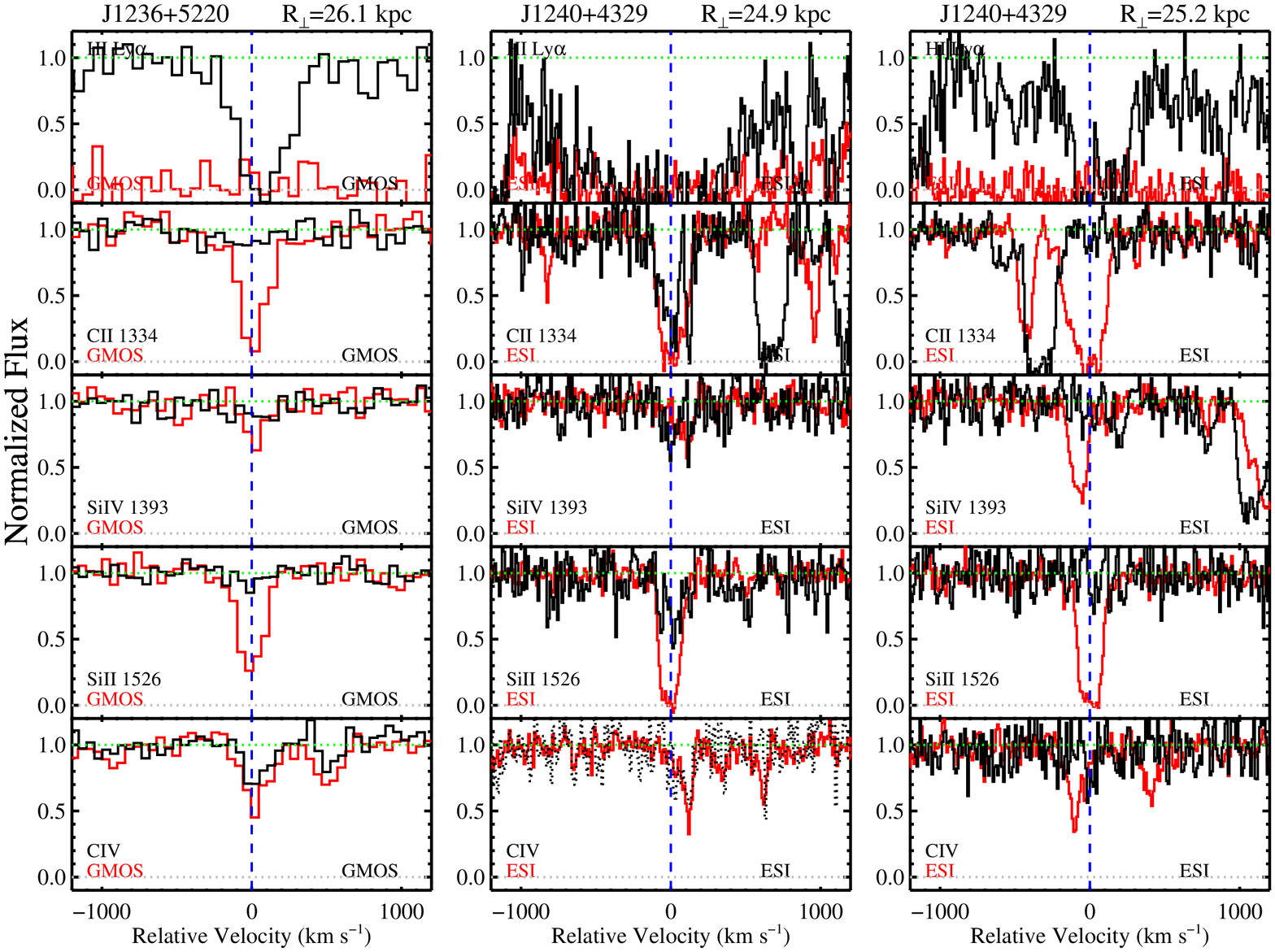}
\includegraphics[angle=90,width=0.8\textwidth]{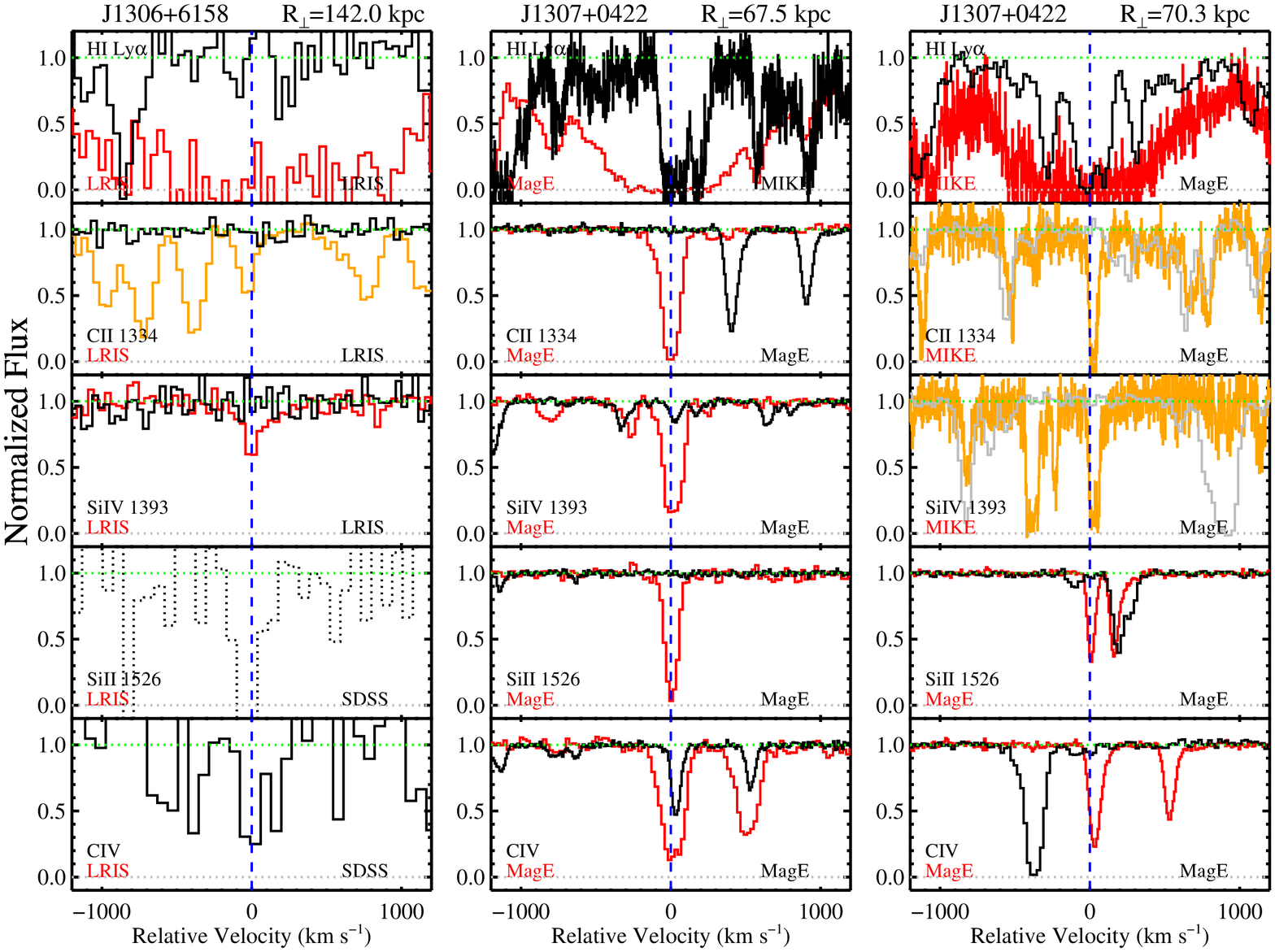}
\caption[]{ -- continued
\label{fig.showspecs5}}
\end{center}
\end{figure*}

\setcounter{figure}{0}
\begin{figure*}%[ht]
\begin{center}
\includegraphics[angle=90,width=0.8\textwidth]{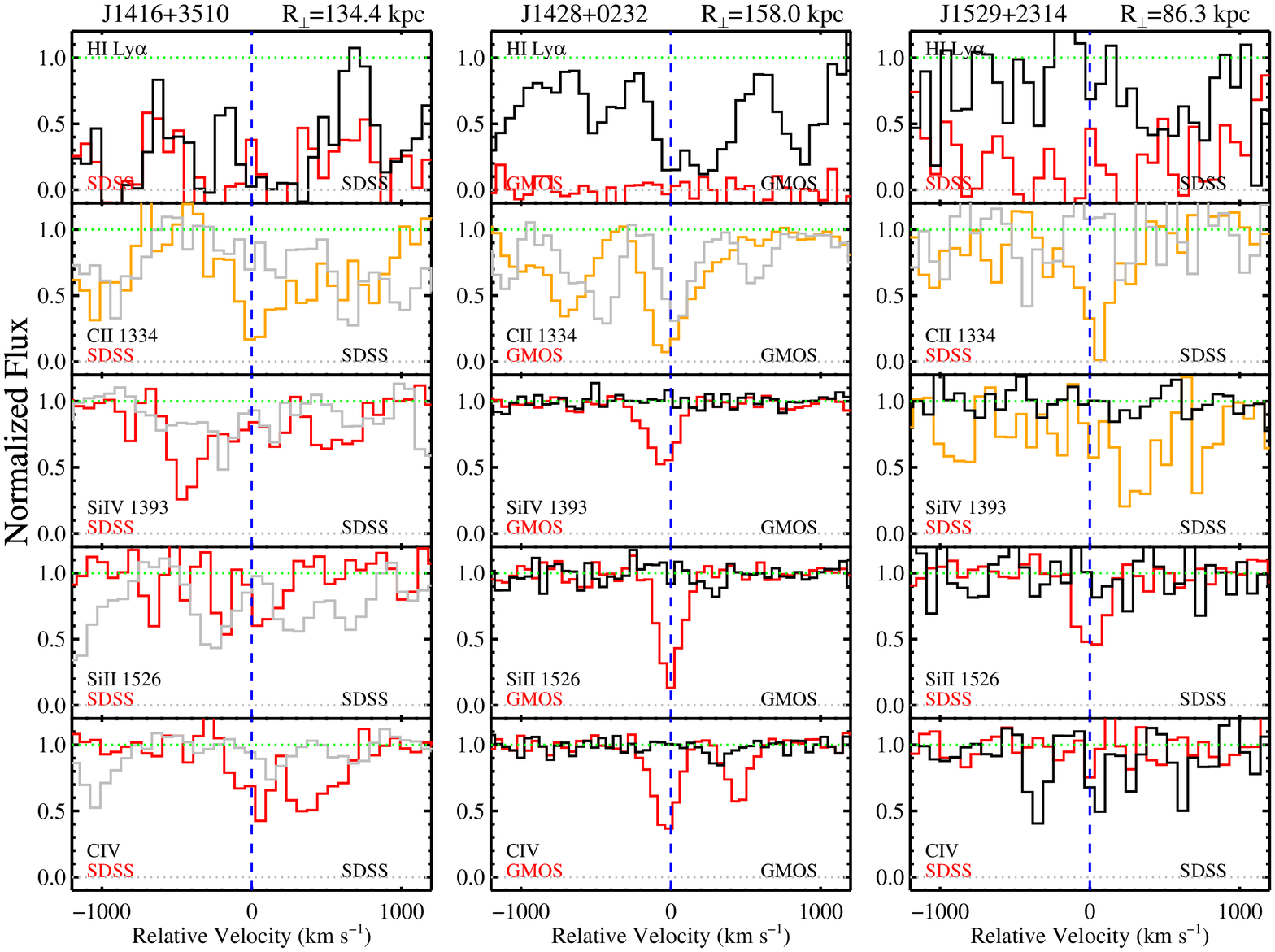}
\includegraphics[angle=90,width=0.8\textwidth]{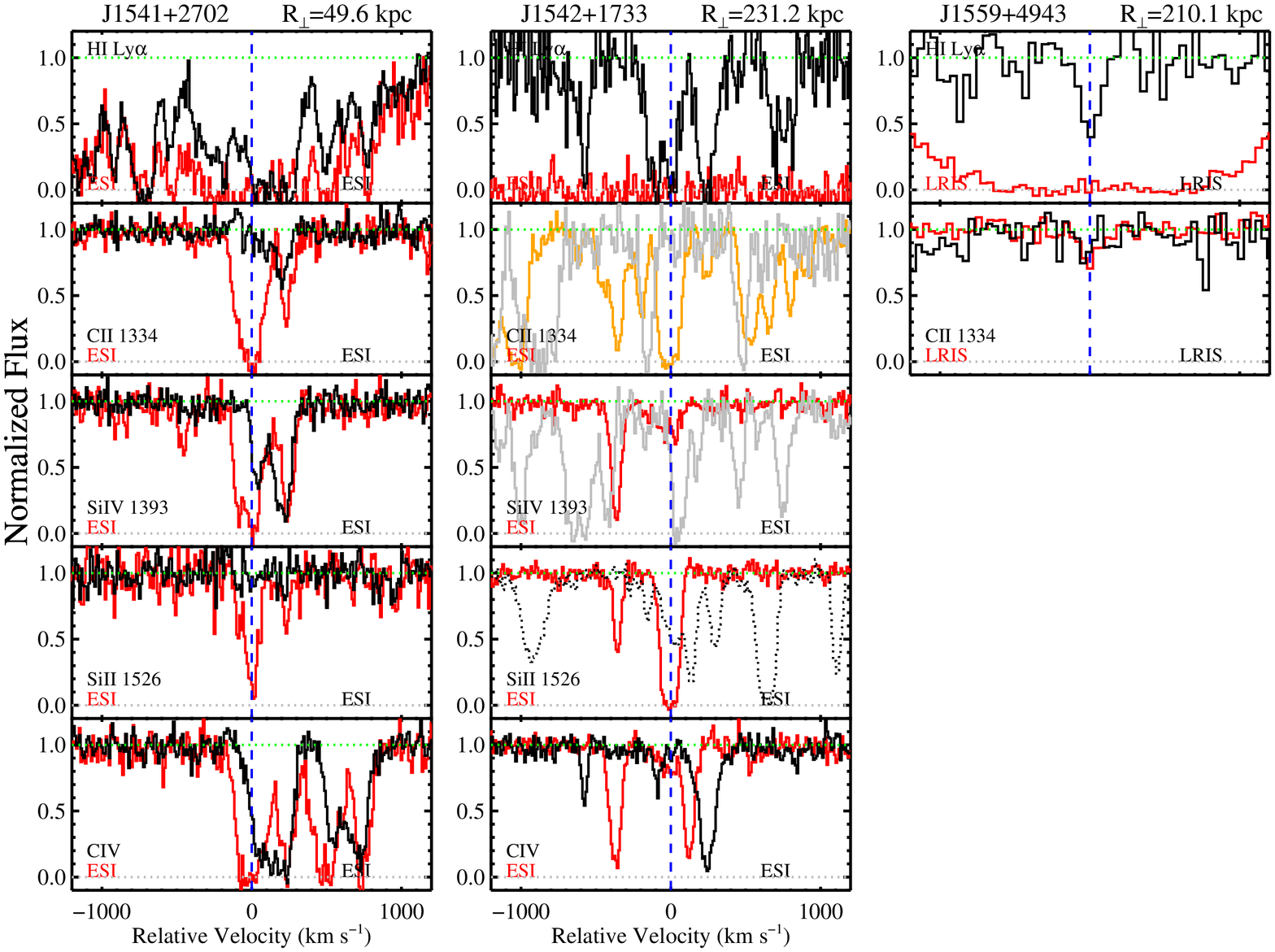}
\caption[]{ -- continued
\label{fig.showspecs6}}
\end{center}
\end{figure*}

\setcounter{figure}{0}
\begin{figure*}%[ht]
\begin{center}
\includegraphics[angle=90,width=0.8\textwidth]{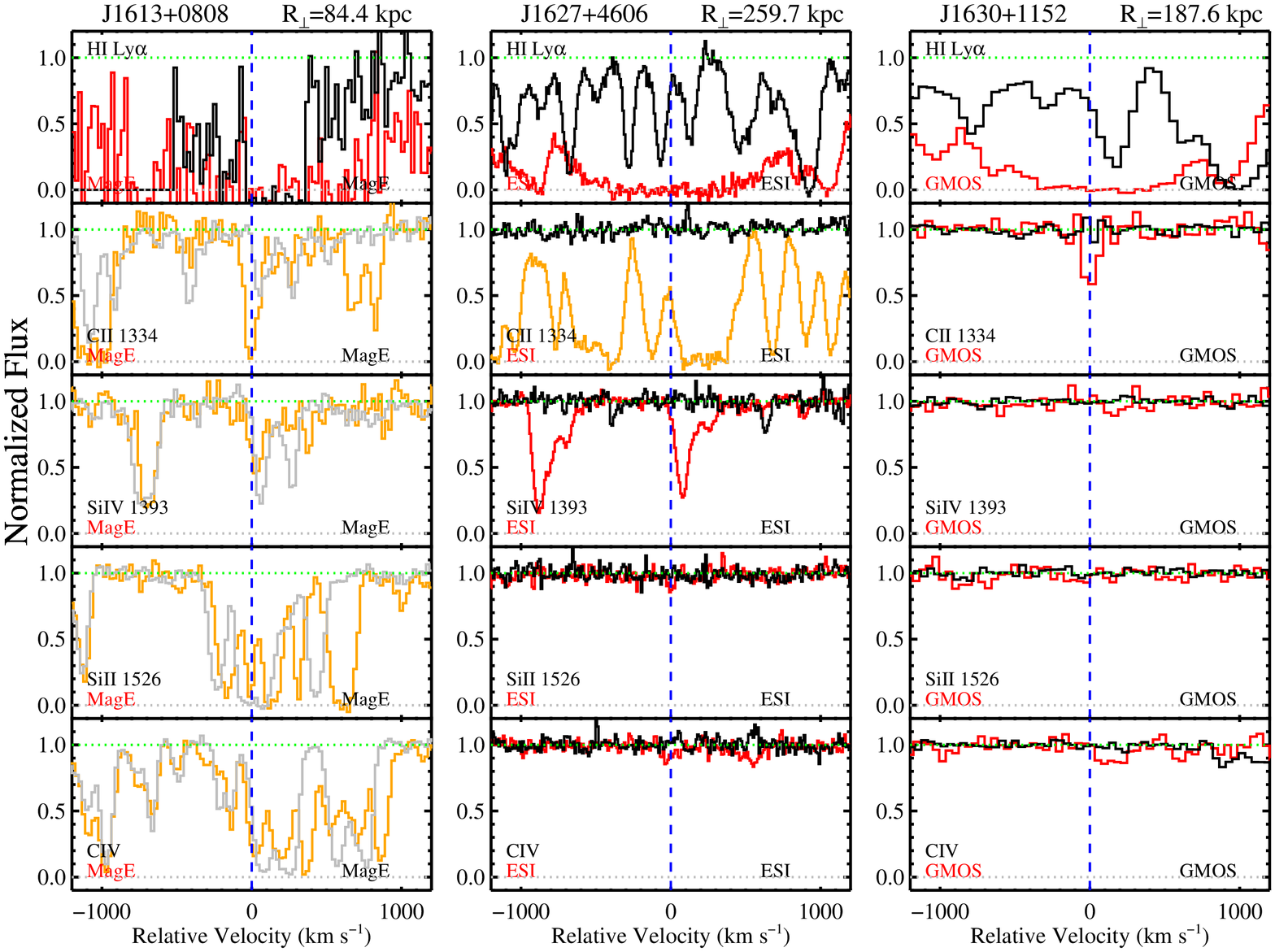}
\includegraphics[angle=90,width=0.8\textwidth]{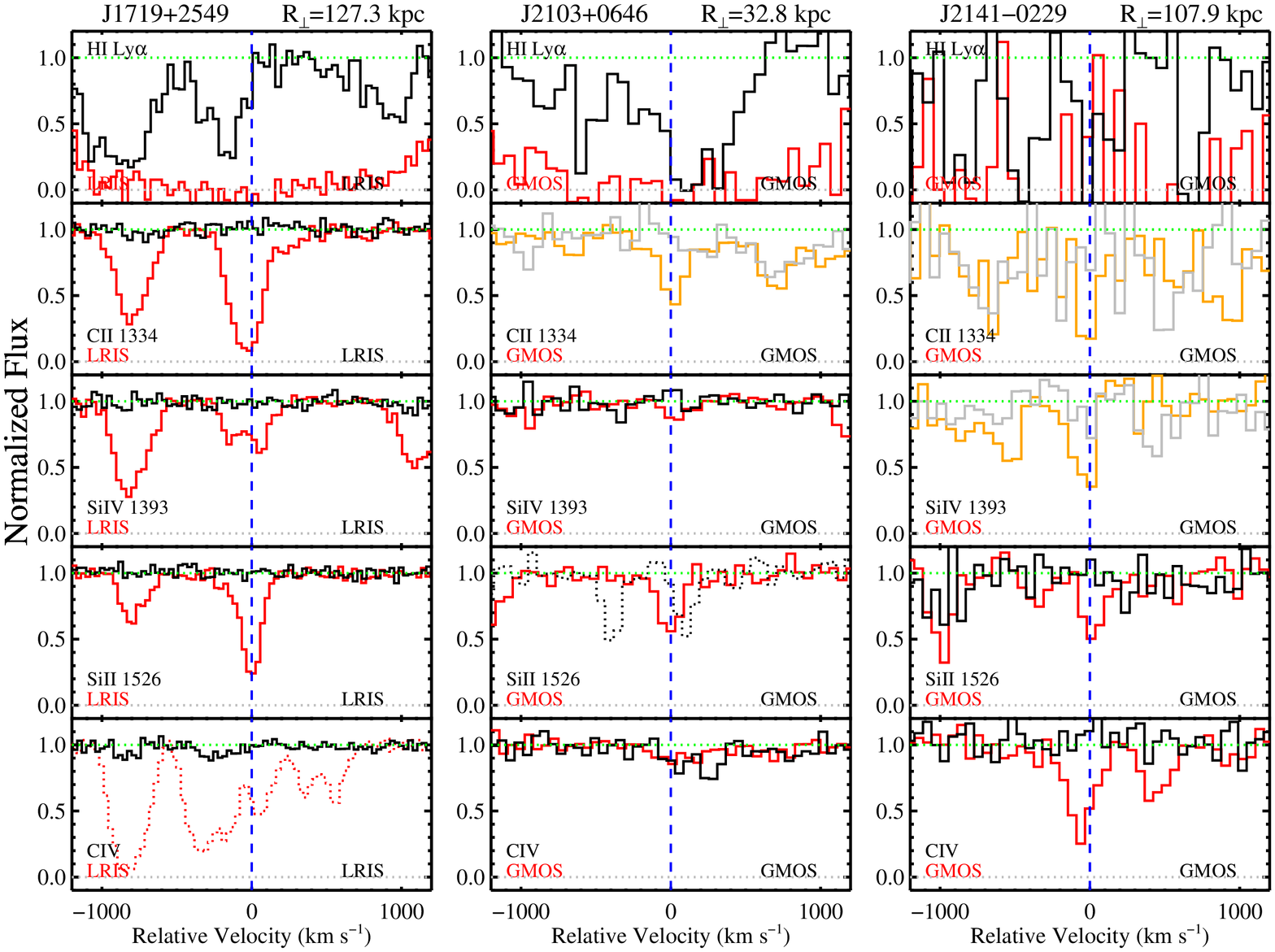}
\caption[]{ -- continued
\label{fig.showspecs7}}
\end{center}
\end{figure*}

\setcounter{figure}{0}
\begin{figure*}%[ht]
\begin{center}
\includegraphics[angle=90,width=0.8\textwidth]{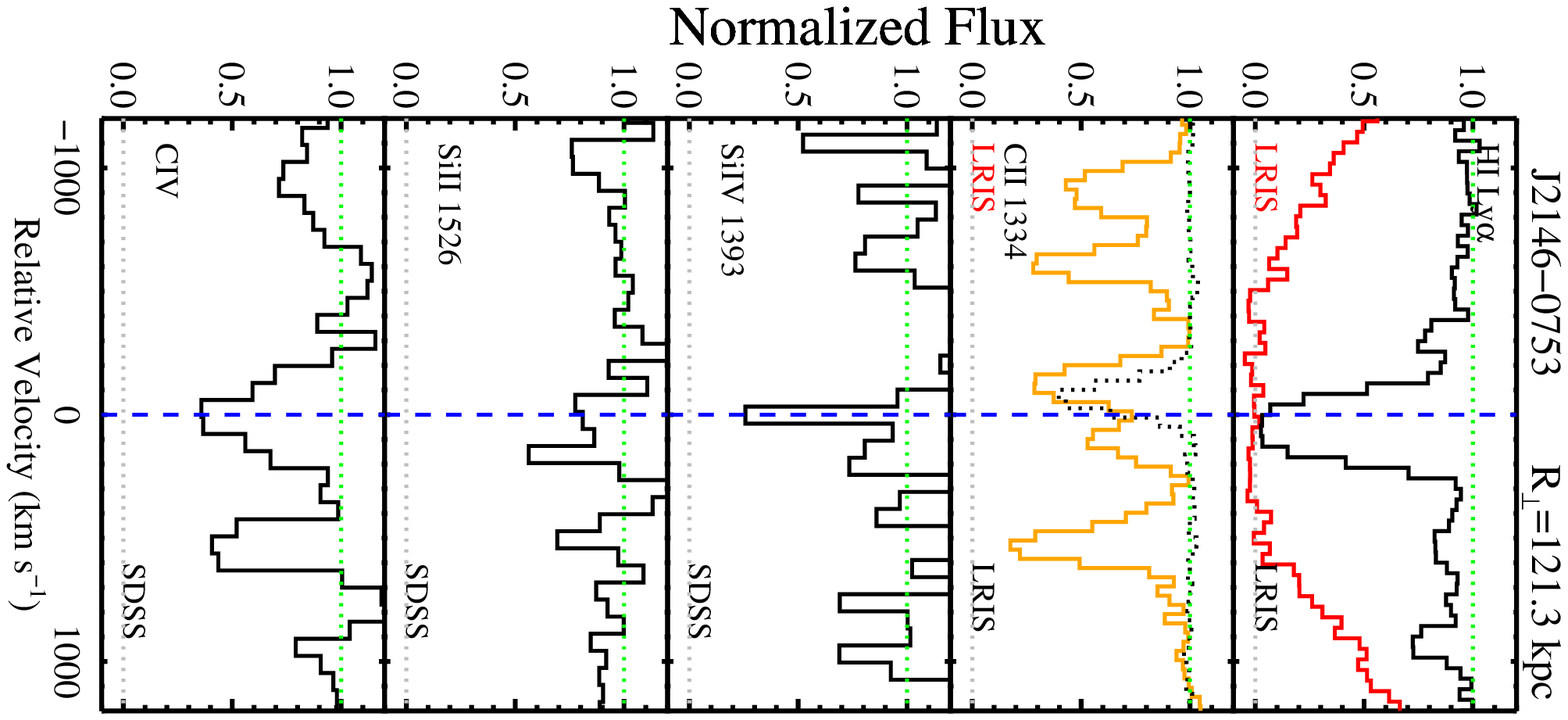}
\caption[]{ -- continued
\label{fig.showspecs8}}
\end{center}
\end{figure*}

\end{document}